\documentclass{aa}

\newcommand{\rev}[1]{\textcolor{black}{#1}}
\newcommand{\revsec}[1]{\textcolor{black}{#1}}
\newcommand{\revthi}[1]{\textcolor{black}{#1}}
\newcommand{\revfif}[1]{\textcolor{black}{#1}}

\newcommand{\revsix}[1]{\textcolor{black}{#1}}
\newcommand{\revsev}[1]{\textcolor{black}{#1}}
\newcommand{\reveig}[1]{\textcolor{black}{#1}}

\newif\iffigure
\figurefalse
\figuretrue

\usepackage{graphicx,natbib,url,twoopt}
\usepackage[varg]{txfonts}           
\usepackage{hyperref}              
\usepackage{pdfcomment}              
\usepackage{acronym}                 
\usepackage{comment}
\usepackage{cases}
\usepackage{empheq}                  
\usepackage{here}
\usepackage{bm}
\usepackage[version=3]{mhchem}
\hypersetup{
 colorlinks=true,%
 linkcolor=blue,
 citecolor=blue,
}

\usepackage{lineno}
\newcommand*\patchAmsMathEnvironmentForLineno[1]{
  \expandafter\let\csname old#1\expandafter\endcsname\csname #1\endcsname
  \expandafter\let\csname oldend#1\expandafter\endcsname\csname end#1\endcsname
  \renewenvironment{#1}
     {\linenomath\csname old#1\endcsname}
     {\csname oldend#1\endcsname\endlinenomath}}
\newcommand*\patchBothAmsMathEnvironmentsForLineno[1]{
  \patchAmsMathEnvironmentForLineno{#1}
  \patchAmsMathEnvironmentForLineno{#1*}}
\AtBeginDocument{
\patchBothAmsMathEnvironmentsForLineno{equation}
\patchBothAmsMathEnvironmentsForLineno{align}
\patchBothAmsMathEnvironmentsForLineno{flalign}
\patchBothAmsMathEnvironmentsForLineno{alignat}
\patchBothAmsMathEnvironmentsForLineno{gather}
\patchBothAmsMathEnvironmentsForLineno{multline}
}

\bibpunct{(}{)}{;}{a}{}{,}    

\makeatletter
\newcommand{\bibnote}[2]{\global\@namedef{#1note}{#2}}
\newcommand{\biblink}[2]{\global\@namedef{#1link}{#2}}
\newcommand{\Tabref}[1]{Table~\ref{#1}}
\newcommand{\Equref}[1]{Eq.~(\ref{#1})}
\newcommand{\Figref}[1]{Fig.~\ref{#1}}
\makeatother

\makeatletter
 \newcommandtwoopt{\citeads}[3][][]{%
   \nonstopmode
   \href{http://adsabs.harvard.edu/abs/#3}%
        {\def\hyper@linkstart##1##2{}%
         \let\hyper@linkend\@empty\citealp[#1][#2]{#3}}
   \biblink{#3}{\href{http://adsabs.harvard.edu/abs/#3}{ADS}}%
   \errorstopmode}            
 \newcommandtwoopt{\citepads}[3][][]{%
   \nonstopmode
   \href{http://adsabs.harvard.edu/abs/#3}%
        {\def\hyper@linkstart##1##2{}%
         \let\hyper@linkend\@empty\citep[#1][#2]{#3}}
   \biblink{#3}{\href{http://adsabs.harvard.edu/abs/#3}{ADS}}%
   \errorstopmode}            
 \newcommandtwoopt{\citetads}[3][][]{%
   \nonstopmode
   \href{http://adsabs.harvard.edu/abs/#3}%
        {\def\hyper@linkstart##1##2{}%
         \let\hyper@linkend\@empty\citet[#1][#2]{#3}}
   \biblink{#3}{\href{http://adsabs.harvard.edu/abs/#3}{ADS}}%
   \errorstopmode}            
 \newcommandtwoopt{\citeyearads}[3][][]{%
   \nonstopmode
   \href{http://adsabs.harvard.edu/abs/#3}%
        {\def\hyper@linkstart##1##2{}%
         \let\hyper@linkend\@empty\citeyear[#1][#2]{#3}}
   \biblink{#3}{\href{http://adsabs.harvard.edu/abs/#3}{ADS}}%
   \errorstopmode}            
\makeatother

\newacro{ADS}{Astrophysics Data System}
\newacro{NLTE}{non-local thermodynamic equilibrium}
\newacro{NASA}{National Aeronautics and Space Administration}

\defcitealias{Kuwahara:2020}{Paper {\rm I}}

\begin{document}
\authorrunning{A. Kuwahara et al.}
\titlerunning{\revsec{Analytic description of the planet-induced gas flow}}
   \title{\revsec{Analytic description of the gas flow around planets embedded in protoplanetary disks}}
   \subtitle{}
      \author{Ayumu Kuwahara\inst{1,2,3}
          \thanks{\email{ayumu.kuwahara@sund.ku.dk}} 
          \and Hiroyuki Kurokawa\inst{\revthi{4,3}}}

   \institute{\revsec{Center for Star and Planet Formation, GLOBE Institute, University of Copenhagen, Øster Voldgade 5-7, 1350 Copenhagen, Denmark}
        \and
              Department of Earth and Planetary Sciences, Tokyo Institute of Technology, \revsev{2-12-1 Ookayama, Meguro-ku, Tokyo 152-8551, Japan}
         \and
             Earth-Life Science Institute, Tokyo Institute of Technology, \revsev{2-12-1 Ookayama, Meguro-ku, Tokyo \reveig{152-8550}, Japan}
         \and Department of Earth Science and Astronomy, Graduate School of Arts and Sciences, The University of Tokyo\revsev{, 3-8-1 Komaba, Meguro-ku, Tokyo 153-8902, Japan}}

   \date{Received September XXX; accepted YYY}

 
  \abstract
 {\revsix{A growing planet embedded in a protoplanetary disk induces three-dimensional gas flow, \revsev{which exhibits} a midplane outflow that can suppress dust accretion onto the planet and form global dust substructures (rings and gaps).}}
   {\revsec{Because analytic formulae for the planet-induced outflow \revsix{are} useful for modeling its influences on local and global dust surface densities and planet accretion, we derive} \rev{the analytic formulae that describe the morphology and velocity of the planet-induced outflow.}}
   {\rev{We first perform three-dimensional, nonisothermal hydrodynamical simulations of the gas flow past a planet, \revsec{which enables} us to introduce a fitting formula describing the morphology of the outflow. We then derive an analytic formula for the outflow speed using Bernoulli’s theorem.}}
   {\revsec{We successfully derived a fitting formula for the midplane outflow morphology (\revsec{the shape of the streamline}), which is valid when \revthi{the dimensionless thermal mass falls below $m\lesssim0.6$.}} \rev{\revsec{The obtained analytic formulae for the outflow, such as the maximum outflow speed and the velocity distributions of the outflow in the radial and vertical directions to the disk, \revthi{show good agreement}} with the numerical results. \revsec{We find} the following trends: (1) the \revsec{maximum} outflow speed increases with the planetary mass and has a peak of $\sim30\text{--}40\%$ of the sound speed when the dimensionless thermal mass is $m\sim0.3$, corresponding to a super-Earth mass planet at $1$ au for the typical steady accretion disk model, and (2) \revthi{the presence of the headwind (namely, the global pressure force acting in the positive radial direction of the disk) enhances (reduces) the outflow toward the outside (inside) of the planetary orbit.}}}
   {\revsec{The planet-induced outflow of the gas affects the dust motion when the dimensionless stopping time of dust falls below ${\rm St}\lesssim\min(10m^2,0.1)$, which could be used for modeling of the dust velocity influenced by the outflow.}}
   

   \maketitle
%

\section{Introduction}\label{sec:Introduction}
\rev{Planets form in protoplanetary disks and interact with the surrounding gas and dust. The disk-planet interaction has been studied for a wide range of planetary mass. \revsix{Planets with typically $\gtrsim15\,M_{\oplus}$ (Earth masses) can open a gas gap around their orbit\revsev{s} \citep[e.g.,][]{duffell2013gap} and form substructures in the distribution of dust in disks, such as dust rings and gaps \citep[e.g.,][]{paardekooper2004planets}}}.



\revsec{Low-mass (typically $\sim0.1\text{--}10\,M_\oplus$) planets do not open a deep gas gap, but those planets embedded in protoplanetary disks also interact with surrounding disk gas and dust.} \revsev{When the physical radius of a planet exceeds the Bondi radius, $R_{\rm Bondi}=GM_{\rm pl}/c_{\rm s}^2$, the planet starts to attract a disk gas, where $G$ is the gravitational constant, $M_{\rm pl}$ is the mass of the planet, and $c_{\rm s}$ is the sound speed. In this study, we define the disk gas within the Bondi radius as the envelope.} Recent three-dimensional (3D) hydrodynamical simulations \revsec{exhibited} the complex structure of the perturbed gas flow \revsev{around the envelope \citep[the planet-induced gas flow; e.g.,][]{Ormel:2015b,Fung:2015}}. 
Gas from the disk flows into the \revsev{envelope} \revsix{through the poles} 
 and returns to the disk as the outflow through the midplane, which is referred to as \revthi{(the) recycling flow} \citep{Ormel:2015b}. \revsev{We note that though the pattern of the recycling flow is characterized by the polar inflow and the midplane outflow in previous studies \citep{Ormel:2015b,Fung:2015,Fung:2019,Lambrechts:2017,Cimerman:2017,Kurokawa:2018,Popovas:2018a,Kuwahara:2019,Bethune:2019,moldenhauer2021steady,moldenhauer2022recycling}, the inflow from the midplane, sometimes accompanied by the polar outflow, has been identified in a specific situation where the planet has a large luminosity due to accretion of solids \citep{Chrenko:2019} or \revsev{experiences the strong headwind of the gas} \citep{Ormel:2015b,Kurokawa:2018,bailey2021three}.} 

\revsec{The morphology and velocity of the planet-induced outflow are crucial to investigate the hydrodynamic effects of the planet-induced outflow. In this study, we \revthi{use} the word morphology to describe the shapes of the streamlines.} \rev{The midplane outflow induced by an embedded planet, which occurs in the radial direction to the disk, affects the motion of $\lesssim$millimeter(mm)- to centimeter(cm)-sized small dust grains. \revthi{Because these small dust grains tend to follow the gas streamlines, their accretion rates} onto an embedded planet with $\sim1\,M_\oplus$ \revthi{are} reduced by orders of magnitude due to the outflow \citep{Popovas:2018a,Popovas:2018b,Kuwahara:2020a,Kuwahara:2020b,okamura2021growth}. The radial drift of $\lesssim$cm-sized dust is disturbed both inside and outside the planetary orbit by the outflow induced by the embedded planet \revsec{$(\gtrsim1\,M_\oplus)$\footnote{\revsec{Precisely the mass range is dependent on parameters such as the orbital radius; see Sect. \ref{sec:Numerical simulations}.}}}, \revthi{which leads} to the formation of the dust ring and gap with \revsix{their} radial extent of $\sim1\text{--}10$ times the gas scale height in a disk \citep{kuwahara2022dust}.}


\revsec{Few studies} investigated the gas flow \rev{speed} perturbed by an embedded planet. \cite{Ormel:2013} considered two-dimensional (2D), isothermal, and invisid disk gas and derived an approximate analytic solution of the gas velocity. Assuming an isothermal gas disk, \cite{Fung:2015} performed 3D hydrodynamical simulations and found that the outflow of the gas induced by \revthi{an} \rev{embedded} planet at the midplane has a speed of \revthi{$\sim20\text{--}40$\%} of sound speed. \cite{Kuwahara:2019} assumed an isothermal, Keplerian gas disk and performed 3D hydrodynamical simulations and investigated the dependence of the outflow speed on the planetary mass. The authors found that the outflow speed increases with the planetary mass. 

\revsec{There are no comprehensive studies investigating the flow speed in 3D for a wide range of parameters, namely the planetary mass and the magnitude of the deviation of the disk gas rotation from Keplerian rotation. The sub-Keplerian motion of the gas due to the global pressure gradient in the disk alters the 3D structure of the planet-induced gas flow \citep{Ormel:2015b,Kurokawa:2018}.} \rev{Three-dimensional hydrodynamical simulations are computationally expensive, making it difficult to incorporate the hydrodynamic effects of the perturbed gas flow into models of planet formation and dust substructure formation. Thus, construction of an analytic model of the planet-induced gas flow is \revthi{demanded} for understanding gas-dust-planet interactions.} 

\rev{In this study, we aim for construction of an analytic model of the outflow induced by an embedded planet, \revsec{guided by numerical simulations for an extensive parameter space}. We first perform 3D, nonisothermal hydrodynamical simulations (Sect. \ref{sec:Numerical simulations}). We then \revsec{derive} the analytic formulae of the flow field based on the hydro-simulations results \revsec{obtained} in Sect. \ref{sec:Derivation of analytic formula}. } 
\revsec{In Sect. \ref{sec:Comparison to numerical results}, we compare our analytic formula for the outflow speed with the numerical results.} In Sect. \ref{sec:Discussion}, we discuss the implications for the dust motion influenced by the planet-induced gas flow. We conclude in Sect. \ref{sec:Conclusions}.  
\revsec{For readers who \revthi{want to} quickly go through our key results, \Figref{fig:vx_with_analytic} in Sect. \ref{sec:Comparison to numerical results} summarizes our main result in which the maximum outflow speed obtained from hydrodynamical simulations and an analytic formula for the maximum outflow speed (\Equref{eq:v_x,out}) are shown.}

\section{\rev{Numerical simulations}}\label{sec:Numerical simulations}

\begin{figure}[htbp]
    \centering
    \includegraphics[width=\linewidth]{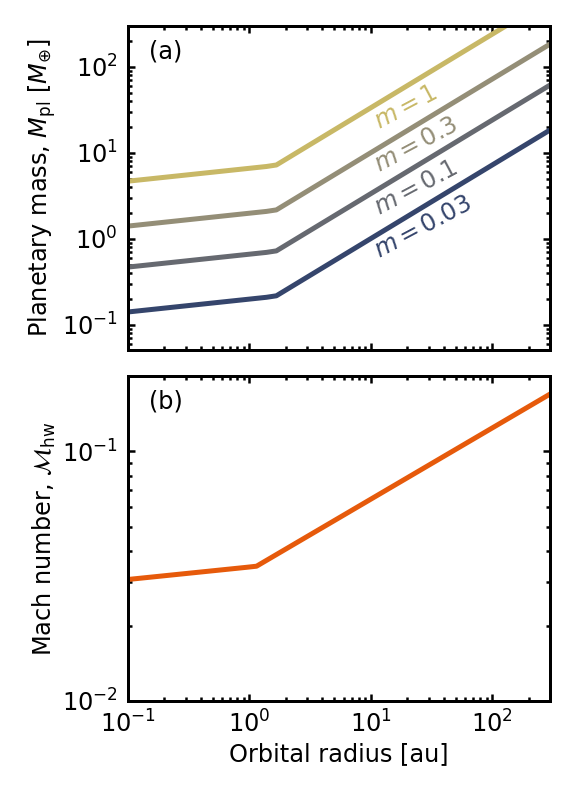}
    \caption{\rev{\textit{Top panel:} relation between the dimensionless dimensionless planetary mass and dimensional ones as a function of the orbital radius. \textit{Bottom panel:} Mach number as a function of the orbital radius. For the conversion, we assumed a typical steady accretion disk model with a \revsec{dimensionless viscous alpha parameter} \citep{Shakura:1973}, $\alpha_{\rm acc}$, including viscous heating and stellar irradiation \citep[e.g.,][]{Ida:2016}. In this model, the aspect ratio is given by $h=\max(0.027\,(a/1\,\text{au})^{1/20},\,0.024\,(a/1\,\text{au})^{2/7})$, where a solar-mass star, a solar luminosity, the typical accretion rate of classical T Tauri stars, $10^{-8}\,M_\odot\text{/yr}$, and $\alpha_{\rm acc}=10^{-3}$ were assumed \revsix{(Appendix \ref{sec:Aspect ratio of the steady accretion disk})}.}}
    \label{fig:physical_quantity2panel}
\end{figure} 

\begin{table*}[htbp]
\caption{List of hydrodynamical simulations. The following columns give the simulation name, the size of the Bondi radius of the planet, the size of the Hill radius of the planet, the size of the inner boundary, the size of the outer boundary, the smoothing length, the Mach number of the headwind, the dimensionless cooling time, the injection time, and the length of the calculation time. The top row, separated by a horizontal line, represents the fiducial runs, and the bottom row represents the additional runs.}
\centering
\scalebox{0.78}{
\begin{tabular}{lcccccccccc}\hline\hline
    Name & $R_{\rm Bondi}$ & $R_{\rm Hill}$ & $r_{\rm inn}$ & $r_{\rm out}$ & $r_{\rm sm}$ & $\mathcal{M}_{\rm hw}$ & $\beta$ & $t_{\rm inj}$ & $t_{\rm end}$\\ \hline
    \texttt{m003-hw0,-hw001,-hw003,-hw01} & 0.03 & 0.22 & 9.32$\times10^{-3}$ & 0.5  & 0.1 $R_{\rm grav}$ & 0, 0.01, 0.03, 0.1 & 0.09 & 0.5 & \revsec{100}   \\
    \texttt{m005-hw0,-hw001,-hw003,-hw01} & 0.05 & 0.26 & 1.11$\times10^{-2}$ & 1 & 0.1 $R_{\rm grav}$   & 0, 0.01, 0.03, 0.1 & 0.25 & 0.5 & 100  \\
    \texttt{m007-hw0,-hw001,-hw003,-hw01} & 0.07 & 0.29 & 1.23$\times10^{-2}$ & 1 &  0.1 $R_{\rm grav}$   & 0, 0.01, 0.03, 0.1 & 0.49 & 0.5 & 100  \\
    \texttt{m01-hw0,-hw001,-hw003,-hw01}  & 0.1  & 0.32 & 1.39$\times10^{-2}$ & 5 & 0.1 $R_{\rm grav}$   & 0, 0.01, 0.03, 0.1 & 1    & 0.5 & 200  \\
    \texttt{m02-hw0,-hw001,-hw003,-hw01}  & 0.2  & 0.41 & 1.75$\times10^{-2}$ & 5 &  0.1 $R_{\rm grav}$   & 0, 0.01, 0.03, 0.1 & 4    & 0.5 & 200  \\
    \texttt{m03-hw0,-hw001,-hw003,-hw01}  & 0.3  & 0.46 & 2$\times10^{-2}$   & 5 & 0.1 $R_{\rm grav}$   & 0, 0.01, 0.03, 0.1 & 9    & 0.5 & 1000  \\
    \texttt{m04-hw0,-hw001,-hw003,-hw01}  & 0.4  & 0.51 & 2.21$\times10^{-2}$ & 5 & 0.1 $R_{\rm grav}$   & 0, 0.01, 0.03, 0.1 & 16   & 0.5 & 1000 \\
    \texttt{m05-hw0,-hw001,-hw003,-hw01}  & 0.5  & 0.55 & 2.38$\times10^{-2}$ & 5 & 0.1 $R_{\rm grav}$   & 0, 0.01, 0.03, 0.1 & 25   & 1   & 1000 \\
    \texttt{m06-hw0,-hw001,-hw003,-hw01}  & 0.6  & 0.58 & 2.53$\times10^{-2}$ & 5 & 0.1 $R_{\rm grav}$   & 0, 0.01, 0.03, 0.1 & 36   & 1   & 1000 \\
    \texttt{m07-hw0,-hw01}  & 0.7  & 0.62 & 2.66$\times10^{-2}$ & 5 & 0.1 $R_{\rm grav}$   & 0,             0.1 & 49   & 1   & 1000 \\
    \texttt{\revsix{m08-hw0}}  & 0.8  & 0.64 & 2.78$\times10^{-2}$ & 5 & 0.1 $R_{\rm grav}$   & \revsix{0} & 64   & 1   & 1000 \\
    \texttt{\revsix{m09-hw0}}  & 0.9  & 0.67 & 2.9$\times10^{-2}$  & 5 & 0.1 $R_{\rm grav}$   & \revsix{0} & 81   & 1   & 1000 \\
    \texttt{\revsix{m1-hw0}}   & 1    & 0.69 & 3$\times10^{-2}$    & 5 & 0.1 $R_{\rm grav}$   & \revsix{0} & 100  & 1   & 1000 \\\hline
    \texttt{m003-hw001-01rinn,-hw003-01rinn,-hw01-01rinn}  & 0.03 & 0.22 & 9.32$\times10^{-4}$ & 0.5  & 0.1 $R_{\rm grav}$ & 0.01, 0.03, 0.1 & 0.09 & 0.5 & 50 \\
    \texttt{m003-hw0-b001,-hw0-b01,-hw0-b10,-hw0-b100} & 0.03 & 0.22 & 9.32$\times10^{-3}$ & 0.5  & 0.1 $R_{\rm grav}$ & 0 & 0.0009, 0.009, 0.9, 9 & 0.5 & \revsec{100}   \\
    \texttt{m005-hw0-b001,-hw0-b01,-hw0-b10,-hw0-b100} & 0.05 & 0.26 & 1.11$\times10^{-2}$ & 1 & 0.1 $R_{\rm grav}$   & 0 & 0.0025, 0.025, 2.5, 25 & 0.5 & 100  \\
    \texttt{m007-hw0-b001,-hw0-b01,-hw0-b10,-hw0-b100} & 0.07 & 0.29 & 1.23$\times10^{-2}$ & 1 &  0.1 $R_{\rm grav}$   & 0 & 0.0049, 0.049, 4.9, 49 & 0.5 & 100  \\
    \texttt{m01-hw0-b001,-hw0-b01,-hw0-b10,-hw0-b100}  & 0.1  & 0.32 & 1.39$\times10^{-2}$ & 5 & 0.1 $R_{\rm grav}$   & 0 & 0.01, 0.1, 10, 100    & 0.5 & 200  \\
    \texttt{m02-hw0-b001,-hw0-b01,-hw0-b10,-hw0-b100}  & 0.2  & 0.41 & 1.75$\times10^{-2}$ & 5 &  0.1 $R_{\rm grav}$   & 0 & 0.04, 0.4, 40, 400    & 0.5 & 200  \\
    \texttt{m03-hw0-b001,-hw0-b01,-hw0-b10,-hw0-b100}  & 0.3  & 0.46 & 2$\times10^{-2}$   & 5 & 0.1 $R_{\rm grav}$   & 0 & 0.09, 0.9, 90, 900    & 0.5 & 1000  \\
    \texttt{m04-hw0-b001,-hw0-b01,-hw0-b10,-hw0-b100}  & 0.4  & 0.51 & 2.21$\times10^{-2}$ & 5 & 0.1 $R_{\rm grav}$   & 0 & 0.16, 1.6, 160, 1600   & 0.5 & 1000 \\
    \begin{tabular}[c]{@{}l@{}}\texttt{m003-hw0-rsm0},\\ \texttt{-hw001-rsm0},\\ \texttt{-hw003-rsm0},\\\texttt{-hw01-rsm0} \end{tabular}  & 0.01 & 0.15 & 6.46$\times10^{-3}$ & 0.5 & 0 & 0, 0.01, 0.03, 0.1 & $9\times10^{-4}$ & 0.5 & 10   \\
    \begin{tabular}[c]{@{}l@{}}\texttt{m01-hw0-rsm0(-rsm3rinn)},\\ \texttt{-hw001-rsm0(-rsm3rinn)},\\ \texttt{-hw003-rsm0(-rsm3rinn)},\\\texttt{-hw01-rsm0(-rsm3rinn)} \end{tabular}  & 0.1  & 0.32 & 1.39$\times10^{-2}$ & 5 & 0, $3\,r_{\rm inn}$   & 0, 0.01, 0.03, 0.1 & 1    & 0.5 & 200  \\
    \begin{tabular}[c]{@{}l@{}}\texttt{m02-hw0-rsm0},\\ \texttt{-hw001-rsm0},\\ \texttt{-hw003-rsm0},\\\texttt{-hw01-rsm0} \end{tabular}  & 0.2  & 0.41 & 1.75$\times10^{-2}$ & 5 & 0                  & 0, 0.01, 0.03, 0.1 & 4    & 0.5 & 200  \\
    \begin{tabular}[c]{@{}l@{}}\texttt{m03-hw0-rsm0},\\ \texttt{-hw001-rsm0},\\ \texttt{-hw003-rsm0},\\\texttt{-hw01-rsm0} \end{tabular}  & 0.3  & 0.46 & 2$\times10^{-2}$   & 5 & 0    & 0, 0.01, 0.03, 0.1 & 9    & 0.5 & 200  \\
    \begin{tabular}[c]{@{}l@{}}\texttt{m04-hw0-rsm0},\\ \texttt{-hw001-rsm0},\\ \texttt{-hw003-rsm0},\\\texttt{-hw01-rsm0} \end{tabular}  & 0.4  & 0.51 & 2.21$\times10^{-2}$ & 5 & 0  & 0, 0.01, 0.03, 0.1 & 16   & 0.5 & 500 \\
    \begin{tabular}[c]{@{}l@{}}\texttt{m05-hw0-rsm0(-rsm3rinn)},\\ \texttt{-hw001-rsm0(-rsm3rinn)},\\ \texttt{-hw003-rsm0(-rsm3rinn)},\\\texttt{-hw01-rsm0(-rsm3rinn)} \end{tabular}  & 0.5  & 0.55 & 2.38$\times10^{-2}$ & 5 & 0 , $3\,r_{\rm inn}$ & 0, 0.01, 0.03, 0.1 & 25   & 1   & 500 \\\hline
\end{tabular}}
\label{tab:hydro simulations}
\end{table*}

\rev{We first performed \rev{hydrodynamical simulations for broad parameter ranges.} The obtained hydro-simulations results were used to construct an \revthi{analytic} model of the flow field (Sect. \ref{sec:Derivation of analytic formula}).} Our methods of hydrodynamical simulations are the same as described in \cite{Kuwahara:2020a,Kuwahara:2020b} and \cite{kuwahara2022dust}, but this study handles the broader parameter \revsec{spaces than these previous works in terms of the planetary mass and the headwind speed caused by the global pressure gradient} \rev{to derive an \revthi{analytic} formula for the outflow speed}. Here we summarize the key points of our \rev{numerical} model.

\subsection{Dimensionless unit system}
In this study, the length, times, velocities, and densities are normalized by the disk gas scale height, $H$, reciprocal of the orbital frequency, $\Omega^{-1}$, \revsix{isothermal} sound speed, $c_{\rm s}$, and unperturbed gas density at the location of the planet, $\rho_\infty$, respectively. We introduce the dimensionless thermal mass of the planet as \citep{Ormel:2015a}:
\begin{align}
    m\equiv \frac{R_{\rm Bondi}}{H}=\frac{M_{\rm pl}}{M_{\rm th}},
\end{align}
where $M_{\rm th}=M_\ast h^3$ is the thermal mass of the planet \citep{goodman2001planetary}, $M_\ast$ is the stellar mass, and $h$ is the aspect ratio of the disk, respectively. \revsix{In this study, $h$ is used only when we convert the dimensionless quantities into dimensional ones. We describe the aspect ratio for a typical steady accretion disk model in Appendix \ref{sec:Aspect ratio of the steady accretion disk}.} The Hill radius is given by $R_{\rm Hill}=(m/3)^{1/3}\,H$. 
We considered $m=0.03\text{--}1$ in this study, \revsec{which corresponds to \revthi{planets} with $M_{\rm pl}\simeq0.2\text{--}6.6\,M_\oplus$ ($M_{\rm pl}\simeq1\text{--}33\,M_\oplus$) orbiting a solar-mass star at $1$ au ($10$ au; \Figref{fig:physical_quantity2panel}a).} We considered the planet on a fixed circular orbit \revsec{\revthi{revolving with} the Keplerian speed, $v_{\rm K}$.}

The planet experiences a headwind of the gas because the disk gas rotates at a different speed from the Keplerian speed due to the global pressure gradient. The deviation of the rotation speed of the gas from Keplerian rotation can be characterized by $-\eta v_{\rm K}$, where $\eta=-1/2(c_{\rm s}/v_{\rm K})^2(\mathrm{d}\ln p/\mathrm{d}\ln \rev{a})$ is a dimensionless quantity characterizing the global pressure gradient of the disk gas \citep{Nakagawa:1986} \revsec{and $a$ is the orbital radius}. We assumed $\mathrm{d}\ln p/\mathrm{d}\ln \rev{a}<0$ across the entire region of the disk. This implies that the disk gas rotates slower than the Keplerian speed (sub-Keplerian motion of the gas). We defined the Mach number of the headwind as:
\revsec{
\begin{align}
    \mathcal{M}_{\rm hw}\equiv\frac{\eta v_{\rm K}}{c_{\rm s}}.\label{eq:Mach number}
\end{align}
}We considered $\mathcal{M}_{\rm hw}=0\text{--}0.1$ in this study. \rev{We plotted \revthi{\Equref{eq:Mach number}} for a typical steady accretion disk model in \Figref{fig:physical_quantity2panel}b.}

\subsection{\rev{Methods of three-dimensional hydrodynamical simulations}}\label{sec:Methods of three-dimensional hydrodynamical simulations}
Assuming a compressible, invisid sub-Keplerian gas disk, we performed nonisothermal hydrodynamical simulations. These simulations were performed in the spherical polar coordinates, $(\rev{r},\theta,\phi)$, co-rotating with a planet. The numerical resolution was $[\log \rev{r},\theta,\phi]=[128,64,128]$. \revsev{The Bondi and Hill radii are resolved by $38\text{--}88$ and $69\text{--}101$ cells in the radial direction in our hydrodynamical simulations.} We used a free-slip condition at the inner boundary, $\rev{r}=r_{\rm inn}$. Assuming the density of the planetary core as $\rho_{\rm pl}=5\,\text{g\,cm}^{-3}$ and the solar-mass star, the size of the inner boundary was set to follow a mass-radius relationship of the planetary core \citep{Kuwahara:2020a}:
\revsec{
\begin{align}
    r_{\rm inn}=\Biggl(\frac{3\,M_{\rm pl}}{4\pi\rho_{\rm pl}}\Biggr)^{1/3}H=3\times10^{-2}\,m^{1/3}\,H\,\Biggl(\frac{a}{0.1\,\text{au}}\Biggr)^{-1}\equiv r_{\rm inn,0}.\label{eq:inner boundary}
\end{align}
\revsix{We adopted $a=0.1$ in \Equref{eq:inner boundary} in our fiducial runs.} We confirmed the size of the inner boundary does not affect our results (Appendix \ref{sec:Dependence on inner boundary}). Thus, when we convert the dimensionless quantity into a dimensional one, we can use an arbitrary orbital radius.} \revsix{As the initial state of the disk, we assumed the vertical stratification of the gas density, $\rho_{\rm g}=\rho_\infty\exp[-(z/2)^2]$. The initial gas velocity \revsev{is} $\bm{v}_{\rm g}=\bm{v}_\infty=(-3x/2-\mathcal{M}_{\rm hw})\bm{e}_y$, where $\bm{v}_{\rm g}=(v_{x,{\rm g}},v_{y,{\rm g}},v_{z,{\rm g}})$ is the gas velocity in the local Cartesian coordinates centered at the planet. The internal and total \revsev{energies} of the gas were given by $U=p/(\gamma-1)$ and $E=U+\rho_{\rm g}v_{\rm g}^2/2$, respectively, where $\gamma=1.4$ is the adiabatic index.} At the outer boundary, $\rev{r}=r_{\rm out}$, we fixed the gas density and velocity, $\rho_{\rm g}=\rho_\infty$ and $\bm{v}_{\rm g}=\bm{v}_\infty$. Hereafter we denote the $x$-, $y$-, and $z$-directions as the radially outward, azimuthal, and vertical directions to the disk, respectively.

We solved the continuity, Euler's, and energy conservation equations using Athena++ code\footnote{https://github.com/PrincetonUniversity/athena} \citep{White:2016,stone2020athena++}. The external force in the Euler's equation includes the Coriolis force, $\bm{F}_{\rm cor}=-2\bm{e}_z\times\bm{v}_{\rm g}$, the tidal force, $\bm{F}_{\rm tid}=3x\bm{e}_x-z\bm{e}_z$, the global pressure force due to the sub-Keplerian motion of the gas,  $\bm{F}_{\rm hw}=2\mathcal{M}_{\rm hw}\bm{e}_x$, and the gravitational force \citep[e.g.,][]{Ormel:2015b}:
\begin{align}
    \bm{F}_{\rm grav}=-\nabla\Biggl(\frac{m}{\sqrt{\rev{r}^2+r_{\rm sm}^2}}\Biggr)\Biggl\{1-\exp\Biggl[-\frac{1}{2}\Biggl(\frac{t}{t_{\rm inj}}\Biggr)^2\Biggr]\Biggr\},
\end{align}
where $\rev{r}=\sqrt{x^2+y^2+z^2}$ is the distance from the planet, $r_{\rm sm}$ is the smoothing length, and $t_{\rm inj}$ is the injection time. We set $r_{\rm sm}=0.1\,R_{\rm grav}$ for our fiducial runs, \rev{where $R_{\rm grav}\equiv\min(R_{\rm Bondi},R_{\rm Hill})$ is the radius of the gravitational sphere of the planet.} We investigated the effect of the smoothing length in Appendix \ref{sec:Dependence on smoothing length}.

Our simulations utilized the $\beta$ cooling model, where the radiative cooling occurs on a finite timescale, $\beta$ \citep{Gammie:2001}. The gas is heated by adiabatic compression. Following \cite{Kurokawa:2018}, we assumed that the typical scale of the temperature perturbation can be estimated by the Bondi radius, $R_{\rm Bondi}=mH$. The dimensionless cooling time can be described by: 
\begin{align}
    \beta=\Biggl(\frac{m}{0.1}\Biggr)^2\equiv\beta_0,\label{eq:beta}
\end{align}
\revsec{which represents the cooling time near the outer edge of the \revsix{envelope}, $\sim$$R_{\rm Bondi}$ \citep{Kurokawa:2018}.} We list our parameter sets in \Tabref{tab:hydro simulations}. We set $\beta=\beta_0$ for our fiducial runs. We investigated the effect of $\beta$ in Sect. \ref{sec:Dependence on cooling time}.
\subsection{\rev{Hydrodynamical simulation results}}\label{sec:Hydrodynamical simulation results}

\rev{This section summarizes the hydrodynamical simulation results. Based on the numerical results, an analytic formulation of the flow field is presented in Sect. \ref{sec:Derivation of analytic formula}.}

\begin{figure*}[htbp]
    \centering
    \includegraphics[width=\linewidth]{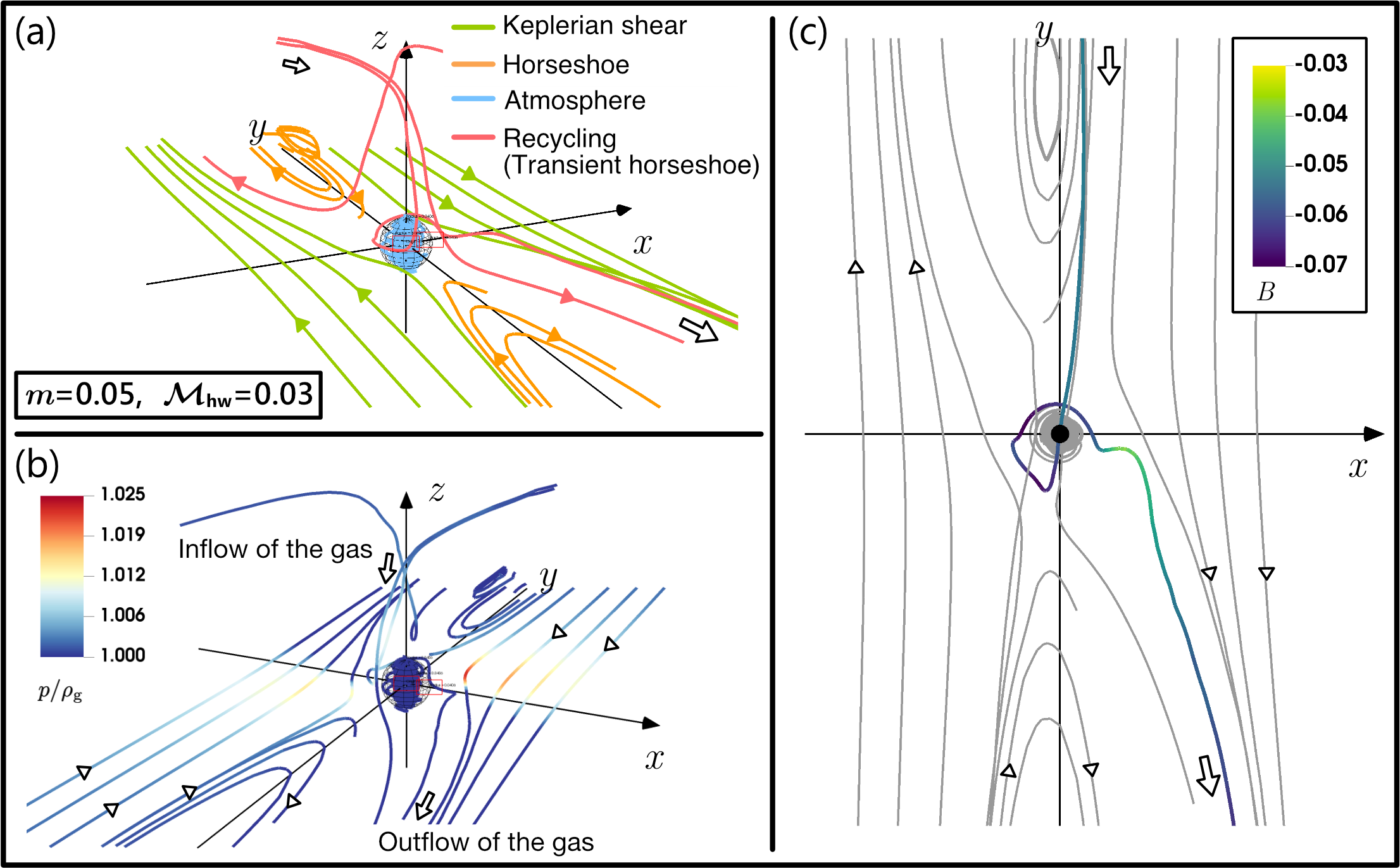}
    \caption{Streamlines of gas flow around the planet with $m=0.05$ and $\mathcal{M}_{\rm hw}=0.03$. The arrows represent the direction of the gas flow. \textit{Panel a}: Perspective view of the flow field. Different colors correspond to different types of streamlines. The size of the atmosphere is $0.04\,H$. \textit{Panel b}: Same as \textit{panel a}, but the color contour represents $p/\rho_{\rm g}$, which is the measure of temperature (Sect. \ref{sec:Bernoulli's theorem}). \textit{Panel c}: $x\text{-}y$ plane viewed from the $+z$-direction. The critical streamline is highlighted by the color contour which represents the Bernoulli's function, $\mathcal{B}$ (Sects. \ref{sec:Bernoulli's theorem} and \ref{sec:Applicability of Bernoulli's theorem}).}
    \label{fig:streamline_m005_Mhw003}
\end{figure*}

\begin{figure*}[htbp]
    \centering
    \includegraphics[width=\linewidth]{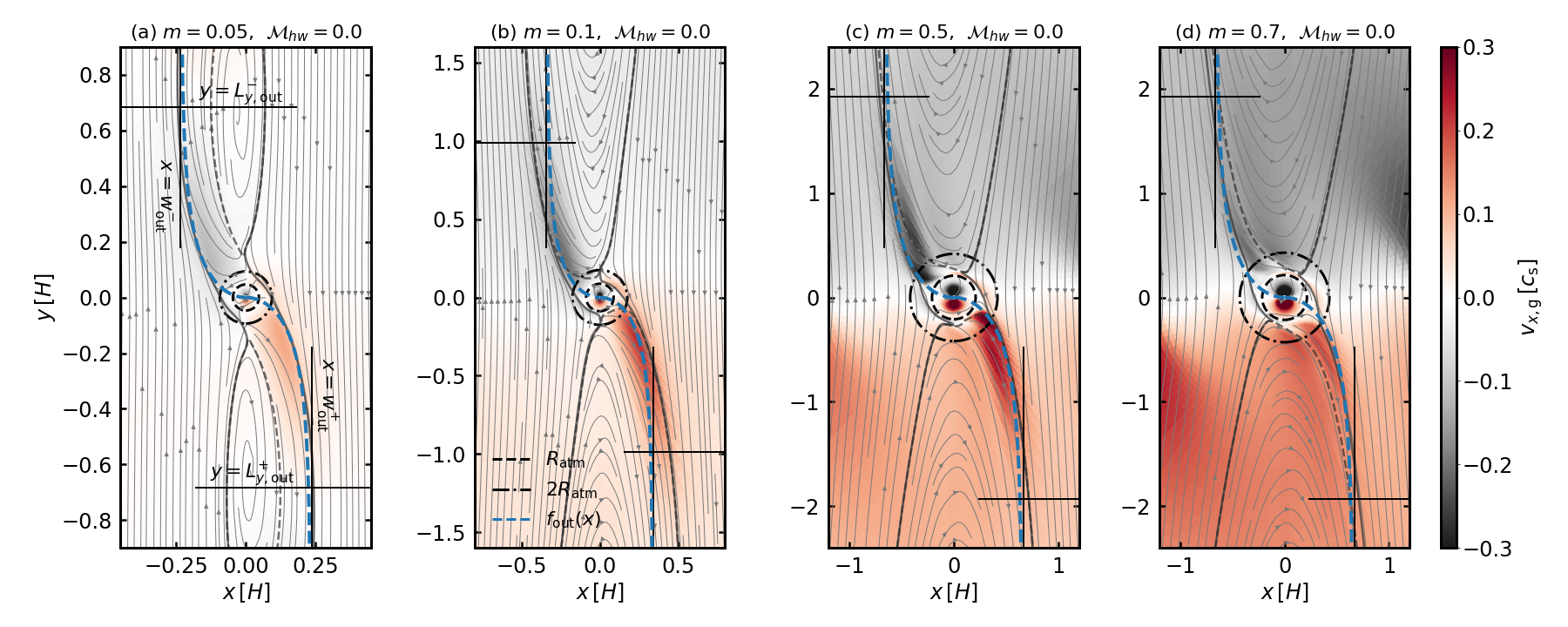}
    \caption{\rev{Dependence of the planet-induced gas flow on the planetary mass at the midplane. We set $\mathcal{M}_{\rm hw}=0$ in all panels.} The results were obtained from \texttt{m005-hw0} (\textit{panel a}), \texttt{m01-hw0} (\textit{panel b}), \texttt{m05-hw0} (\textit{panel c}), and \texttt{m07-hw0} (\textit{panel d}), respectively. The \revsev{thin} gray solid lines are gas streamlines. \revsev{The thick solid and dashed gray lines are the widest horseshoe streamline and the shear streamline passing closest to the planet, respectively (Appendix \ref{sec:Widest horseshoe streamline and the shear streamline passing closest to the planet}).} The color contour represents the flow speed in the $x$-direction. \revsec{The black dashed and dashed-dotted circles denote the atmospheric radius (\Equref{fig:Ratm}) and the twice the atmospheric radius where the outflow speed is measured.} The blue dashed lines are the curves representing  \rev{the fitting formula for the critical recycling streamline} (\Equref{eq:outflow shape in cartesian}; Sect. \ref{sec:Morphology of the critical recycling streamline at the midplane}). The vertical lines \rev{are the asymptotes of \Equref{eq:outflow shape in cartesian}, representing the extent of the outflow region in the $x$-direction} ((\Equref{eq:w_out}); Sect. \ref{sec:Width of the outflow region}). The horizontal lines denote the $y$-coordinates of the characteristic length of the outflow region in the $y$-direction ((\Equref{eq:Ly_out}); Sect. \ref{sec:Length of the outflow region}).}
    \label{fig:vx_m_dependence}
\end{figure*}

\begin{figure*}[htbp]
    \centering
    \includegraphics[width=\linewidth]{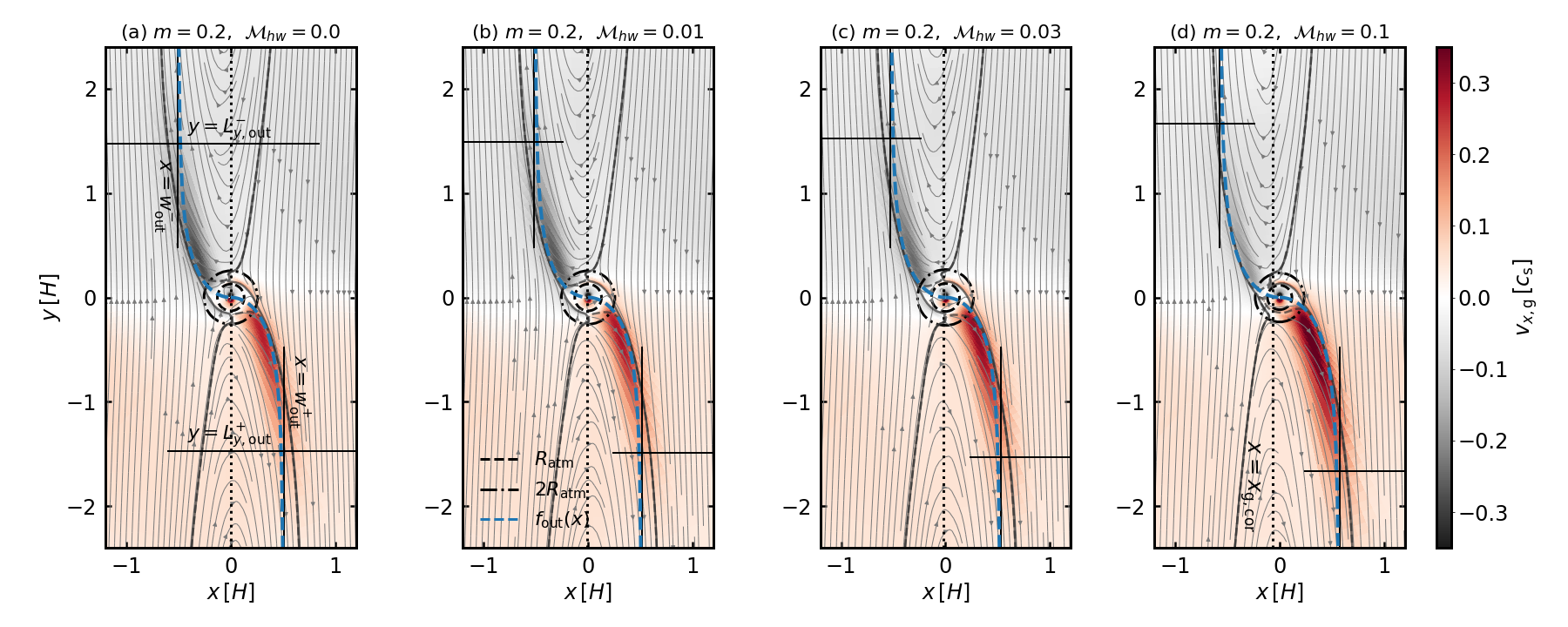}
    \caption{Same as \Figref{fig:vx_m_dependence}, but \rev{this figure shows the dependence of the planet-induced gas flow on the Mach number of the headwind. We set $m=0.2$ in all panels. The results were obtained from} \texttt{m02-hw0} (\textit{panel a}), \texttt{m02-hw001} (\textit{panel b}), \texttt{m02-hw003} (\textit{panel c}), and \texttt{m02-hw01} (\textit{panel d}), respectively. The vertical dotted line corresponds to the corotation radius for the gas.}
    \label{fig:vx_Mhw_dependence}
\end{figure*}

\subsubsection{\rev{Summary of three-dimensional planet-induced gas flow}}\label{sec:Summary of three-dimensional planet-induced gas flow}
\revsec{First, we summarize the general properties of the 3D gas flow as reported in previous studies.} Planets embedded in disks perturb the surrounding gas and induce the 3D gas flow \citep[\Figref{fig:streamline_m005_Mhw003}; e.g.,][]{Ormel:2015b,Fung:2015}. The fundamental features of the gas flow around the planet are as follows: (1) the Keplerian shear flow extends inside and outside the orbit of the planet (the green lines of \Figref{fig:streamline_m005_Mhw003}a). (2) The horseshoe flow exists in the anterior-posterior direction of the orbital path of the planet (the orange lines of \Figref{fig:streamline_m005_Mhw003}a). The horseshoe streamlines have a columnar structure in the vertical direction \citep{Fung:2015,Masset:2016}. (3) \revsev{An isolated inner envelope forms} within the gravitational sphere of the planet, $\rev{r}\lesssim R_{\rm grav}$, due to the buoyancy barrier \citep{Kurokawa:2018}. We refer to this \revsev{isolated envelope} as an atmosphere. (4) Gas from the disk enters the gravitational sphere of the planet (inflow) and exits (outflow; the red line of \Figref{fig:streamline_m005_Mhw003}a). Following \cite{Ormel:2015b} and \cite{Kuwahara:2019}, we refer to this flow as the recycling flow\footnote{This flow also referred to as the transient horseshoe flow in \cite{Fung:2015}.}. The recycling flow passes the planet, tracing the surface of the atmosphere.

The outflow of the gas occurs in the radial direction to the disk \rev{along the recycling streamlines.} \rev{The following sections (Sects. \ref{sec:Gas flow speed at the midplane: Dependence on planetary mass} and \ref{sec:Gas flow speed at the midplane: Dependence on headwind}) show the gas flow speed perturbed by the planet at the midplane. We focus especially on the gas flow in the $x$-direction (the radial gas flow with respect to the disk), which \revthi{affects} both the dust accretion onto the planet and the radial drift of dust.} \revsec{In the following paragraphs, we show detailed analysis of gas flow speed at the midplane enabled by our extensive parameter studies.}

\subsubsection{\rev{Gas flow speed at the midplane: Dependence on planetary mass}}\label{sec:Gas flow speed at the midplane: Dependence on planetary mass}
Figure \ref{fig:vx_m_dependence} shows the structure of the planet-induced gas flow at the midplane for different planetary masses, $m=0.05,\,0.1,\,0.5$, and $0.7$. We set $\mathcal{M}_{\rm hw}=0$. The outflow occurs along the inner-leading and outer-trailing \revthi{recycling streamline} (in the second and fourth quadrants of the $x\text{-}y$ plane). \revsix{The streamlines \revsev{(the thin gray lines in \Figref{fig:vx_m_dependence})} were plotted \revsev{with} the \texttt{matplotlib.pyplot.streamplot} \revsev{library}, where we used the midplane gas velocities interpolated by the bilinear interpolation method (Appendix \ref{sec:Interpolation of the gas velocity}). We highlighted the widest horseshoe streamline and the shear streamline passing closest to the planet \revsev{(the thick solid and dashed gray lines in \Figref{fig:vx_m_dependence}; Appendix \ref{sec:Widest horseshoe streamline and the shear streamline passing closest to the planet}).}} Since we \rev{assumed no headwind in \Figref{fig:vx_m_dependence}}, the flow \revthi{morphology} \revsec{(the shapes of streamlines)} are common in all panels. However, the gas velocity changes with the planetary mass. In particular, the $x$-component of the gas velocity \revthi{along the inner-leading and outer-trailing recycling streamline} \revthi{(in other words, the outflow speed in the $x$-direction, $v_{x,{\rm out}}$, which is the main focus of this study)} shows the complex dependence on the planetary mass. When $m\lesssim0.5$, \revthi{$v_{x,{\rm out}}$} increases with the planetary mass (Figs. \ref{fig:vx_m_dependence}a--c), but it seems to decrease for a higher-mass planet (\Figref{fig:vx_m_dependence}d). From the \revthi{analytic} point of view, we discuss again the dependence of \revthi{$v_{x,{\rm out}}$} on the planetary mass later in Sect. \ref{sec:Revisiting the dependence on planetary mass}.

\subsubsection{\rev{Gas flow speed at the midplane: Dependence on headwind}}\label{sec:Gas flow speed at the midplane: Dependence on headwind}
Figure  \ref{fig:vx_Mhw_dependence} shows the dependence of the gas flow structure on the headwind of the gas for a fixed planetary mass, $m=0.2$. The headwind of the gas breaks a rotational symmetric structure of the planet-induced gas flow \revsec{\citep{Ormel:2015b,Kurokawa:2018}}. 
The \revsix{corotation} radius for the gas, \revthi{$x_{\rm cor,g}=-2\mathcal{M}_{\rm hw}/3$,} shifts to the negative direction in the $x$-axis as $\mathcal{M}_{\rm hw}$ increases, and the horseshoe region which formed around $x=x_{\rm cor,g}$ also shifts to the negative $x$-direction. As shown in \Figref{fig:vx_Mhw_dependence}, the outflow speed toward the inside (outside) of the planetary orbit decreases (increases) as $\mathcal{M}_{\rm hw}$ increases. From the \revthi{analytic} point of view, we discuss again the dependence of \revthi{$v_{x,{\rm out}}$} on the Mach number of the headwind later in Sect. \ref{sec:Revisiting the dependence on headwind}.


\section{\rev{Derivation of analytic formula}}\label{sec:Derivation of analytic formula}
\rev{Based on the numerical results \revsec{obtained} in Sect. \ref{sec:Hydrodynamical simulation results}, here we \revsec{derive} the analytic formulae describing \revthi{the flow morphology, the maximum outflow speed, and the distributions of the velocity component of the outflow in the radial and vertical directions to the disk.}}

\begin{figure}[htbp]
    \centering
    \includegraphics[width=\linewidth]{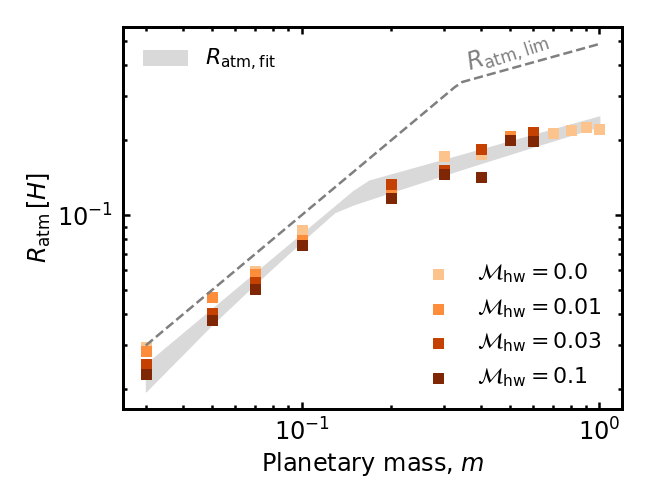}
    \caption{Atmospheric radius as a function of the planetary mass and the Mach number of the headwind of the gas. The square symbols were obtained from hydrodynamical simulations. Different colors correspond to different Mach numbers, $\mathcal{M}_{\rm hw}$. \rev{The gray-shaded region is} given by the fitting formula for the atmospheric radius, $R_{\rm atm,fit}$. \revthi{The gray dashed line is given by \Equref{eq:Ratm lim}.}}
    \label{fig:Ratm}
\end{figure} 

\begin{figure}[htbp]
    \centering
    \includegraphics[width=\linewidth]{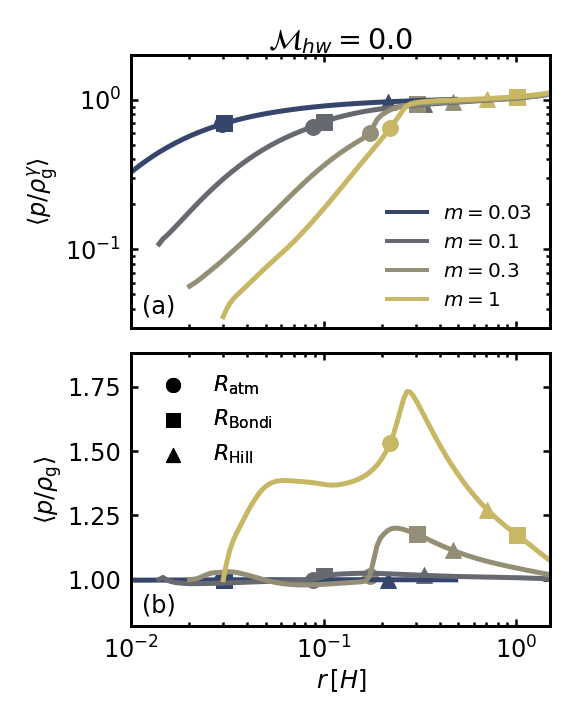}
    \caption{\revsec{Shell-averaged value of $p/\rho_{\rm g}^\gamma$ \revthi{(\textit{panel a}) and $p/\rho_{\rm g}$ (\textit{panel b}), which are the measure of the entropy and the temperature}, as a function of the distance from the planet. \revthi{We set $\gamma=1.4$.} Different colors correspond to different planetary masses. The filled circle, the square, and the triangle symbols denote the atmospheric radius, the Bondi radius, and the Hill radius, respectively. We set $\mathcal{M}_{\rm hw}=0$.}}
    \label{fig:entropy_temp_rad_m_shellavg}
\end{figure} 

\begin{figure*}[htbp]
    \centering
    \includegraphics[width=\linewidth]{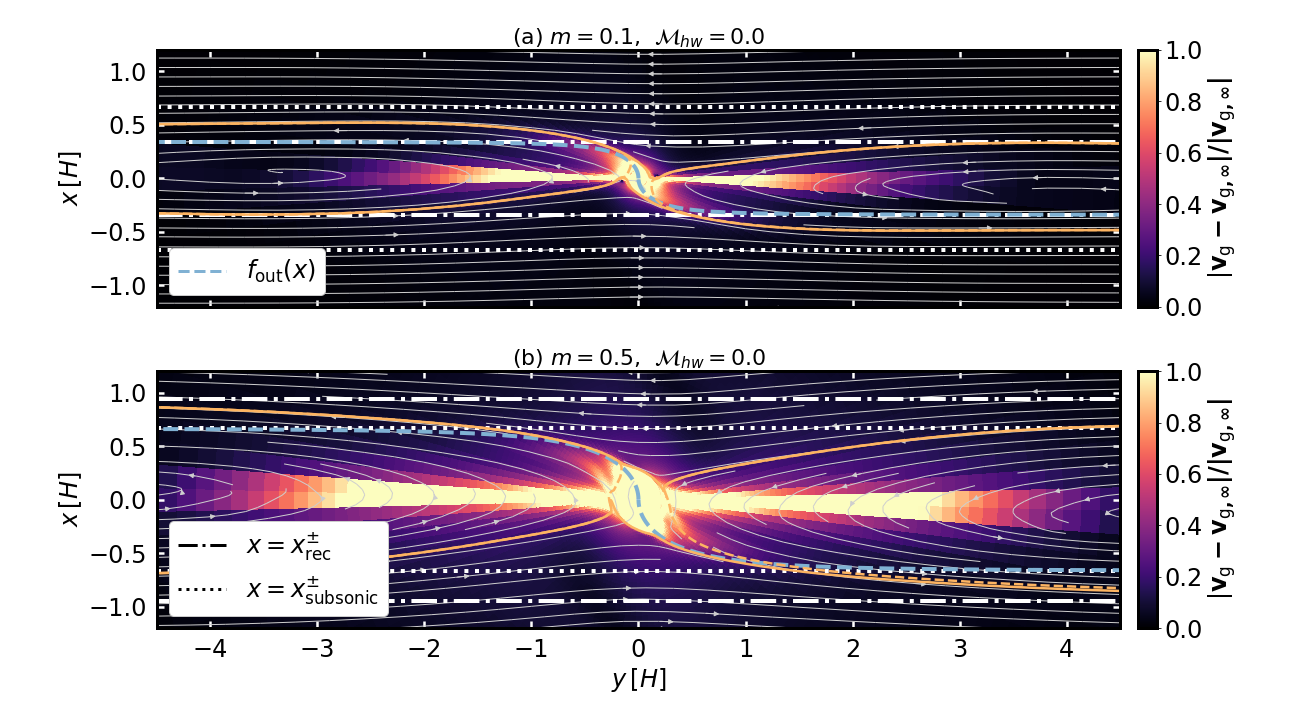}
    \caption{\revsec{Difference of the flow speed from the unperturbed \revthi{Keplrian} shear. The results were obtained from \texttt{m01-hw0} (\textit{panel a}) and \texttt{m05-hw0} (\textit{panel b}). The blue dashed, the white dashed-dotted, and the white dotted lines denote the fitting formula for the critical recycling streamline, the $x$-coordinate of the critical recycling streamline at the midplane in the far field, and the $x$-coordinate of the edge of the subsonic region, respectively. \revsix{Same as \Figref{fig:vx_m_dependence}, we highlighted the widest horseshoe streamline and the shear streamline passing closest to the planet with the orange solid and dased lines, respectively.}}}
    \label{fig:horizontal_m01_m05}
\end{figure*} 

\begin{figure*}[htbp]
    \centering
    \includegraphics[width=\linewidth]{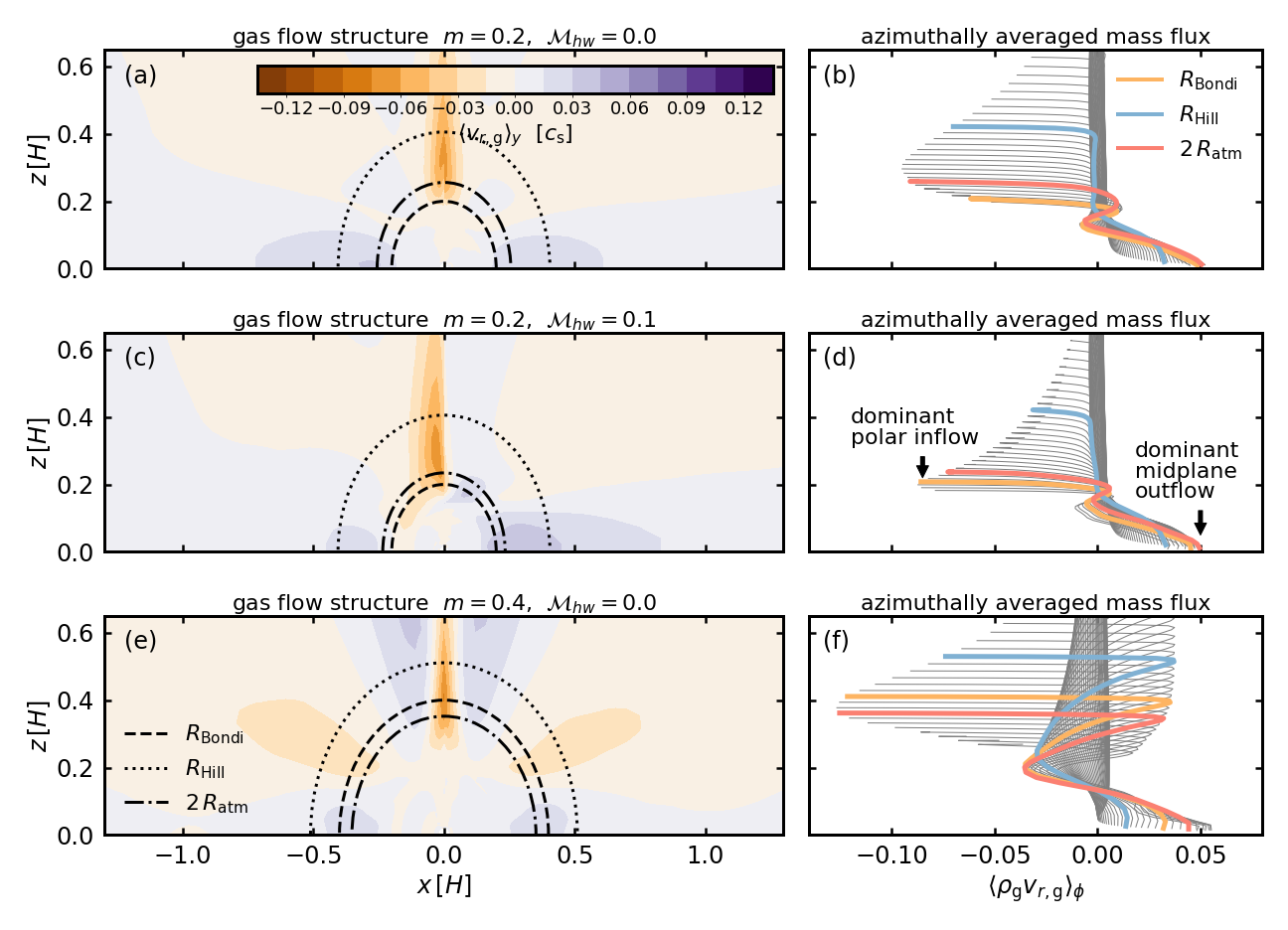}
    \caption{Gas flow structure and azimuthally averaged mass flux. The results were obtained from \texttt{m02-hw0} (\textit{panels a and b}), \texttt{m02-hw01} (\textit{panels c and d}), and \texttt{m04-hw0} (\textit{panels e and f}). \textit{Left column:} Gas flow structure at the meridian plane. The color contour represents the $r$-component of the gas velocity in the spherical polar coordinates centered at the planet averaged in the $y$-direction within the calculation domain of hydrodynamical simulation. \revfif{The dashed, the dotted, and the dashed-dotted circles denote the Bondi radius, the Hill radius, and the twice the atmospheric radius, respectively.} \textit{Right column:} Mass flux of the gas averaged in the aziumthal direction in the spherical polar coordinates centered at the planet, $\langle\rho_{\rm g}v_r\rangle_{\phi}$. Each solid line represents the changes of $\langle\rho_{\rm g}v_r\rangle_{\phi}$; altitude is varied along with a certain radius. We highlight the important radii with yellow ($R_{\rm Bondi}$), blue ($R_{\rm Hill}$), and red ($2\,R_{\rm atm}$) solid lines, respectively.}
    \label{fig:vr_and_massflux}
\end{figure*} 

\begin{figure*}[htbp]
    \centering
    \includegraphics[width=\linewidth]{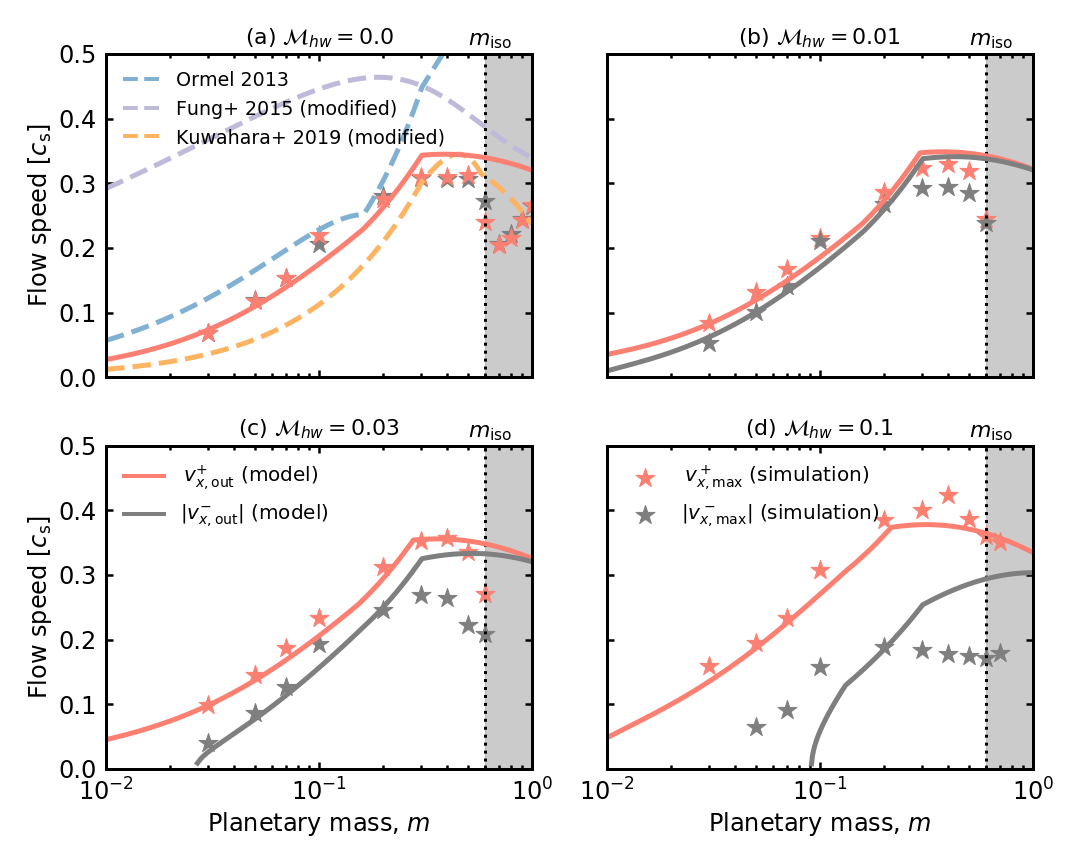}
    \caption{Outflow speed at the midplane as a function of the dimensionless planetary mass and the Mach number of the headwind of the gas. The star symbols were obtained from hydrodynamical simulations, corresponding to the maximum flow speeds in the $x$-direction toward the outside (red) and inside (black) of the planetary orbit within the limited region, $w_{\rm out}^-\leq x\leq w_{\rm out}^+$, $L_{\rm out}^-\leq y\leq L_{\rm out}^+$, and $\sqrt{x^2+y^2}\geq R_{\rm atm}$. \revthi{The last condition avoids mis-sampling the large $v_{x,{\rm g}}$ within the atmosphere.} The red and black solid lines correspond to the analytic formulae for the $x$-\rev{components of the outflow speeds} (\Equref{eq:v_x,out}). The dashed lines correspond to the analytic formulae for the outflow speed in the $x$-direction obtained by \cite{Ormel:2013} (blue; \Equref{eq:v_x,out O13}), \cite{Fung:2015} (purple; \Equref{eq:v_x,out fung}), and \cite{Kuwahara:2019} (yellow; \Equref{eq:v_x,out K19}), respectively. The gray-shaded region represents the region where the planetary mass exceeds the pebble isolation mass, $m_{\rm iso}$ (\Equref{eq:m_iso}). }
    \label{fig:vx_with_analytic}
\end{figure*} 

\begin{figure*}[htbp]
    \centering
    \includegraphics[width=\linewidth]{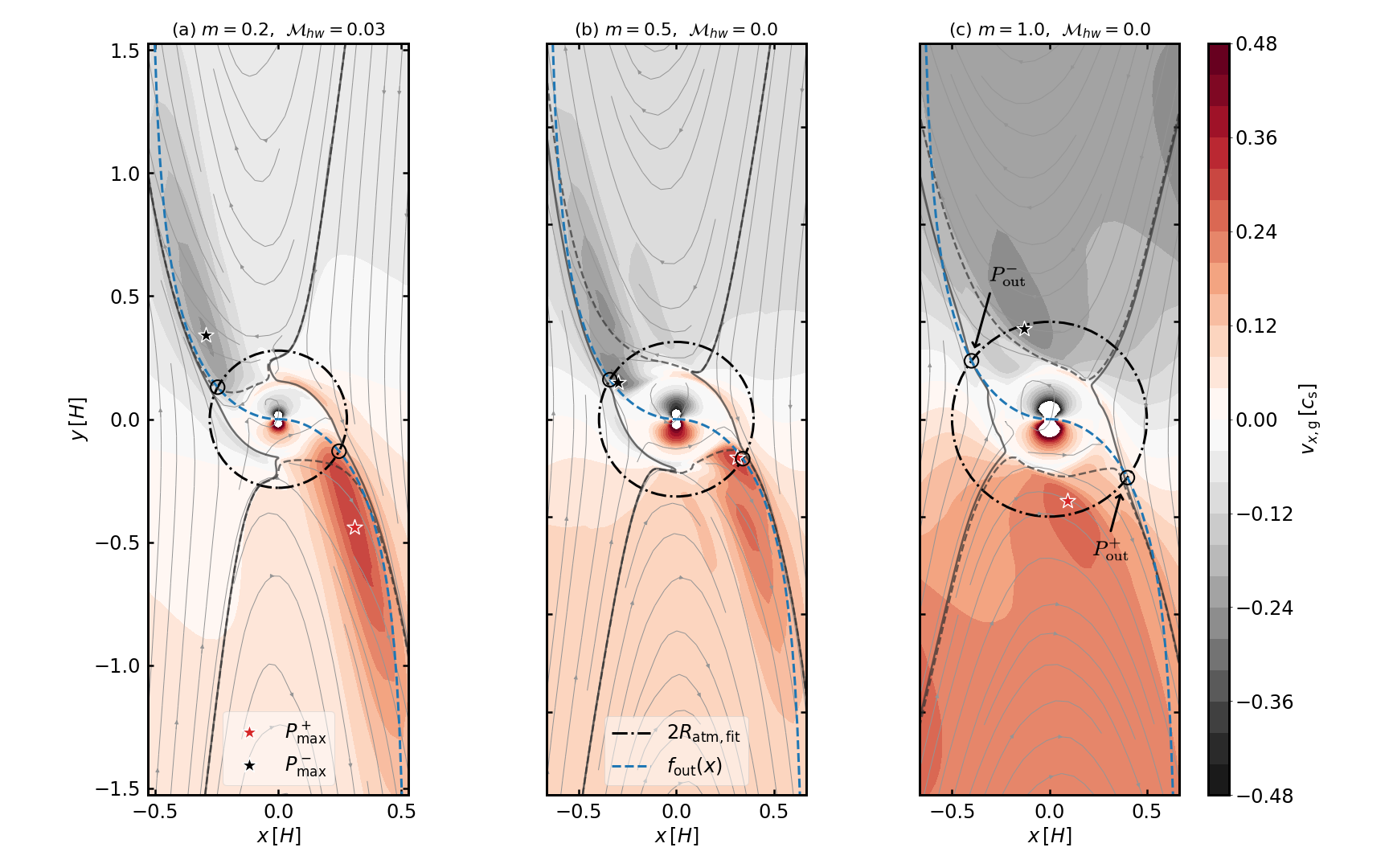}
    \caption{Comparison of the point where \revthi{$v_{x,{\rm g}}$} has the maximum value at the midplane and the assumed outflow point, $P_{\rm out}^\pm$ \revsix{(the open circle on the dashed-dotted circle)}. The results were obtained from \texttt{m02-hw003} (\textit{panel a}), \texttt{m05-hw0} (\textit{panel b}), and \texttt{m1-hw0} (\textit{panel c}). The star symbols denote the points where the flow speed in the positive (the red star) and the negative (the black star) $x$-direction has the maximum value in the limited region, $w_{\rm out}^-\leq x\leq w_{\rm out}^+$, $L_{\rm out}^-\leq y\leq L_{\rm out}^+$ (the ranges of the $x$- and $y$-axes of these panels), and $\sqrt{x^2+y^2}\geq R_{\rm atm}$. The flow speed at the star symbol was plotted in \Figref{fig:vx_with_analytic}a with the same color star symbol.}
    \label{fig:vx_max_point}
\end{figure*} 

\begin{figure}[htbp]
    \centering
    \includegraphics[width=\linewidth]{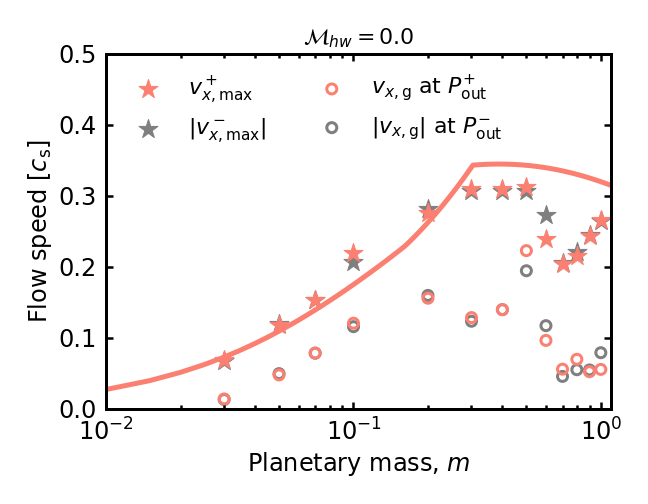}
    \caption{\revsix{Comparison of the maximum flow speed in the $x$-direction, $v_{x,{\rm max}}^\pm$, and $v_{x,{\rm g}}$ at the analytically-assumed outflow point, $P_{\rm out}^\pm$. We set $\mathcal{M}_{\rm hw}=0$.}}
    \label{fig:vx_hw0}
\end{figure} 

\begin{figure}[htbp]
    \centering
    \includegraphics[width=\linewidth]{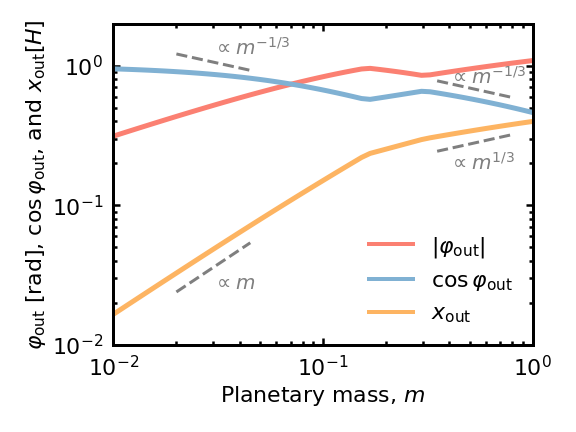}
    \caption{Angle of the outflow at the outflow point, $\varphi_{\rm out}$ (\Equref{eq:varphi out}), $\cos\varphi_{\rm out}$, and the $x$-coordinate of the outflow point, $x_{\rm out}^\pm$ (\Equref{eq:x_out}) as a function of the planetary mass. We set $\mathcal{M}_{\rm hw}=0$.}
    \label{fig:outflow_angle}
\end{figure} 

\begin{figure}[htbp]
    \centering
    \includegraphics[width=\linewidth]{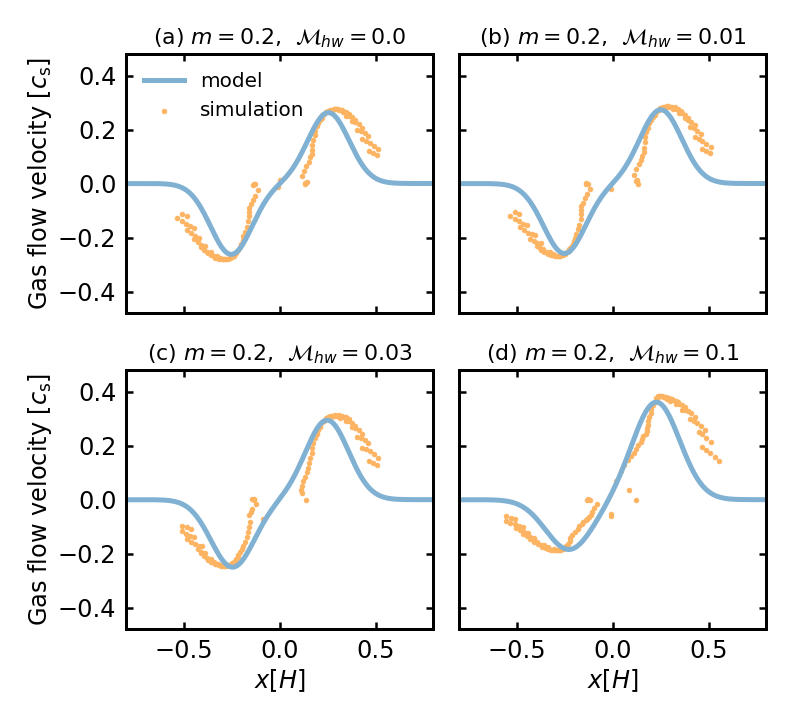}
    \caption{\revsec{Distribution of the $x$-component of the gas velocity} in the $x$-direction at the midplane, \revthi{$v_{x,{\rm max}}(x)$}. The blue solid line is given by \Equref{eq:vx_g_Gaussian}. The yellow dots were obtained from \texttt{m02-hw0} (\textit{panel a}), \texttt{m02-hw001} (\textit{panel b}), \texttt{m02-hw003} (\textit{panel c}), and \texttt{m02-hw01} (\textit{panel d}). We sampled the gas velocity at each radial grid of hydrodynamical simulations where the absolute value of the gas velocity was the maximum within the limited region, $w_{\rm out}^-\leq x\leq w_{\rm out}^+$, $L_{\rm out}^-\leq y\leq L_{\rm out}^+$, and $\sqrt{x^2+y^2}\geq R_{\rm atm}$.}
    \label{fig:vx_dist_m02}
\end{figure}

\begin{figure*}[htbp]
    \centering
    \includegraphics[width=\linewidth]{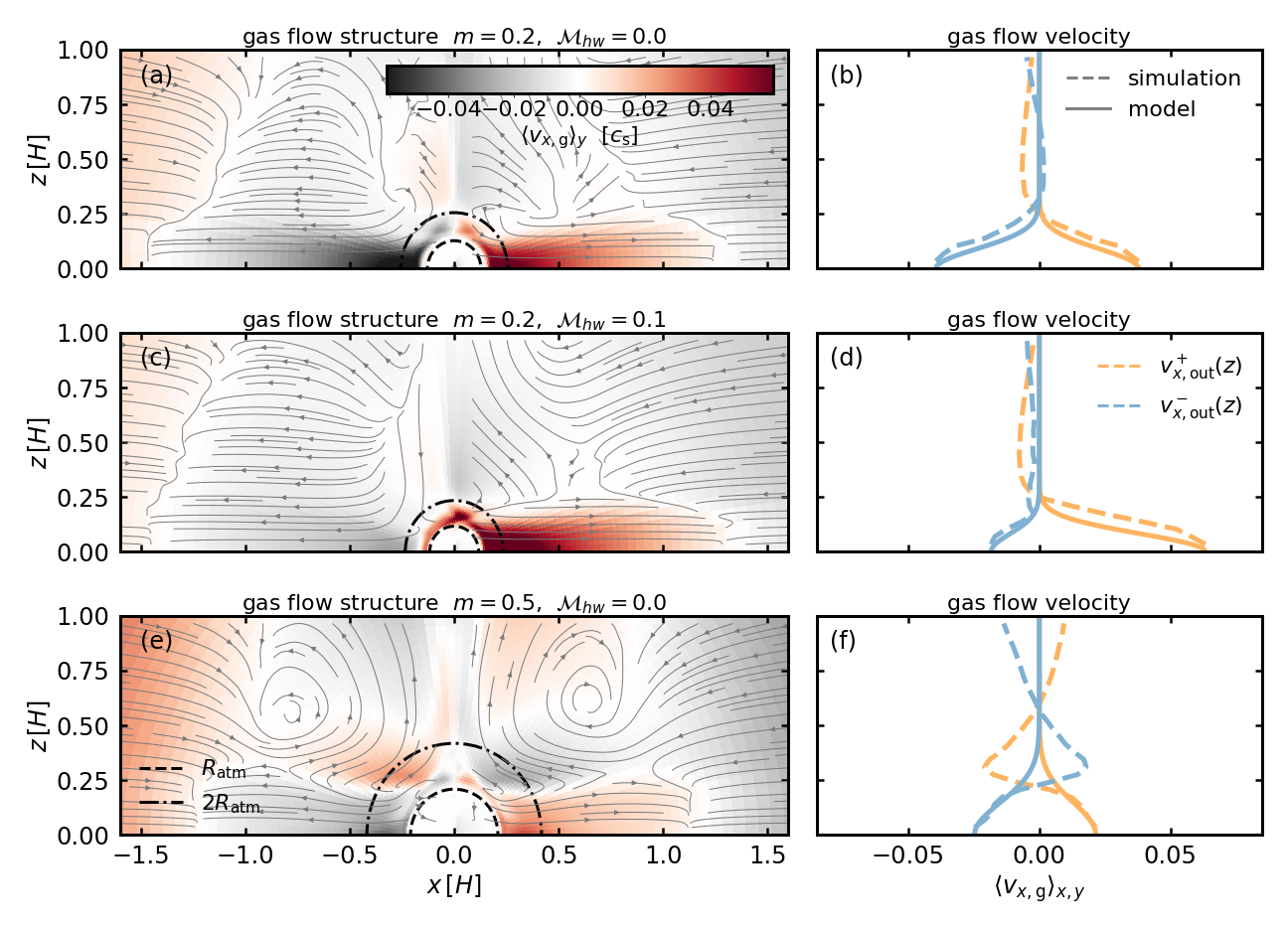}
    \caption{\rev{Vertical structure of the gas flow.} The results were obtained from \texttt{m02-hw0} (\textit{panels a and b}), \texttt{m02-hw01} (\textit{panels c and d}), and \texttt{m05-hw0} (\textit{panels e and f}). \textit{Left column:} Same as the left column of \Figref{fig:vr_and_massflux}, but the color contour represents \revthi{$v_{x,{\rm g}}$} averaged in the $y$-direction within the calculation domain of hydrodynamical simulation. \textit{Right column:} The $x$-component of the gas flow velocity averaged in the $x$- and $y$-directions within the calculation domain of hydrodynamical simulation, $\langle v_{x,{\rm g}}^+\rangle_{x,y}$ (yellow) and $\langle v_{x,{\rm g}}^-\rangle_{x,y}$ (blue). The dashed lines were obtained from hydrodynamical simulations. The solid lines are the fitting formula (\Equref{eq:vx_gas_z_fit}).}
    \label{fig:vx_z_dependence}
\end{figure*}

\subsection{Atmospheric radius}\label{sec:Atmospheric radius}
The size of an atmosphere \rev{affects} gas dynamics around an embedded planet, because the recycling streamlines trace the surface of the atmosphere (\Figref{fig:streamline_m005_Mhw003}). \rev{Hereafter we define the size of the atmosphere as the atmospheric radius, the maximum radius where the azimuthal gas velocity is dominant in the midplane region \citep{moldenhauer2022recycling}}:
\begin{align}
    R_{\rm atm}\equiv\max\left(\rev{r}\,|\,v_{\phi,{\rm g}}|_{\text{midplane}}>\max\left(v_{r,{\rm g}},v_{\theta,{\rm g}}\right)\right),\label{eq:atmospheric radius}
\end{align}
where $\bm{v}_{\rm g}=(v_{r,{\rm g}},v_{\theta,{\rm g}},v_{\phi,{\rm g}})$ is the gas velocity in the spherical polar coordinates centered at the planet. To estimate the atmospheric radius, we used the final state of the hydro-simulations data, $t=t_{\rm end}$, where the flow field seems to have reached a steady state (Appendix \ref{sec:Numerical convergence}).

Figure \ref{fig:Ratm} shows the dependence of the atmospheric radius on the planetary mass and the Mach number of the headwind. We found that the atmospheric radius scales with the Bondi or Hill radii. \revsix{The headwind of the gas reduces the atmospheric radius for a fixed planetary mass, which is consistent with  previous studies \citep{Ormel:2013,moldenhauer2022recycling}. Assuming that the atmospheric radius decreases linearly as $\mathcal{M}_{\rm hw}$ increases,} we introduce a fitting formula for the \revsix{dimensionless} atmospheric radius as:
\revsix{
\begin{align}
    R_{\rm atm,fit}\equiv\min\left(C_1R_{\rm Bondi}-D_1\mathcal{M}_{\rm hw},\,C_2R_{\rm Hill}-D_2\mathcal{M}_{\rm hw}\right),\label{eq:Ratm fit}
\end{align}}where \revsix{$C_1,\,C_2,\,D_1$, and $D_2$} are the fitting coefficients. \revsix{\revsev{With} the least-squares method, we determined the values of these coefficients: $C_1=0.84$, $C_2=0.36$, $D_1=0.056$, and $D_2=0.22$.} \revsev{The atmospheric radii are resolved by $28\text{--}52$ cells in the radial direction in our hydrodynamical simulations.}

\revthi{The atmospheric radius can be constrained by the thermodynamical condition or the planet's gravity, namely, the Bondi radius or the Roche lobe radius of the planet, whichever is smaller \citep{dangelo2008evolution,DAngelo:2013}. The Roche lobe is surrounded by the equipotential surface of the so-called Roche potential. Although the Roche lobe is not spherical, the radius of a sphere having the same volume as the Roche lobe is given by the following fitting formula \citep{eggleton1983approximations}:
\begin{align}
    R_{\rm Roche}=\frac{0.49m^{2/3}h}{0.6m^{2/3}h^2+\ln(1+m^{1/3}h)}H.\label{eq:roche lobe radius}
\end{align}
For a typical steady accretion disk model \citep[e.g.,][]{Ida:2016}, the disk aspect ratio has on the order of $h\sim0.01\text{--}0.1$ (\revsev{Appendix \ref{sec:Aspect ratio of the steady accretion disk}}). The value of $R_{\rm Roche}$ hardly changes for the range of $h$. We set $h=h_0=0.05$ as a nominal value in \Equref{eq:roche lobe radius}. Since $m^{1/3}h\ll1$ in our parameter sets, \Equref{eq:roche lobe radius} can be approximated as $R_{\rm Roche}\simeq0.49m^{1/3}\,H\sim2R_{\rm Hill}/3$ \citep{paczynski1971evolutionary}, where we used $\ln(1+m^{1/3}h)\simeq m^{1/3}h$ and $(1+0.6m^{1/3}h)\simeq1$. We plotted the following equation in \Figref{fig:Ratm}:
\begin{align}
    R_{\rm atm,lim}=\min(R_{\rm Bondi},\,R_{\rm Roche}). \label{eq:Ratm lim}
\end{align}
Equation (\ref{eq:Ratm lim}) predicts that the scaling law changes at $m\sim0.3$ and the atmospheric radius scales with $\sim2R_{\rm Hill}/3$ when $m\gtrsim0.3$. \revfif{These predictions are in qualitative agreement with the numerical result where the atmospheric radius scales with a fraction of the Hill radius when $m\gtrsim0.2$.}}

\revthi{The gas is heated by adiabatic compression and then the \revsix{$\beta$ cooling} occurs within the gravitational sphere of the planet, leading to the formation of the low-entropy region at $r\lesssim R_{\rm grav}$ \citep[\revthi{\Figref{fig:entropy_temp_rad_m_shellavg}a;}][]{Kurokawa:2018}. The buoyancy force characterized by the positive entropy gradient isolates the atmosphere from the surrounding disk gas \citep[][]{Kurokawa:2018}. We found that the temperature is nearly uniform around the atmosphere \revsix{due to our choice of short cooling time when $m\lesssim0.1$ ($\beta\leq1$;} \Figref{fig:entropy_temp_rad_m_shellavg}b). The amplitude of the temperature fluctuation increases with the planetary mass, which is as high as $\sim75\%$ when $m=1$.}

\subsection{Morphology of the outflow region}\label{sec:Morphology of the outflow region}
\rev{This section introduces the fitting formulae for the morphology of the midplane outflow \revsec{(the shape of the outflow streamline)}, describing the region where the $x$-component of the gas velocity, $v_{x,{\rm g}}$, is dominantly perturbed by the embedded planet (the deep red and black contours in Figs. \ref{fig:vx_m_dependence} and \ref{fig:vx_Mhw_dependence}). We refer to this region as the outflow region. The obtained formulae \revthi{will} be used in a future study to \revsec{determine} the extent of the region in which the hydrodynamic effects of the outflow on the dust motion would appear. Hereafter} \revthi{we} restrict our attention to the limited region, $|x|\lesssim H$ and $z=0$.

\subsubsection{\rev{\revthi{Morphology} of the critical recycling streamline at the midplane}}\label{sec:Morphology of the critical recycling streamline at the midplane}
\rev{The outflow occurs along the recycling streamline, \revthi{which suggests} that the outflow region can be determined by the shape of the recycling streamline. The shape of the streamline is also important to estimate \revthi{the outflow speed in the $x$-direction, $v_{x,{\rm out}}$}, which affects the dust motion \revsec{\citep{Kuwahara:2020a,Kuwahara:2020b}}, because \revthi{$v_{x,{\rm out}}$} can be obtained by considering the angle of the flow velocity with respect to the $x$-axis.}

\rev{Here we introduce a representative recycling streamline: a critical recycling streamline. We define the critical recycling streamline as the separatrix streamline which divides the Keplerian shear flow and the inner-leading (outer-trailing) recycling flow at the midplane.} We found that the inner-leading and outer-trailing critical recycling streamlines at the midplane can be outlined by the following fitting formulae: 
\begin{align}
\left\{
\begin{array}{ll}
    \rev{r}=&w^+_{\rm out}\tan\theta\quad(3\pi/2\leq\theta\leq2\pi),\\
    \rev{r}=&w^-_{\rm out}\tan\theta\quad(\pi/2\leq\theta\leq\pi),\label{eq:outflow shape in polar}
\end{array}
\right.
\end{align}
in the polar coordinates centered at the planet, \rev{where $w_{\rm out}^{\pm}$ is the $x$-coordinate of the edge of the outflow region (described later in Sect. \ref{sec:Width of the outflow region}). The dust motion can be disturbed by the outflow in the region where $w_{\rm out}^-\leq x\leq w_{\rm out}^+$.} Hereafter the plus and minus signs for an arbitrary quantity, $X^\pm$, denote the quantity outside and inside the planetary orbit. Equation (\ref{eq:outflow shape in polar}) represents the tangential curves with $w_{\rm out}^\pm$ as the asymptotes. In the local Cartesian coordinates co-rotating with the planet, \Equref{eq:outflow shape in polar} can be written by:
\begin{align}
    y=&-\frac{x^2}{\sqrt{(w^+_{\rm out})^2-x^2}}\mathcal{H}(x)+\frac{x^2}{\sqrt{(w^-_{\rm out})^2-x^2}}\mathcal{H}(-x)\equiv f_{\rm out}(x),\label{eq:outflow shape in cartesian}
\end{align}
where $\mathcal{H}(x)$ is the Heaviside step function. 

We plotted \Equref{eq:outflow shape in cartesian} in Figs. \ref{fig:vx_m_dependence} and \ref{fig:vx_Mhw_dependence}. Equation (\ref{eq:outflow shape in cartesian}) can trace the \rev{critical recycling streamline at the midplane} for a wide range of planetary mass and Mach number of the headwind. \revsix{In most of the panels in Figs. \ref{fig:vx_m_dependence} and \ref{fig:vx_Mhw_dependence}, $f_{\rm out}(x)$ is sandwiched between the widest horseshoe streamline and the shear streamline passing closest to the planet.} \revsec{Using \Equref{eq:outflow shape in cartesian}, we \revthi{estimate} the outflow speed analytically near the atmospheric radius, $R_{\rm atm}\lesssim0.1\,H$, where the dominant outflow occurs (Figs. \ref{fig:vx_m_dependence} and \ref{fig:vx_Mhw_dependence}; described later in Sect. \ref{sec:Positions of inflow and outflow}). We \revthi{use} the analytically estimated outflow speed for the estimation of the outflow effect on the dust motion. Thus, the agreement between \Equref{eq:outflow shape in cartesian} and the \revsev{numerically-obtained} critical recycling streamline in the region close to the planet ($r\sim R_{\rm atm}$) is \revthi{useful}. Although \Equref{eq:outflow shape in cartesian} does not coincide with the \revsev{numerically-obtained} critical recycling streamline in the distant region far from the planet, $|y|\gtrsim H$, the mismatch between \Equref{eq:outflow shape in cartesian} and the \revsev{numerically-obtained} critical recycling streamline in \revthi{the far region} does not affect an \revthi{analytic} derivation of the outflow speed.}

Equation (\ref{eq:outflow shape in cartesian}) can be applied to trace the critical recycling streamline near the atmospheric radius for the entire range of our parameter sets ($m\in[0.03,1]$). When \revthi{$m\gtrsim0.6$}, however, we found that the \revthi{dominant perturbation to $v_{x,{\rm g}}$ occurs} along the midplane horseshoe streamlines rather than the streamlines characterized by \Equref{eq:outflow shape in cartesian} (e.g., \Figref{fig:vx_m_dependence}d; described later in Sect. \ref{sec:Comparison to numerical results}). 
When the planetary mass reaches the so-called pebble isolation mass, \revthi{$M_{\rm iso}=25(h/0.05)^3[0.34(3/\log\alpha_{\rm acc})^4+0.66]\,M_{\oplus}$, a growing planet opens a shallow gas gap} \citep{Lambrechts:2014,Bitsch:2018,ataiee2018much}. In our dimensionless unit, the pebble isolation mass can be described by:
\revthi{\begin{align}
    m_{\rm iso}\simeq0.6\,\Bigg[0.34\Bigg(\frac{3}{\log\alpha_{\rm acc}}\Bigg)^4+0.66\Bigg]\Bigg(\frac{M_\ast}{M_\odot}\Bigg)^{-1}.\label{eq:m_iso}
\end{align}
}\revthi{We assumed $\alpha_{\rm acc}=10^{-3}$ and $M_\ast=M_\odot$ in \Equref{eq:m_iso} as nominal values. Our local simulation cannot handle the gas-gap opening in the high-mass regime ($m\gtrsim m_{\rm iso}$). Thus,}  \revthi{\Equref{eq:outflow shape in cartesian} is valid as a model for the outflow streamline for $m\lesssim m_{\rm iso}\simeq0.6$} to investigate the hydrodynamic effect of the gas flow on the dust.

\subsubsection{Width of the outflow region}\label{sec:Width of the outflow region}
\rev{In Sect. \ref{sec:Morphology of the critical recycling streamline at the midplane}, we introduced a fitting formula for the critical recycling streamline, $y=f_{\rm out}(x)$, which represents the \revthi{edge of the outflow region whose $x$-coordinate approaches asymptotically to} $x=w_{\rm out}^{\pm}$. This section describes the width of the outflow region in the $x$-direction (the radial direction to the disk) and introduces \revthi{an analytic description for $w_{\rm out}^{\pm}$}.}  


We found that the width of the outflow region \revthi{is limited by} the width of the subsonic region \revthi{and} the $x$-coordinate of the recycling streamline at the midplane characterized by the half-width of the horseshoe streamline. 
We defined the subsonic region as the region where the \revsec{unperturbed gas velocity in the local frame} is always subsonic, $|\bm{v}_{{\rm g},\infty}|\leq c_{\rm s}$. The $x$-coordinate of the edge of the subsonic region for the gas is given by:
\begin{align}
    x^\pm_{\rm subsonic}\equiv\pm\frac{2}{3}H-|x_{\rm cor,g}|\quad\text{(double-sign corresponds)}.\label{eq:x_subsonic}
\end{align}
Hereafter we omit the notation of the double-sign corresponds in each equation, unless otherwise specified. 

Next, we introduce the $x$-coordinate of a critical recycling streamline in the far field ($|y|\rightarrow\infty$) \rev{at the midplane}. 
Let us start \revthi{from considering} an endmember case, $\mathcal{M}_{\rm hw}=0$. The gas from the disk enters the gravitational sphere of the planet and flows back to the disk through the region between the Keplerian shear flow and the inner-leading (outer-trailing) horseshoe flow. Thus, the width of the outflow region could be constrained by the width of the horseshoe flow. \revthi{When $m\lesssim0.1$, the half-width of the horseshoe streamline can be estimated by the linear theory, which is given by \citep{Masset:2016}:
\revsix{\begin{align}
     w_{\rm HS}^{\rm MB16}=1.05\sqrt{m},\label{eq:w_HS MB16}
\end{align}
}combined with the numerical factor. When $m\gtrsim0.1$, the half-width of the horseshoe streamline is proportional to the Hill radius of the planet due to the onset of nonlinear effects in the horseshoe region, which can be fitted by $w_{\rm HS}\simeq2.4R_{\rm Hill}\simeq1.7m^{1/3}$ \citep{masset2006migration}. \cite{jimenez2017improved} introduced the following fitting formula for the half-width of the horseshoe streamline, which can be applied to a wide range of planetary masses:}
\revsix{
\begin{align}
    w_{\rm HS}^{\rm JM17}=\frac{1.05\sqrt m+3.4m^{7/3}}{1+2m^2}\equiv w_{\rm HS}.\label{eq:w_HS}
\end{align}
}\revsix{We found that \Equref{eq:w_HS} is a better indicator than \Equref{eq:w_HS MB16} in describing the width of the outflow region for practical use, though it overestimates the widths of the outer-leading and inner-trailing parts of the horseshoe flows (Appendix \ref{sec:Numerically calculated widths of the horseshoe and outflow region}).}

When \revthi{the nonzero Mach number of the headwind is assumed, }
the horseshoe region shifts to the negative direction in the $x$-axis by $-|x_{\rm cor,g}|$. We found that the disk gas along the recycling streamline flows into an empty space \revthi{formed} by the shift of the horseshoe region (\Figref{fig:streamline_m005_Mhw003}a). The critical recycling streamline traces this replenished flow. The $x$-coordinate of the critical recycling streamline at the midplane in the far field \revthi{is} given by: 
\begin{align}
    x_{\rm rec}^\pm\equiv\pm w_{\rm HS}\pm |x_{\rm cor,g}|.\label{eq:x_rec}
\end{align}

\revsec{We found that the width of the outflow region in the $x$-direction can be characterized by \revthi{$x_{\rm subsonic}$ (\Equref{eq:x_subsonic}) or $x_{\rm rec}$ (\Equref{eq:x_rec}), whichever was smaller:
\begin{align}
    w^{\pm}_{\rm out}=\pm\min\left(|x^{\pm}_{\rm subsonic}|,\,|x_{\rm rec}^\pm|\right).\label{eq:w_out}
\end{align}
We plotted \Equref{eq:w_out} in Figs. \ref{fig:vx_m_dependence} and \ref{fig:vx_Mhw_dependence}, which can serve as a useful guide for the extent of the region where $v_{x,{\rm g}}$ is dominantly perturbed by the embedded planet.}}

\revsec{Figure \ref{fig:horizontal_m01_m05} shows the difference of the flow speed from the unperturbed sub-Keplerian speed, where $m=0.1$ and $0.5$, \revsev{and $\mathcal{M}_{\rm hw}=0$} were assumed.} When $|x_{\rm rec}|<|x_{\rm subsonic}|$ ($m\lesssim0.3$; for simplicity, we omit the plus and minus signs), the gas velocity approaches the \revsec{unperturbed} sub-Keplerian shear when $x\rightarrow x_{\rm rec}$, leading to $v_{x,{\rm g}}\rightarrow 0$ \revsec{(\Figref{fig:horizontal_m01_m05}a)}. When $|x_{\rm rec}|>|x_{\rm subsonic}|$ ($m\gtrsim0.3$), the \revsev{numerically-obtained} critical recycling streamline can intrude the supersonic region because its width is determined by $x_{\rm rec}$ \revsev{($=|w_{\rm HS}|>|x_{\rm subsonic}|$; the orange solid lines in \Figref{fig:horizontal_m01_m05}b). However, $v_{x,{\rm g}}$ along the numerically-obtained critical recycling streamline approaches instantaneously to $v_{x,{\rm g}}\rightarrow 0$ in the subsonic region (\Figref{fig:horizontal_m01_m05}b). Since we defined the outflow region as the region where $v_{x,{\rm g}}$ is dominantly perturbed by the planet, we considered that the outflow region is constrained by $w_{\rm out}=|x_{\rm subsonic}|$, regardless of the $x$-coordinate of the numerically-obtained critical recycling streamline at $|y|\gtrsim L_{y,{\rm out}}\simeq \,H$. Although a mismatch between the numerically-obtained critical recycling streamline and $f_{\rm out}(x)$ can be seen at $|y|\gtrsim H$, again, it does not affect an analytic derivation of the outflow speed.}

\subsubsection{Length of the outflow region}\label{sec:Length of the outflow region}
\rev{In this section}, we \revthi{evaluate} the characteristic length of the outflow region in the $y$-direction (the orbital direction of the planet). \rev{The extent of the outflow region in the $y$-direction can be used when we investigate the outflow effects on the dust motion during the encounter with the planet.} Based on \Equref{eq:outflow shape in cartesian}, we \revthi{define} the $y$-coordinate at a position close enough to the asymptotes, $x=w_{\rm out}^\pm$, as the characteristic length of the outflow region:
\begin{align}
    L^\pm_{y,{\rm out}}\equiv f_{\rm out}(x_{\rm crit}^{\pm}),\label{eq:Ly_out}
\end{align}
where\revsix{
\begin{align}
    x_{\rm crit}^{\pm}=C_{\rm crit} \times w_{\rm out}^{\pm}.
\end{align}
We set $C_{\rm crit}=0.95$} as a fitting parameter. We plotted \Equref{eq:Ly_out} in Figs. \ref{fig:vx_m_dependence} and \ref{fig:vx_Mhw_dependence}, which \revthi{shows good agreement with numerical simulations}.

\subsection{\rev{Bernoulli's theorem}}\label{sec:Bernoulli's theorem}
The outflow speed can be estimated from Bernoulli’s theorem \citep{Fung:2015,Kuwahara:2019}. For a steady-state barotropic flow, a Bernoulli's function conserves along a streamline \citep{Ormel:2013,Fung:2015,Kuwahara:2019}:
\begin{align}
    \mathcal{B}=\frac{v_{\rm g}^2}{2}+\mathcal{\rev{P}}-2\mathcal{M}_{\rm hw}x+\Phi_{\rm eff},\label{eq:Bernoulli_invariant}
\end{align}
where $\rev{\mathcal{P}=\int\mathrm{d}p/\rho_{\rm g}}$ is the \rev{pressure function} and $\Phi_{\rm eff}$ is the effective potential. \revthi{Here we assume an isothermal flow along a recycling streamline, $p=\rho_{\rm_ g}c_{\rm s}^2$. The} \rev{pressure function} \revthi{is} described by $\mathcal{\rev{P}}=c_{\rm s}^2\ln\rho_{\rm g}$. The effective potential \revthi{is} described by: 
\begin{align}
    \Phi_{\rm eff}=-\frac{3}{2}x^2+\frac{z^2}{2}-\frac{m}{\rev{r}},\label{eq:effective potential}
\end{align}
where the first and second terms in the right-hand side of \Equref{eq:effective potential} were originated from the tidal potential. 

It should be noted that the Bernoulli's function does not strictly conserve along the streamline as shown in \Figref{fig:streamline_m005_Mhw003}c, because the \revsix{$\beta$ cooling} occurs on a finite timescale, $\beta=(m/0.1)^2$. However, we found that Bernoulli's theorem is still useful to estimate the outflow speed because \rev{the gas flow along a recycling streamline can be considered as an isothermal flow} \revthi{due to the efficient \revsix{$\beta$ cooling} (\Figref{fig:entropy_temp_rad_m_shellavg}b). As discussed in Sect. \ref{sec:Atmospheric radius}, when $m\lesssim0.1$ the temperature around the atmosphere is nearly uniform (\Figref{fig:entropy_temp_rad_m_shellavg}b). Although the amplitude of the temperature fluctuation around the atmosphere is as high as $\sim40\,\%$ at most when $m\sim m_{\rm iso}$, we stick to an isothermal assumption regardless of the planetary mass. Again, our model for the outflow speed is not valid in the higher-mass regime ($m\gtrsim m_{\rm iso}$; Sect. \ref{sec:Morphology of the critical recycling streamline at the midplane}).} We discuss the applicability of Bernoulli's theorem later in Sect. \ref{sec:Applicability of Bernoulli's theorem}.

\subsection{\revsix{Positions of inflow and outflow}}\label{sec:Positions of inflow and outflow}
To estimate the outflow speed from Bernoulli's theorem, following \cite{Kuwahara:2019}, we \revsix{select} a recycling streamline and set the inflow and outflow points on it, $P_{\rm in}$ and $P_{\rm out}$, where the gas chiefly flows in and out. \revsix{These points are determined by the analysis of the mass flux of the gas.}



The left column of \Figref{fig:vr_and_massflux} shows the velocity field of the gas at the meridian plane, in which the color contour corresponds to the radial velocity in the spherical polar coordinates centered at the planet. The gas seems to be \revsix{dominantly} flowing in and out of the sphere of radius \rev{$\gtrsim2\,R_{\rm atm}$}. The right column of \Figref{fig:vr_and_massflux} shows the azimuthally averaged mass flux for a certain spherical shell of radius $\rev{r}$, $\langle\rho_{\rm g}v_{r,{\rm g}}\rangle_{\phi}$, as a function of the altitude, $z$. Gas flows in where $\langle\rho_{\rm g}v_{r,{\rm g}}\rangle_{\phi}<0$ and flows out where $\langle\rho_{\rm g}v_{r,{\rm g}}\rangle_{\phi}>0$, respectively. We found that gas chiefly flows in and out of the twice the atmospheric radius, $2\,R_{\rm atm}$.

Based on the above findings, we set the inflow and outflow points as $P_{\rm in}=(0,0,z_{\rm in})$ and $P^\pm_{\rm out}=(x^\pm_{\rm out}, y^\pm_{\rm out}, 0)$, where $z_{\rm in}=2\,R_{\rm atm,fit}$ and $\sqrt{(x_{\rm out}^\pm)^2+(y_{\rm out}^\pm)^2}=2\,R_{\rm atm,fit}$. Here we used a fitting formula for the atmospheric radius (\Equref{eq:Ratm fit}). The $x$- and $y$-coordinates of the outflow points were determined by those of the intersections of the \rev{fitting formula for the critical recycling streamline at the midplane} (\Equref{eq:outflow shape in cartesian}) and the circle of the twice the atmospheric radius, $2\,R_{\rm atm,fit}$. They are given by:
\begin{align}
    x_{\rm out}^\pm&=\pm2\,R_{\rm atm,fit}\cos\Biggl(\tan^{-1}\Biggl(\frac{2\,R_{\rm atm,fit}}{w_{\rm out}^\pm}\Biggr)\Biggr),\\
    y_{\rm out}^\pm&=f_{\rm out}(x_{\rm out}^\pm).\label{eq:x_out}
\end{align}


\subsection{Analytic formula of the outflow speed}\label{sec:Analytic formula of the outflow speed}
An analytic formula of the outflow speed was derived from Bernoulli’s theorem. We first consider the inflow point. From the analysis of the kinetic energy of gas flow in our simulations, we found that the first term in the right-hand side of \Equref{eq:Bernoulli_invariant} was negligible at the inflow point, $P_{\rm in}$ \citep{Fung:2015,Kuwahara:2019}. The second and fourth terms in the right-hand side of \Equref{eq:Bernoulli_invariant} \revthi{are canceled with each other} because the stellar gravitational potential energy balances the \revsec{pressure function} in hydrostatic equilibrium. Thus, the Bernoulli’s function at the inflow point \revthi{is} described by: 
\begin{align}
    \mathcal{B}_{\rm in}=-\frac{m}{2\,R_{\rm atm,fit}}.\label{eq:B in}
\end{align}
The Bernoulli’s function at the outflow point \revthi{is} described by: 
\begin{align}
    \mathcal{B}^\pm_{\rm out}=\frac{(v_{\rm out}^\pm)^2}{2}-\frac{3}{2}(x_{\rm out}^\pm)^2-2\mathcal{M}_{\rm hw}x^\pm_{\rm out}-\frac{m}{2\,R_{\rm atm,fit}}.\label{eq:B out}
\end{align}

From Eqs. (\ref{eq:B in}) and (\ref{eq:B out}), we \revthi{obtain} the outflow speed at the outflow point:
\begin{align}
    |v^\pm_{\rm out}|=\sqrt{3(x^\pm_{\rm out})^2+4\mathcal{M}_{\rm hw}x^\pm_{\rm out}}.\label{eq:|v_out|}
\end{align}
The first and second terms in the square root in the right-hand side of \Equref{eq:|v_out|} \revthi{originate} from the tidal potential \citep{Kuwahara:2019} and the global pressure gradient of the disk gas, respectively.

From \Equref{eq:|v_out|}, we can estimate the flow speed in the $x$-direction (the radial flow speed with respect to the disk), which \rev{affects both the dust accretion onto the planet and} the radial drift of dust. The $x$-component of the outflow speed is given by:
\begin{align}
    v_{x,{\rm out}}^{\pm}&=|v_{\rm out}^\pm|\cos\varphi_{\rm out}^\pm,\label{eq:v_x,out}
\end{align}
where $\varphi_{\rm out}^\pm$ is an angle of the outflow to the $x$-axis at the outflow point,
\begin{align}
    \varphi^\pm_{\rm out}=&\tan^{-1}\left.\left(\frac{\mathrm{d}f_{\rm out}(x)}{\mathrm{d}x}\right)\right|_{x=x_{\rm out}^\pm},\label{eq:varphi out}
\end{align}
\revsec{which can be described by:
\begin{empheq}
    [left={\empheqlbrace}]{alignat=2}
    \varphi_{\rm out}^+&=\tan^{-1}\left[-\frac{x_{\rm out}^+\left(2(w_{\rm out}^+)^2-(x_{\rm out}^+)^2\right)}{\left((w_{\rm out}^+)^2-(x_{\rm out}^+)^2\right)^{3/2}}\right],\\
    \varphi_{\rm out}^-&=\tan^{-1}\left[\frac{x_{\rm out}^-\left(2(w_{\rm out}^-)^2-(x_{\rm out}^-)^2\right)}{\left((w_{\rm out}^-)^2-(x_{\rm out}^-)^2\right)^{3/2}}\right].
\end{empheq}
We \revthi{compare} \Equref{eq:v_x,out} with the numerical results later in Sect. \ref{sec:Maximum outflow speed at the midplane}.}

\subsection{\revsec{Analytic formula of the velocity distribution of the outflow}}\label{sec:Analytic formula of the velocity distribution of the outflow}
\revthi{We derive analytic formulae for $v_{x,{\rm out}}(x)$ and $v_{x,{\rm out}}(z)$.} These formulae \revthi{are useful for modeling the influence of the outflow on} the radial drift of dust in disks for the following reasons: (1) \revthi{Dust} particles drift from the outside to the inside of the planetary orbit. The radial drift of dust can be disrupted by the gas flow, leading to the formation of a dust substructure such as a dust ring and a gap \citep{kuwahara2022dust}. (2) The dust is stirred up by the gas turbulence, leading to an increase in the dust scale height \revsec{\citep[e.g.,][]{Youdin:2007}}. Thus, the influence of the gas flow on the dust motion depends on the vertical structure of the planet-induced gas flow.

\subsubsection{\revsec{Radial direction to the disk}}\label{sec:Radial direction to the disk}
The outflow region has a spread in the $x$-direction with the width of $w_{\rm out}^\pm$. Here we aim to model the \revsec{distribution of \revthi{$v_{x,{\rm out}}$}} in the $x$-direction at the midplane, \revthi{$v_{x,{\rm out}}(x)$}. We \revthi{found} that \revthi{$v_{x,{\rm out}}(x)$} \revthi{can} be modeled by a Gaussian function with peaks at $x=x_{\rm out}^\pm$. We introduce the following formula:
\revsix{
\begin{align}
    v_{x,{\rm out}}(x)=v_{x,{\rm out}}^+ \exp\left[-\frac{(x-x_{\rm out}^+)^2}{{2(D^{+})}^2}\right]
-v_{x,{\rm out}}^- \exp\left[-\frac{(x-x_{\rm out}^-)^2}{{2(D^{-})}^2}\right],\label{eq:vx_g_Gaussian}
\end{align}}where we assumed that the full width at half maximum (FWHM) corresponds to the half-width of the outflow region, $\text{FWHM}=2\sqrt{2\ln2}\,D^{\pm}=w_{\rm out}^{\pm}/2$. \revsec{We \revthi{compare} \Equref{eq:vx_g_Gaussian} with the numerical results later in Sect. \ref{sec:Velocity distribution of the outflow}.} 

\subsubsection{\revsec{Vertical direction to the disk}}\label{sec:Vertical direction to the disk}
\revsec{The outflow has a vertical extent as shown in the left column of \Figref{fig:vr_and_massflux}. The outflow speed \revthi{decreases} sharply at $z\gtrsim R_{\rm atm}$. Assuming that \revthi{$v_{x,{\rm out}}(z)$} decreases exponentially when $z>R_{\rm atm}$, we introduce the following formula:
\revthi{
    \begin{align}
    v^\pm_{x,{\rm out}}(z)&=\langle v^\pm_{x,{\rm g}}\rangle_{x,y}|_{z=0}\times\exp\left[-\left(\frac{z}{R_{\rm atm}}\right)^2\right],\nonumber\\
    &=\left.\frac{\int\int\rho_{\rm g}v^\pm_{x,{\rm g}}\mathrm{d}x\mathrm{d}y}{\int\int\rho_{\rm g}\mathrm{d}x\mathrm{d}y}\right|_{\rm z=0}\times\exp\left[-\left(\frac{z}{R_{\rm atm}}\right)^2\right].\label{eq:vx_gas_z_fit}
\end{align}
We used the numerical results to compute the integral in the right-hand side of \Equref{eq:vx_gas_z_fit}.} We \revthi{compare} \Equref{eq:vx_gas_z_fit} with the numerical results later in Sect. \ref{sec:Velocity distribution of the outflow}.}

\section{\revthi{Comparison to numerical results}}\label{sec:Comparison to numerical results}
\revthi{In this section, we compare our analytic formulae for the outflow speed obtained in Sect. \ref{sec:Derivation of analytic formula} with the numerical results.}

\subsection{\revsec{Maximum outflow speed at the midplane}}\label{sec:Maximum outflow speed at the midplane}
Figure \ref{fig:vx_with_analytic} compares our analytic formula for the $x$-component of the outflow speed (\Equref{eq:v_x,out}) with the maximum flow speed in the $x$-direction obtained from hydrodynamical simulations, \revfif{$v_{x,{\rm max}}$}. Equation (\ref{eq:v_x,out}) reproduces the results obtained from hydrodynamical simulations for a wide range of planetary mass and Mach number of the headwind of the gas. 


It should be noted that the point where $|v_{x,{\rm g}}|$ has the maximum value, \revthi{$P^\pm_{\rm max}$,} does not necessarily coincide with the assumed outflow point, $P^\pm_{\rm out}$, \revfif{which is determined by the analysis of the azimuthally averaged mass flux} (\Figref{fig:vx_max_point}). \revsix{Figure \ref{fig:vx_hw0} shows the mismatch between $v_{x,{\rm max}}$ and $v_{x,{\rm g}}$ at the assumed outflow point, $P_{\rm out}^\pm$, meaning that flow speed at the exact location as the analytically-predicted outflow point does not take the maximum value.} This is because the fitting formula for the \rev{critical recycling streamline}, \revfif{$f_{\rm out}(x)$,} does not trace a streamline which gives the maximum flow speed at the midplane region. \revsix{The fitting formula, $f_{\rm out}(x)$, traces the edge of the the outflow region where $v_{x,{\rm g}}$ is dominantly perturbed (e.g., \Figref{fig:vx_max_point}a). It almost coincides with the shear streamline passing closest to the planet near the atmosphere. \revsev{The assumed outflow point, $P_{\rm out}^\pm$, is located at the point very close to the intersection of the shear streamline and the circle of the twice the atmospheric radius, where the perturbation to $v_{x,{\rm g}}$ along the shear streamline is small (\Figref{fig:vx_max_point}).} Thus, the flow speed obtained from hydrodynamical simulations at $P_{\rm out}^\pm$ deviates from the maximum value (\Figref{fig:vx_hw0}).}

\revthi{We found that $P_{\rm max}^\pm$ lies on a circle of radius $\sim1.6\text{--}4R_{\rm atm}$ \revfif{with no clear trend in the location of $P_{\rm max}^\pm$ as a function of the planetary mass and the Mach number of the headwind \revsix{(\Figref{fig:vx_max_point})}. We found that the error between the maximum value of $v_{x,{\rm g}}$ on a circle of radius $2R_{\rm atm}$ and $v_{x,{\rm max}}$ is less than $30\%$. Thus, the twice the atmospheric radius can still be considered as the radius where the dominant outflow occurs. \revsev{For simplicity},} we stick to setting $\sqrt{(x_{\rm out}^\pm)^2+(y_{\rm out}^\pm)^2}=2R_{\rm atm,fit}$ for the derivation of outflow speed.}

\subsubsection{Revisiting the dependence on planetary mass}\label{sec:Revisiting the dependence on planetary mass}
Based on \Figref{fig:vx_with_analytic}, we revisit the dependence of \revthi{$v_{x,{\rm out}}$} on the planetary mass as mentioned in Sect. \ref{sec:Gas flow speed at the midplane: Dependence on planetary mass}. The outflow speed in the $x$-direction, \revthi{$v_{x,{\rm out}}$}, has a peak when $m\sim0.3$, corresponding to a super-Earth mass planet at $1$ au for a typical steady accretion disk model. From the \revthi{analytic} point of view, this trend can be understood by considering the dependence of the outflow angle on the planetary mass (\Figref{fig:outflow_angle}). In this section, we only consider the shear only case ($\mathcal{M}_{\rm hw}=0$) for simplicity. 

When $m\lesssim0.3$, the outflow speed is proportional to $|v_{\rm out}|\propto x_{\rm out}\propto \,R_{\rm atm,fit}\propto m$ (\Equref{eq:|v_out|}). We found that the cosine term in \Equref{eq:v_x,out}, $\cos\varphi_{\rm out}$, is proportional to $\sim m^{-1/3}$ (\Figref{fig:outflow_angle}). Thus, the outflow speed in the $x$-direction is almost proportional to $v^\pm_{x,{\rm out}}\propto m^{2/3}$. When $m\gtrsim0.3$, the dependence of the outflow speed on the planetary mass is slightly lower than the power of $1/3$ due to the cosine term in \Equref{eq:x_out}, $\cos(\tan^{-1}(2\,R_{\rm atm,fit}/w_{\rm out}^\pm))$ (\Figref{fig:outflow_angle}). The cosine term in \Equref{eq:v_x,out}, $\cos\varphi_{\rm out}$, is proportional to $\sim m^{-1/3}$ (\Figref{fig:outflow_angle}).
Thus, \revthi{$v_{x,{\rm out}}$} decreases as the planetary mass increases.

Our analytic formula deviates from the results obtained by hydrodynamical simulations when $m\gtrsim0.6$ (\Figref{fig:vx_with_analytic}).  For such higher-mass planets, we found that \revthi{$P^{\pm}_{\rm max}$} is located at the region close to the planetary orbit (\Figref{fig:vx_max_point}b). In \Figref{fig:vx_max_point}b, \revthi{$v_{x,{\rm g}}$} has the maximum value on the midplane horseshoe streamline, not on the critical recycling streamline. Thus, the flow speed cannot be estimated by \Equref{eq:v_x,out}. \revthi{As discussed in Sect. \ref{sec:Morphology of the critical recycling streamline at the midplane}, our model for the critical recycling streamline can be applied when $m\lesssim m_{\rm iso}\simeq0.6$. Thus, our analytic formula for the outflow speed is also valid when $m\lesssim m_{\rm iso}$. We plotted $m_{\rm iso}$ (\Equref{eq:m_iso}) in \Figref{fig:vx_with_analytic}.} 

\subsubsection{Revisiting the dependence on headwind}\label{sec:Revisiting the dependence on headwind}
Based on \Figref{fig:vx_with_analytic}, we revisit the dependence of \revthi{$v_{x,{\rm out}}$} on the Mach number of the headwind, $\mathcal{M}_{\rm hw}$, as mentioned in Sect. \ref{sec:Gas flow speed at the midplane: Dependence on headwind}. The outflow speed in the $x$-direction toward the inside (outside) of the planetary orbit decreases (increases) as $\mathcal{M}_{\rm hw}$ increases. This is caused by the second term in the square root in the right-hand side of \Equref{eq:|v_out|}, which \revsix{comes from} the global pressure force due to the sub-Keplerian motion of the gas, $\bm{F}_{\rm hw}=2\mathcal{M}_{\rm hw}\bm{e}_x$. Since the gas receives the global pressure force in the positive direction in the $x$-axis, the outflow toward the inside (outside) of the planetary orbit is reduced (enhanced).

When $\mathcal{M}_{\rm hw}=0.1$, the strongest Mach number considered in this study, our analytic formula deviates from the results obtained by hydrodynamical simulations inside the planetary orbit. We speculate that the deviation was caused by \revthi{the changes of the actual location of the inflow point, $P_{\rm in}$}. To obtain the analytic formula for the outflow speed, we set \revthi{$P_{\rm in}$} at the pole of the spherical shell of the radius $2\,R_{\rm atm, fit}$ regardless of the assumed $\mathcal{M}_{\rm hw}$. In reality, \revthi{$P_{\rm in}$} shifts \revthi{to} the negative direction in the $x$-axis as $\mathcal{M}_{\rm hw}$ increases (\Figref{fig:vr_and_massflux}b), because the horseshoe flow shifts to the negative direction in the $x$-axis. However, as long as we consider the influence of the outflow on the radial drift of dust, the outflow toward the inside of the planetary orbit \revthi{is} less important than that toward the outside the planetary orbit.

\subsection{Velocity distribution of the outflow}\label{sec:Velocity distribution of the outflow}
\revsec{Figure \ref{fig:vx_dist_m02} shows \revthi{$v_{x,{\rm out}}(x)$} at the midplane obtained from both our analytic formula (\Equref{eq:vx_g_Gaussian}) and hydrodynamical simulations. We set $m=0.2$ and varied the Mach number of the headwind, $\mathcal{M}_{\rm hw}=0\text{--}0.1$ in \Figref{fig:vx_dist_m02}. We found that \Equref{eq:vx_g_Gaussian} can reproduce the trends of the hydro-simulations results: the peak speed, the full width at half maximum, and the $x$-coordinate of the peak position.}

\revsec{Figure \ref{fig:vx_z_dependence} shows \revthi{$v_{x,{\rm out}}(z)$} for different planetary masses, $m=0.2,\text{and }0.5$, and Mach numbers, $\mathcal{M}_{\rm hw}=0,\text{and }0.1$.} The left column of \Figref{fig:vx_z_dependence} shows the vertical structure of the gas flow field at the meridian plane \revsec{obtained from hydrodynamical simulations.} The gas flow velocity was averaged in the $y$-direction within the calculation domain of hydrodynamical simulation. The right column of \Figref{fig:vx_z_dependence} \revsec{shows \revthi{$v_{x,{\rm out}}(z)$}} obtained from both the numerical result and our analytic formula. \revfif{The numerical result was obtained by averaging the simulation data, as:}
\revthi{
\begin{align}
    \displaystyle v^{\rm num}_{x,{\rm out}}(z)\equiv\langle v^\pm_{x,{\rm g}}(z)\rangle_{x,y}=\frac{\int\int\rho_{\rm g}v^\pm_{x,{\rm g}}(z)\mathrm{d}x\mathrm{d}y}{\int\int\rho_{\rm g}\mathrm{d}x\mathrm{d}y}.\label{eq:vx gas at midplane}
\end{align}}The interval of integration in the $y$-direction was $|y|\leq r_{\rm out}$. In the $x$-direction, the interval of integration was $0\leq x\leq w_{\rm out}^+$ for $\langle v^+_{x,{\rm g}}\rangle_{x,y}$ and $w_{\rm out}^-\leq x\leq 0$ for $\langle v^-_{x,{\rm g}}\rangle_{x,y}$.

\revsix{Our fitting law for $v_{x,{\rm out}}(z)$ introduced in \Equref{eq:vx_gas_z_fit} is indeed a practical representation of $v^{\rm num}_{x,{\rm out}}(z)$ for the lower-mass planets}, $m\lesssim0.4$ (Figs. \ref{fig:vx_z_dependence}b and d). For the higher-mass planets, $m\gtrsim0.5$, \Equref{eq:vx_gas_z_fit} reproduces a decreasing trend of \revthi{$v^{\rm num}_{x,{\rm out}}(z)$} only in the range of $z<R_{\rm atm}$. For the higher-mass planets, the meridional circulation of the gas above the outflow leads to the complex behavior of the flow speed (\Figref{fig:vx_z_dependence}e). \revthi{As discussed in Sects. \ref{sec:Morphology of the critical recycling streamline at the midplane} and \ref{sec:Revisiting the dependence on planetary mass}, our analytic formula for \revthi{$v_{x,{\rm out}}(z)$} cannot be applied to the higher-mass regime ($m\gtrsim m_{\rm iso}$).}

\section{Discussion}\label{sec:Discussion}

\begin{figure}[htbp]
    \centering
    \includegraphics[width=\linewidth]{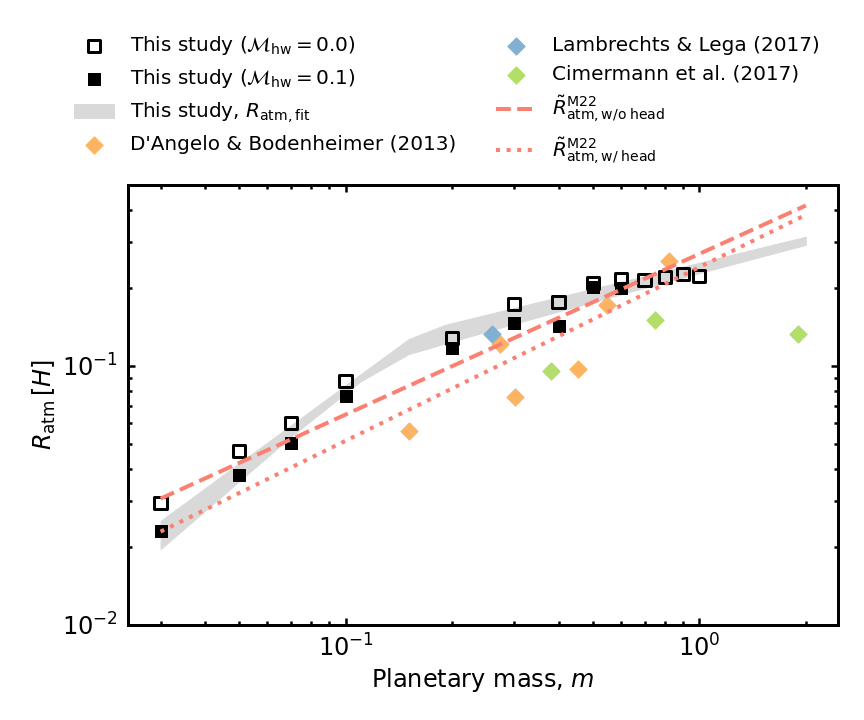}
    \caption{Comparison of the atmospheric radii obtained in our study and those in previous studies. \rev{The open and filled square symbols were obtained from our hydrodynamical simulations, $\mathcal{M}_{\rm hw}=0$ and $0.1$, respectively. The gray-shaded region is given by the fitting formula for the atmospheric radius, $R_{\rm atm,fit}$ (\Equref{eq:Ratm fit}). The red dashed and dotted lines are given by \Equref{eq:R atm M22} \citep{moldenhauer2022recycling}. \revthi{We extrapolated \Equref{eq:R atm M22} to the range of $m\leq0.3$. The yellow, blue, and green diamonds corresponds to the results of \cite{DAngelo:2013}, \cite{Lambrechts:2017}, and \cite{Cimerman:2017}, respectively. \revthi{To plot the results of \cite{DAngelo:2013}, the atmospheric radii were calculated by $R_{\rm atm}/H=C^{\rm DB13}a/H=C^{\rm DB13}/h$. The values of $C^{\rm DB13}=2.6\times10^{-3}\text{--}9.6\times10^{-3}$ were given in Table 3 of \cite{DAngelo:2013}. \revsev{We used $h=\max(0.027\,(a/1\,\text{au})^{1/20},\,0.024\,(a/1\,\text{au})^{2/7})$ (Appendix \ref{sec:Aspect ratio of the steady accretion disk}).}}}}}
    \label{fig:Ratm_with_previous_studies}
\end{figure}

\begin{figure}[htbp]
    \centering
    \includegraphics[width=\linewidth]{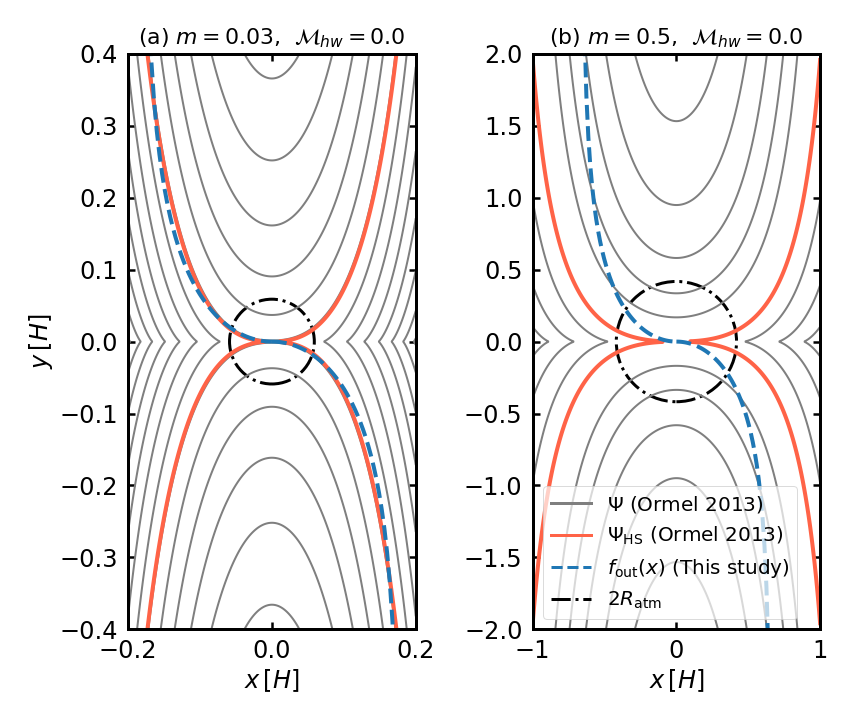}
    \caption{Isocontours of the stream function (the gray solid lines), $\Psi$, and the fitting formula of the \rev{critical recycling streamline at the midplane} (\Equref{eq:outflow shape in cartesian}; the blue dashed lines). The widest horseshoe streamlines characterized by $\Psi_{\rm HS}$ are highlighted with red. Isocontours of $\Psi=0.02\text{--}0.1$ (\textit{panel a}) and $\Psi=0.42\text{--}1.7$ (\textit{panel b}) are shown.}
    \label{fig:vx_and_stream_func_hydrooff}
\end{figure}

\begin{figure}[htbp]
    \centering
    \includegraphics[width=\linewidth]{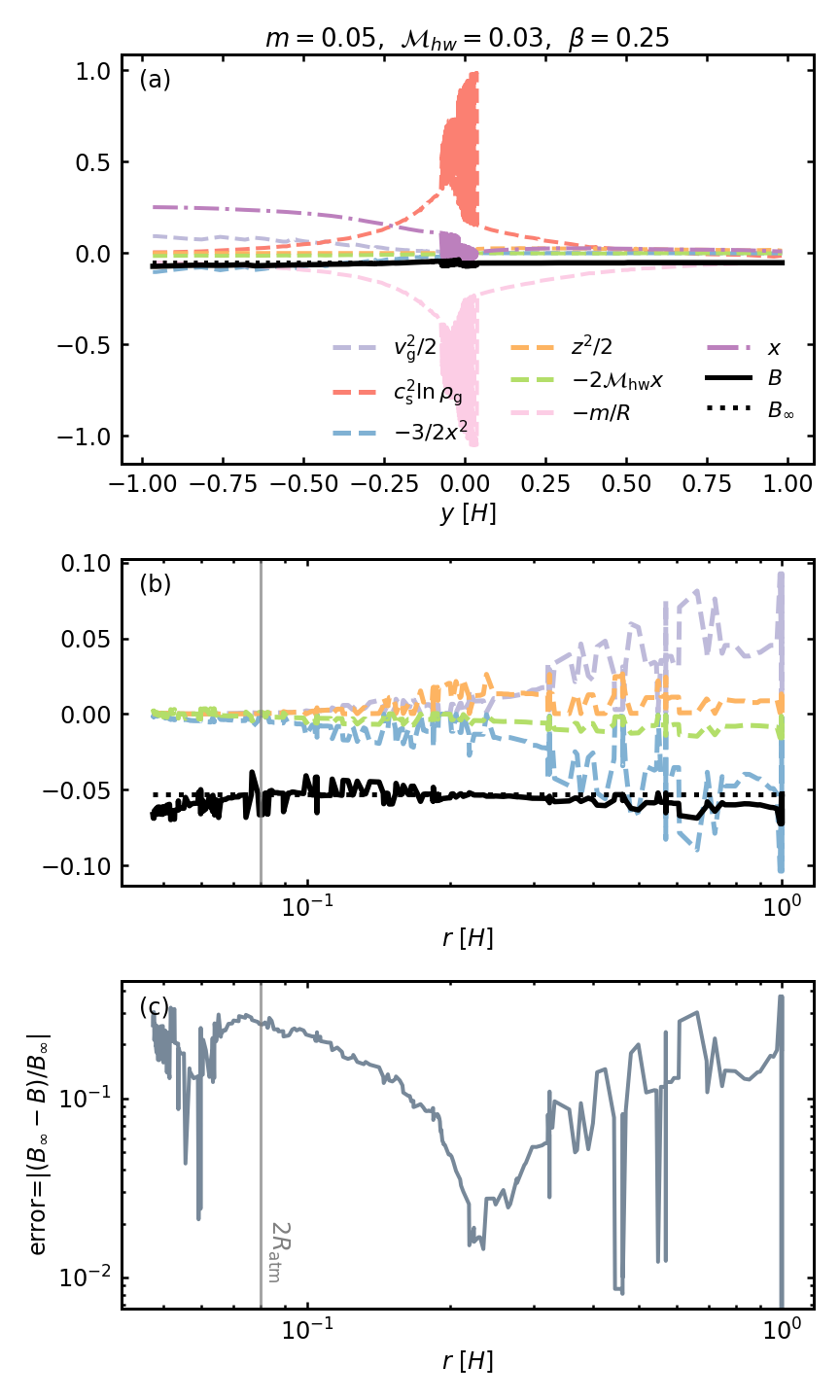}
    \caption{\textit{Panels a and b:} Quantities along a streamline through $(x,y,z)=(0.001, 0.01, 0.2)$ (the colored solid line shown in \Figref{fig:streamline_m005_Mhw003}c). The results were obtained from \texttt{m005-hw003}. The black solid and dotted lines represent the Bernoulli’s function, $\mathcal{B}$, and that in the far field, $\mathcal{B}_{\infty}$. The colored dashed lines correspond to the first to sixth terms in the Bernoulli’s function. The purple dashed-dotted line in \textit{panel a} corresponds to the $x$-coordinate of the streamline.   \textit{Panel c}: Relative error between $\mathcal{B}$ and $\mathcal{B}_\infty$. The vertical solid lines in \textit{panels b and c} denote the twice the atmospheric radius. This figure should be compared to \Figref{fig:streamline_m005_Mhw003}c.}
    \label{fig:B_m005}
\end{figure}

\begin{figure}[htbp]
    \centering
    \includegraphics[width=\linewidth]{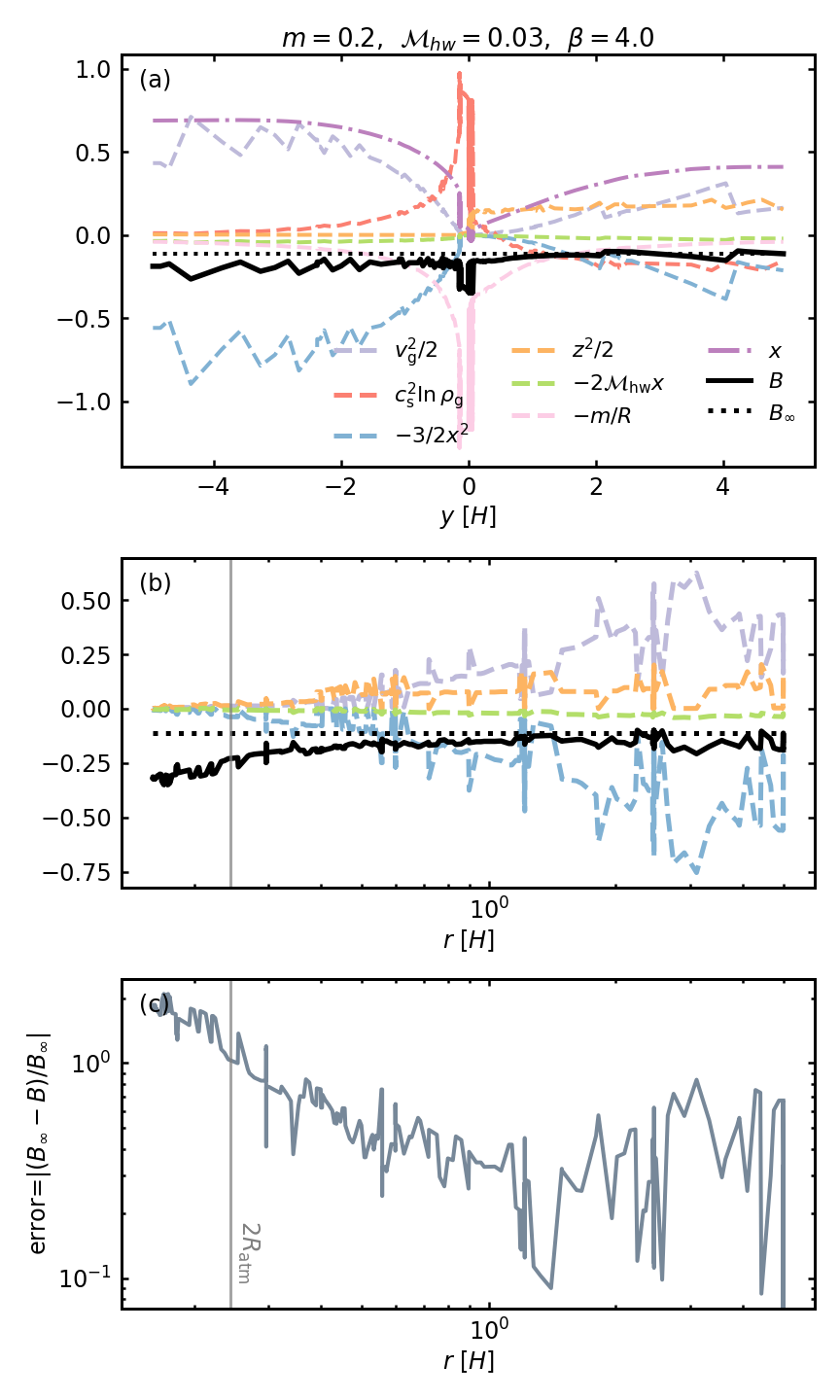}
    \caption{Same as \Figref{fig:B_m005}, but the results were obtained from \texttt{m02-003}. We picked up a streamline through $(x,y,z)=(0.001, 0.01, 0.44)$.}
    \label{fig:B_m02}
\end{figure}

\begin{figure}[htbp]
    \centering
    \includegraphics[width=\linewidth]{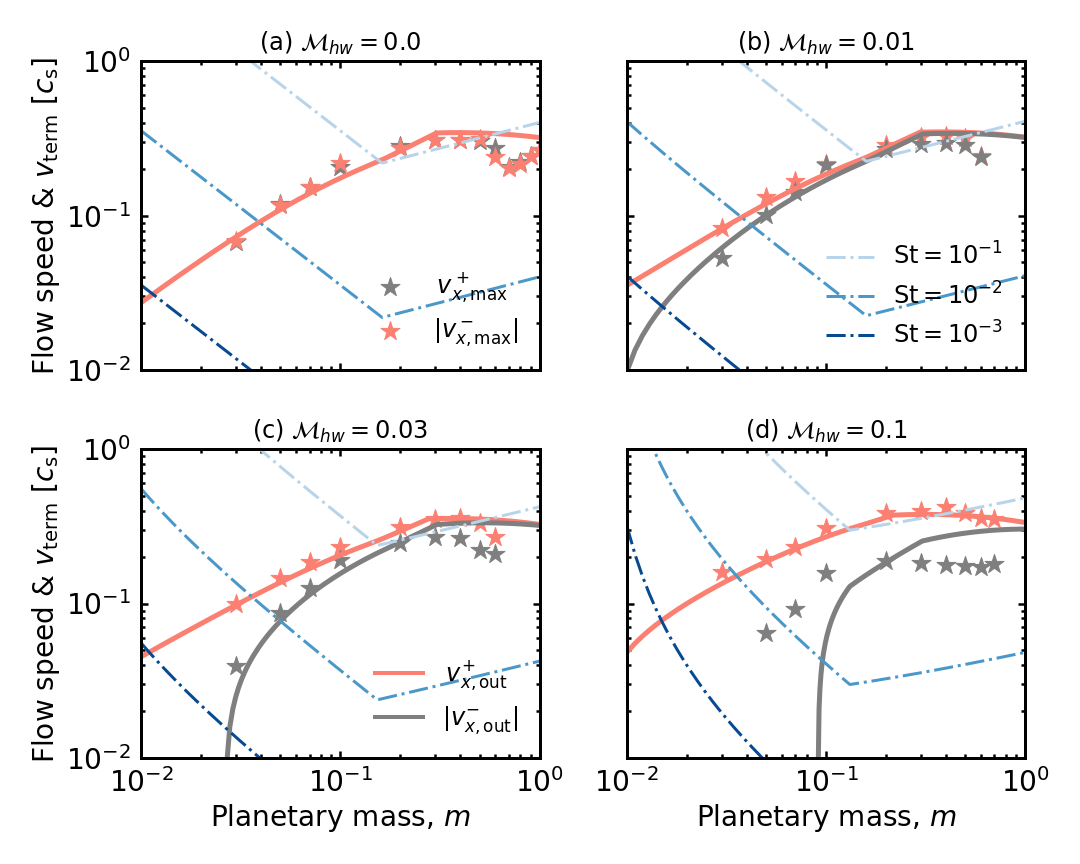}
    \caption{Same as \Figref{fig:vx_with_analytic}, but with the terminal velocity of dust (\Equref{eq:terminal speed of dust}; blue dashed-dotted lines). Different colors correspond to different Stokes numbers. The vertical axis is on a log scale.}
    \label{fig:vx_with_analytic_vpeb}
\end{figure}

\begin{figure*}[htbp]
    \centering
    \includegraphics[width=\linewidth]{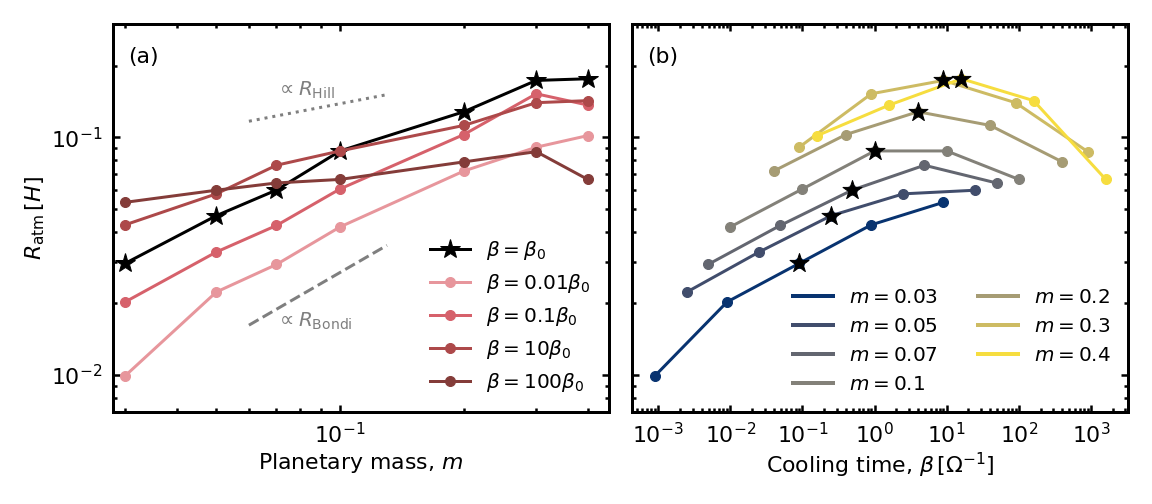}
    \caption{Atmospheric radius as a function of the planetary mass and the cooling time, $\beta$. We set $\mathcal{M}_{\rm hw}=0$. The star symbols correspond to the results obtained from the fiducial runs (\Figref{fig:Ratm}).}
    \label{fig:Ratm_beta_2panel}
\end{figure*}

\begin{figure*}[htbp]
    \centering
    \includegraphics[width=\linewidth]{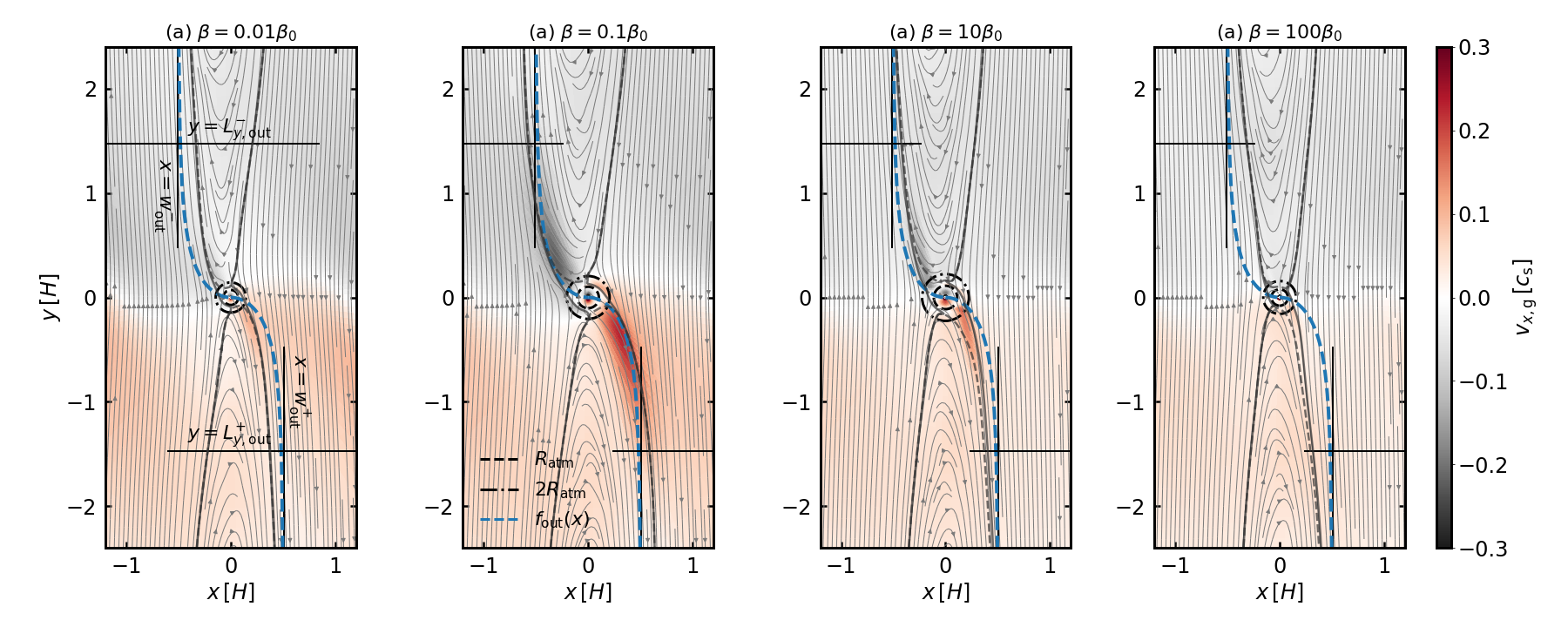}
    \caption{Same as \Figref{fig:vx_m_dependence}, but \rev{this figure shows the dependence of the outflow speed on the cooling time. We set $m=0.2$ in all panels. The results were obtained from} \texttt{m02-hw0-b001} (\textit{panel a}), \texttt{m02-hw0-b01} (\textit{panel b}), \texttt{m02-hw0-b10} (\textit{panel c}), and \texttt{m02-hw0-b100} (\textit{panel d}), respectively.}
    \label{fig:vx_beta_dependence_m02}
\end{figure*}

\begin{figure}[htbp]
    \centering
    \includegraphics[width=\linewidth]{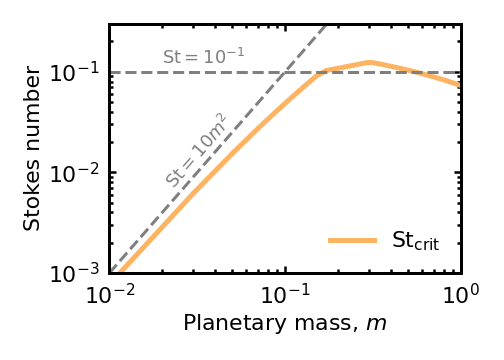}
    \caption{\revsec{Upper limit of the Stokes number of dust that affected by the outflow as a function of the planetary mass (\Equref{eq:St criteria}).}}
    \label{fig:Stcrit}
\end{figure}  

\subsection{Comparison to previous studies: Atmospheric radius}\label{Comparison to previous studies: Atmospheric radius}
An atmospheric radius is crucial to understand gas dynamics around an embedded planet. Several previous studies reported the different scaling laws for the atmospheric radius which are summarized in \Figref{fig:Ratm_with_previous_studies}. In this study, we estimated the atmospheric radius based on the definition given by \cite{moldenhauer2022recycling} and obtained a fitting formula (\Equref{eq:Ratm fit}). It should be noted that \Equref{eq:Ratm fit} differs from the scaling law obtained by \cite{moldenhauer2022recycling}. \cite{moldenhauer2022recycling} performed 3D radiation-hydrodynamics simulations in a local frame centered at the planet with $M_{\rm pl}=1\text{--}10\,M_\oplus$ at $0.1$ au (corresponding to $m\sim0.3\text{--}3$ in our dimensionless unit). Based on \Equref{eq:atmospheric radius}, the authors estimated the atmospheric radius and introduced the following fitting formula: $R_{\rm atm}^{\rm M22}=C^{\rm M22}(M_{\rm pl}/M_{\oplus})^n\,R_{\oplus}$, where $R_\oplus$ is the Earth radius, $C^{\rm M22}=4.7\pm0.6$ and $n=0.67\pm0.08$ when the headwind is included, and $C^{\rm M22}=5.7\pm0.3$ and $n=0.62\pm0.04$ without the headwind. In our dimensionless unit, this can be rewritten as:
\begin{align}    
    \tilde{R}_{\rm atm}^{\rm M22}\equiv\frac{R_{\rm atm}^{\rm M22}}{H}=&C^{\rm M22}\frac{(3.3\times10^5)^n}{2.3\times10^3}\Biggl(\frac{M_\ast}{1\,M_\odot}\Biggr)^n\Biggl(\frac{\rev{a}}{0.1\,\text{au}}\Biggr)^{-1}\,m^nh^{3n-1},\nonumber\\\simeq&
    \begin{cases}
        0.24m^{0.67}\equiv\tilde{R}^{\rm M22}_{\rm atm,w/\;head},\\
        0.27m^{0.62}\equiv\tilde{R}^{\rm M22}_{\rm atm,w/o\;head},\label{eq:R atm M22}
    \end{cases}
\end{align}
where we assumed $h\simeq0.024$ for a typical steady accretion disk at $0.1$ au with a solar-mass star, the typical accretion rate of classical T Tauri stars, $\dot{M}_\ast=10^{-8}\,M_\odot\text{/yr}$, and $\alpha_{\rm acc}=10^{-3}$ \citep{Ida:2016}. \cite{moldenhauer2022recycling} suggested that the atmospheric radius increases approximately proportional to $\sim R_{\rm Hill}^2$.

\cite{DAngelo:2013} performed global 3D radiation-hydrodyamic simulations for the embedded planets with \revthi{$M_{\rm pl}=5,\,10,\,\text{and }15\,M_{\oplus}$ at $5$ au (corresponding to $m\simeq0.27,\,0.55,\,\text{and }0.82$) and $10$ au ($m\simeq0.15,\,0.3,\,\text{and }0.45$ in our dimensionless unit, respectively}; \rev{\Figref{fig:physical_quantity2panel}}). In \cite{DAngelo:2013}, a realistic equation of state and a dust opacity were considered. \revthi{Using passive tracers to determine the size of the bound atmosphere, the authors estimated the atmospheric radius as $\sim R_{\rm Hill}/5\text{--}R_{\rm Hill}/3$ \citep[see Table 3 of][for details]{DAngelo:2013}.}

\cite{Lambrechts:2017} performed 3D global radiation-hydrodynamic simulations with an embedded planet, $M_{\rm pl}=5\,M_{\oplus}$ at $5.2$ au (corresponding to $m\simeq0.26$ in our dimensionless unit; \rev{\Figref{fig:physical_quantity2panel}}). The authors considered the dust opacity and the luminosity from the accretion of solid. From the analysis of the mass flux within the Hill sphere of the planet, the authors estimated the atmospheric radius as $\sim0.3\,R_{\rm Hill}$. \cite{Cimerman:2017} performed 3D radiation-hydrodynamic simulations in a local frame co-rotating with the planet. The authors defined the atmosphere as the region where the azimuthally averaged ratio of radial to azimuthal velocities falls below a threshold of 10\%, $\langle |v_r/v_\phi|\rangle_{\phi}\leq0.1$. They reported that the atmospheric radii were $0.25\,R_{\rm Bondi},\,0.2\,R_{\rm Bondi},\,\text{and}\,0.07\,R_{\rm Bondi}$ for the planets with $m=0.38,\,0.75,$ and $1.9$, respectively.

Figure \ref{fig:Ratm_with_previous_studies} compares the atmospheric radii obtained from above-mentioned previous studies with those obtained from our study. We emphasize that the definition of the atmospheric radius differs among previous studies. Although a direct comparison of this study with previous studies is difficult due to different \rev{simulation} settings among these studies, our fiducial results agree with the results of previous studies within a factor of $\sim2$. We note that the atmospheric radius depends on the cooling time, $\beta$. \revsix{The atmospheric radius changes by a factor of $\sim2\text{--}3$ when we varied $\beta$ by $\sim1\text{--}2$ orders of magnitude.} 
We discuss the dependence of the atmospheric radius on the cooling time later in Sect. \ref{sec:Dependence on cooling time}.

\subsection{Comparison to previous studies: Outflow speed}\label{Comparison to previous studies: Outflow speed}
Several previous studies proposed analytic formulae of the outflow speed which were plotted in \Figref{fig:vx_with_analytic}a. \cite{Ormel:2013} considered 2D, isothermal, inviscid, and compressible flow past a planet. The author derived an approximate solution of the flow pattern around the planet in the linear regime in terms of the stream function:
\begin{align}
    \Psi=\Psi_\infty+\Psi_{y0},\label{eq:ormel 1}
\end{align}
where the first and second terms in the right-hand side of \Equref{eq:ormel 1} correspond to the unperturbed and perturbed components. Each term can be described by:
\begin{align}
    \Psi_\infty&=\mathcal{M}_{\rm hw}x+\frac{3}{4}x^2,\label{eq:ormel 2}\\
    \Psi_{y0}&=\frac{\left(\pi|y|-2y\,{\rm Si}(\tilde{y})\right)\cos\tilde{y}+2y\,{\rm Ci}(\tilde{y})\sin\tilde{y}}{\sqrt{2}|y|},\label{eq:ormel 3}
\end{align}
where $\tilde{y}=y/\sqrt{2}$. In the right-hand side of \Equref{eq:ormel 3}, ${\rm Si}(y)$ and ${\rm Ci}(y)$ are the sine and cosine integral functions. 

The flow speed in the $x$-direction (the radial direction to the disk) is given by $v_{x,{\rm g}}\approx\partial \Psi_{y0}/\partial y$ in the linear regime. \cite{Ormel:2013} obtained the following formula:
\begin{align}
    v_{x,{\rm out}}^{\rm O13}=m\Biggl[{\rm Ci}(\tilde{y})\cos\tilde{y}-\frac{1}{2}\left(\pi-2\,{\rm Si}(\tilde{y})\right)\sin\tilde{y}\Biggr],\label{eq:v_x,out O13}
\end{align}
where $y>0$. In a procedure of the derivation of \Equref{eq:v_x,out O13}, the contribution of the headwind was neglected. Equation (\ref{eq:v_x,out O13}) gives \revthi{$v_{x,{\rm g}}$} within the horseshoe region as a function of $y$ in the shear only case ($\mathcal{M}_{\rm hw}=0$). In the absence of the headwind, the 2D flow filed has a symmetric structure with respect to the $x$- and $y$-axes. Owing to the symmetry, here we can only consider $x<0$ and $y>0$. Assuming that the widest horseshoe streamline can be outlined by our fitting formula (\Equref{eq:outflow shape in cartesian}), we substituted
\begin{align}
    y=\frac{(x_{\rm out}^-)^2}{\sqrt{(w_{\rm out}^-)^2-(x_{\rm out}^-)^2}},
\end{align}
into \Equref{eq:v_x,out O13}. This gives us \revthi{$v_{x,{\rm g}}$} at $(x_{\rm out}^-,\,f_{\rm out}(x_{\rm out}^-))$ on the widest inner-leading horseshoe streamline. The result was plotted in \Figref{fig:vx_with_analytic}a with the blue dashed line. 

\cite{Ormel:2013} considered the 2D flow, where the 3D recycling streamlines do not appear. Nevertheless, \Equref{eq:v_x,out O13} agrees with our results when $m\lesssim0.1$ (\Figref{fig:vx_with_analytic}a). We speculate that this comes from the close relation between the outflow region and the horseshoe streamlines. Figure \ref{fig:vx_and_stream_func_hydrooff} shows isocontours of the stream function, $\Psi$, and the fitting formula for the outflow region, $f_{\rm out}(x)$. In \Figref{fig:vx_and_stream_func_hydrooff}a, where we set $m=0.03$, we found that our fitting formula is in excellent agreement with the widest horseshoe streamline characterized by $\Psi=\Psi_{\rm HS}=\pi m/\sqrt{2}$ \citep{Ormel:2013}. This may lead to the agreement in the results of \revthi{$v_{x,{\rm out}}$} between our study and \cite{Ormel:2013}. The linearlized approximation breaks down for the higher-mass planets. In \Figref{fig:vx_and_stream_func_hydrooff}b, where we set $m=0.5$, we found a significant mismatch between $f_{\rm out}(x)$ and $\Psi$. Thus, \Equref{eq:v_x,out O13} does not match with \revthi{$v_{x,{\rm out}}$} obtained in this study when $m\gtrsim0.3$.

\cite{Fung:2015} considered isothermal and viscous flow. The authors performed 3D hydrodynamical simulation for the planet with $m=0.56$ and found that the outflow speed at the midplane region was $\sim0.2\text{--}0.4c_{\rm s}$. \cite{Fung:2015} applied Bernoulli’s theorem along with a transient horseshoe streamline and derived the outflow speed as:
\begin{align}
    |v_{\rm out}^{F15}|=\sqrt{v_{\rm in}^2+2m\left(\frac{1}{D_{\rm out}}-\frac{1}{\sqrt{D_{\rm in}^2+D_{\rm out}^2}}\right)},\label{eq:v_out fung}
\end{align}
where $v_{\rm in}$ is the inflow speed, $D_{\rm in}$ and $D_{\rm out}$ are the distance of the inflow and outflow points from the planet. Following \cite{Fung:2015}, we substituted $D_{\rm in}\simeq H$ and $D_{\rm out}\simeq w_{\rm HS}$ into \Equref{eq:v_out fung}, and neglected $v_{\rm in}^2$. Additionally, here we assume that the outflow occurs with an angle of $\sim\pm\pi/4$ to the $x$-axis \citep[e.g., Fig.8 of][]{Fung:2015}. In \Figref{fig:vx_with_analytic}a, we plotted the following modified formula for the outflow speed in the $x$-direction with the purple dashed line:
\begin{align}
    v_{x,{\rm out}}^{\rm F15}=\sqrt{2m\left(\frac{1}{w_{\rm HS}}-\frac{1}{\sqrt{H^2+w_{\rm HS}^2}}\right)}\times\cos(\pi/4).\label{eq:v_x,out fung}
\end{align}
As discussed in \cite{Fung:2015}, \Equref{eq:v_out fung} should be considered an upper limit because the height of the inflow point from the midplane assumed in \cite{Fung:2015} was higher than ours. 
\revsix{One can see that \Equref{eq:v_x,out fung} overestimates the outflow speed compared to our results.}

\cite{Kuwahara:2019} considered isothermal and inviscid flow and performed 3D hydrodynamical simulations for the planets with $m=0.01\text{--}2$. \cite{Kuwahara:2019} proposed that the outflow speed can be described by:
\begin{align}
    |v_{{\rm out}}^{\rm K19}|=\sqrt{3 \,R_{\rm grav}^2}.\label{eq:|v_out K19|}
\end{align}
Equation (\ref{eq:|v_out K19|}) represents the flow speed at the intersections of a line of $y=-x$ and a circle of radius $R_{\rm grav}$ in the shear only case.  Assuming that the critical recycling streamline at the midplane can be outlined by \Equref{eq:outflow shape in cartesian}, we plotted the following modified formula for the $x$-component of the outflow speed in \Figref{fig:vx_with_analytic}a with the yellow dashed line:
\begin{align}
    v_{x,{\rm out}}^{\rm K19}&=\sqrt{3 \,R_{\rm grav}^2}\cos\varphi_{\rm out},\nonumber\\
    &=\sqrt{3 \,R_{\rm grav}^2}\cos\left[\tan^{-1}\left.\left(\frac{\mathrm{d}f_{\rm out}(x)}{\mathrm{d}x}\right)\right|_{x=x_{\rm out}^{K19}}\right],\label{eq:v_x,out K19}
\end{align}
where we assumed $x_{\rm out}^{\rm K19}=R_{\rm grav}\cos(\pi/4)$ \citep{Kuwahara:2019}. \revsix{Equation (\ref{eq:v_x,out K19}) underestimates the outflow speed compared to our results for a wide range of planetary masses, $m\lesssim0.3$.} This is caused by the following assumptions: (1) \cite{Kuwahara:2019} considered an isothermal flow. (2) \cite{Kuwahara:2019} set the inflow and outflow points on the sphere of radius $R_{\rm grav}$.

In an isothermal case, there is no clear boundary \revsix{demarcating} the bound atmosphere from the disk gas \citep{Ormel:2015b}. The inflow of the gas intrudes deep inside the gravitational sphere of the planet. Thus, in the isothermal case, the inflow and outflow points were located at the points closer to the planetary core than in the nonisothermal case, leading to an underestimation of our results. The atmospheric radius positively correlated to the cooling timescale, $\beta$, and the \revsix{atmosphere} disappear in the isothermal limit, $\beta\rightarrow0$ \citep{Kurokawa:2018}.

\subsection{Applicability of Bernoulli's theorem}\label{sec:Applicability of Bernoulli's theorem}
In Sect. \ref{sec:Analytic formula of the outflow speed}, assuming an isothermal flow along a recycling streamline, we estimated the outflow speed using Bernoulli’s theorem. The applicability of Bernoulli’s theorem depends on the thermodynamics of the gas along the streamline. Our hydrodynamical simulations utilized the $\beta$ cooling model in which the \revsix{$\beta$ cooling} occurs on a finite timescale. Thus, the Bernoulli's function, $\mathcal{B}$, does not strictly conserve along the streamline (\Figref{fig:streamline_m005_Mhw003}c).

Let us begin with a qualitative discussion. When a parcel of the gas approaches to the planet and reaches the region where $\rev{r}\lesssim R_{\rm grav}$, the temperature of the parcel increases due to the adiabatic compression. The temperature of the gas relaxes toward the background temperature with the dimensionless timescale $\beta=(m/0.1)^2$ (\Equref{eq:beta}). Assuming that the gas parcel has the shear velocity, the crossing time on which the gas crosses the gravitational sphere of the planet can be estimated by the shear timescale, \revsix{$\tau_{\rm cross}\sim\Omega^{-1}$}. Thus, the isothermal assumption would be valid when $\beta\lesssim1$. \revsev{We note that $\tau_{\rm cross}$ \revsev{can} be longer than the shear timescale, because the gas parcel along the recycling streamline sometimes circulates the atmosphere, which leads to an increase in $\tau_{\rm cross}$ (\Figref{fig:streamline_m005_Mhw003}a).}

Figures \ref{fig:B_m005} and \ref{fig:B_m02} show the quantitative description of the Bernoulli's function along a certain recycling streamline. Panels a and b in Figs. \ref{fig:B_m005} and \ref{fig:B_m02} represent the changes of $\mathcal{B}$ and each term of $\mathcal{B}$ as a function of $y$ and \revthi{$r$}. Panel c of Figs. \ref{fig:B_m005} and \ref{fig:B_m02} shows a relative error between the Bernoulli's function and the unperturbed one, $\mathcal{B}_\infty$. We set $\mathcal{B}_\infty=\mathcal{B}|_{r=r_{\rm out}}$, where $\mathcal{B}|_{r=r_{\rm out}}$ corresponds to the value of $\mathcal{B}$ at the edge of the computational domain of hydrodynamical simulations, i.e., the starting point of the streamline.

In \Figref{fig:B_m005}, where we set $m=0.05$ and $\beta=0.25$, the Bernoulli's function is almost constant. The relative error between $\mathcal{B}$ and $\mathcal{B}_\infty$ is on the order of $\sim10^{-2}\text{--}10^{-1}$ and increases as the gas approaches to the planet. Since the cooling time depends on the planetary mass, one would expect that the amplitude of the fluctuation of \revthi{$\mathcal{B}$} increases as the planetary mass increases. This can be seen in \Figref{fig:B_m02}, where we set $m=0.2$ and $\beta=4$. Overall trends of \Figref{fig:B_m02} are similar to those of \Figref{fig:B_m005}, but the relative error between $\mathcal{B}$ and $\mathcal{B}_\infty$ for $m=0.2$ is $\sim10^{-1}\text{--}10^0$ which is larger than that for $m=0.05$. We found that the relative error between $\mathcal{B}$ and $\mathcal{B}_\infty$ is more larger for the higher-mass planets, $m>0.2$ ($\beta\gtrsim10$). This could lead to a mismatch between our analytic formula for the outflow speed and the numerical results in particular for a higher-mass planet (\Figref{fig:vx_with_analytic}).


\subsection{Dependence on cooling time}\label{sec:Dependence on cooling time}
\revsec{So far we only showed the results} obtained from the fiducial runs in which we considered a fixed cooling time, $\beta_0$, for a given planetary mass (\Equref{eq:beta}). \revsev{To determine the $\beta_0$ value, we first considered that an inflow of a disk gas into an envelope would be heated by adiabatic compression (Sect. \ref{sec:Methods of three-dimensional hydrodynamical simulations}). We expect that the length scale of the temperature perturbation to the inflow of the disk gas can be estimated by the size of the gravitational sphere of the planet, $R_{\rm Bondi}$ or $R_{\rm Hill}$. In this study, following \cite{Kurokawa:2018}, we assumed that the length scale of the temperature perturbation is equal to $R_{\rm Bondi}$. The $\beta_0$ value represents the cooling time near the outer edge of the atmosphere. As long as we focus on the recycling flow, we consider that $\beta_0$ can be used as the typical value based on the following reasons: (1) the recycling flow traces the outer edge of the atmosphere. (2) When the dimensionless planetary mass falls below $m<0.6$, which is the important mass range in this study, we always have $R_{\rm grav}=R_{\rm Bondi}$.}

\revsev{However, the $\beta$ value could vary within the atmosphere, which becomes larger in the dense region of the atmosphere. The $\beta$ value also depends on the dust opacity in the atmosphere, which could be changed by orders of magnitude \citep{movshovitz2008opacity,ormel2014atmospheric,mordasini2014grain}.} In this section, we discuss the dependence of our results on the cooling time, \revthi{$\beta$}. An atmospheric radius, an important quantity for gas dynamics around an embedded planet, depends on $\beta$. Qualitatively, in both the isothermal ($\beta\rightarrow0$) and adiabatic ($\beta\rightarrow\infty$) limits, a bound atmosphere does not appear \citep{Kurokawa:2018,Popovas:2018b} \revthi{or is very small \citep[$<0.2\,R_{\rm Bondi}$;][]{Fung:2019}}. Figure \ref{fig:Ratm_beta_2panel} shows the dependence of the atmospheric radius on $\beta$. \revsix{We found that the atmospheric radius can be scaled with the Bondi or Hill radius of the planet even when we varied $\beta$ by an order of magnitude (\Figref{fig:Ratm_beta_2panel}a).} The atmospheric radius has a peak when $\beta\sim1\text{--}10$ (\Figref{fig:Ratm_beta_2panel}b). \revsix{We found that $R_{\rm atm}$ is not scaled with $R_{\rm Bondi}$ or $R_{\rm Hill}$ when $\beta=0.01\,\beta_0$ and $100\,\beta_0$. These endmember cases approach the isothermal and adiabatic conditions, where the bound atmosphere would disappear. Thus, in the limits of $\beta\rightarrow0$ and $\beta\rightarrow\infty$, there is no guarantee that $R_{\rm atm}$ will scale with $R_{\rm Bondi}$ or $R_{\rm Hill}$.}

The outflow speed in the $x$-direction depends on $\beta$. In \Figref{fig:vx_beta_dependence_m02}, where we set $m=0.2,\,\mathcal{M}_{\rm hw}=0$, and $\beta_0=4$, \revthi{the perturbation to $v_{x,{\rm g}}$} is the most prominent when $\beta=0.1\beta_0=0.4$. We found that the width of the horseshoe streamline would positively correlate with the atmospheric radius.
Our fitting formula for the \rev{critical recycling streamline at the midplane}, \revthi{$f_{\rm out}(x)$} (\Equref{eq:outflow shape in cartesian}), cannot apply for the endmember cases, $\beta\ll\beta_0$ and $\beta\gg\beta_0$ (\Figref{fig:vx_beta_dependence_m02}). From a series of hydrodynamical simulations, we found that \revthi{$f_{\rm out}(x)$} and \revthi{$v_{x,{\rm out}}$} (\Equref{eq:v_x,out}) can be used when the moderate cooling time was considered, $\beta\sim10^{-1}\text{--}10^{1}$. 
\revsev{Again, the $\beta$ value could be changed by orders of magnitude. Thus, }
further studies are needed to include the dependence of gas dynamics on the cooling time, such as the atmospheric radius, the width of the horseshoe region, and the outflow speed.

\subsection{Angular momentum of the gas}\label{sec:Angular momentum of the gas}
Figures \ref{fig:B_m005}a and \ref{fig:B_m02}a show the changes of the $x$-coordinate of the gas along the recycling streamline passing outside the planetary orbit. We found that the $x$-coordinates of the streamline at the edges of the calculation domain of the hydrodynamical simulation do not match each other, implying that the angular momentum of the gas along the recycling streamline increases after the gravitational interaction with the planet. The detailed investigation of the net torque on the planet is 
beyond the scope of this study, but this result indicates that the gas flow around the planet could affect the migration rate of the planet as suggested by \cite{Fung:2015}. Since our hydrodynamical simulations were performed in the local domain, our results may need to be compared with the results obtained by global hydrodynamical simulations.

\subsection{\revsec{Influence of the outflow on the growth of protoplanets}}\label{sec:Influence of the outflow on the growth of protoplanets}
Based on \revsec{the analytic formula for the outflow speed (\Equref{eq:v_x,out}), here we} discuss the \revthi{influence} of the outflow on the \revsev{growth of protoplanets}. \revsec{In this section, we \revthi{assume} that dust settle to the midplane for simplicity. \revthi{More quantitatively, we assume that} the vertical extent of the outflow is larger than the dust scale height. The vertical extent of the outflow can be estimated by the atmospheric radius, $R_{\rm atm}\lesssim0.2\,H$ (Sect. \ref{sec:Vertical direction to the disk}). The dust scale height is given by \citep{Youdin:2007}:
\begin{align}
    H_{\rm d}=\Biggl(1+\frac{\rm St}{\alpha_{\rm diff}}\frac{1+2{\rm St}}{1+{\rm St}}\Biggr)^{-1/2}\,H,\label{eq:dust scale height}
\end{align}
where $\alpha_{\rm diff}$ is the dimensionless turbulent parameter. Equation (\ref{eq:dust scale height}) gives $H_{\rm d}\lesssim0.3\,H$ when ${\rm St}\gtrsim10^{-3}$ and $\alpha_{\rm diff}\lesssim10^{-4}$ were assumed, which means that, under weak turbulence, the outflow has a potential to affect the dust motion.}

Following \cite{Kuwahara:2019}, we \revthi{compare} \revthi{$v_{x,{\rm out}}$ (\Equref{eq:v_x,out})} with the terminal speed of a dust particle at a distance of $\rev{r}=2\,R_{\rm atm,fit}$ from the planet, where the dominant outflow of the gas occurs. Given the force balance between the gas drag and the gravity of the planet acting on the dust particle at $\rev{r}=2\,R_{\rm atm,fit}$, the terminal speed of the dust particle can be described by: 
\begin{align}
    v_{\rm term}=\frac{m\,{\rm St}}{4\,R_{\rm atm,fit}^2},\label{eq:terminal speed of dust}
\end{align}
in our dimensionless unit, where ${\rm St}$ is the Stokes number (the dimensionless stopping time) of dust. We plotted \Equref{eq:terminal speed of dust} with different Stokes numbers, ${\rm St}=10^{-3},\,10^{-2},\,\text{and }10^{-1}$ in \Figref{fig:vx_with_analytic_vpeb} and compared the terminal speed of the dust particle with the outflow speed. The outflow speed in the $x$-direction can exceed the terminal speed of the dust particle.

We \revthi{derive} the upper limit of the Stokes number of dust that \revsix{can be} affected by the outflow by the order-of-magnitude estimate. For simplicity, we \revthi{neglect} the effect of the headwind of the gas, \revsix{$\mathcal{M}_{\rm hw}=0$,} and only \revsix{consider} $x\geq0$. 
From Eqs. (\ref{eq:v_x,out}) and (\ref{eq:terminal speed of dust}), the outflow affects the motion of dust when \revsec{the Stokes number falls below the critical Stokes number:}
\revsec{
\begin{align}
    {\rm St}\lesssim\frac{8\sqrt3\,R_{\rm atm,fit}^3}{m}\cos\varphi_{\rm out}^+\cos\left(\frac{2\,R_{\rm atm,fit}}{w_{\rm out}^+}\right)\equiv{\rm St}_{\rm crit}.\label{eq:St criteria}
\end{align}
We plotted \Equref{eq:St criteria} in \Figref{fig:Stcrit}.}

When \revsix{$m\lesssim0.1\,(R_{\rm atm,fit}=0.84\,R_{\rm Bomdi})$}, the product of the cosine terms in the right-hand side of \Equref{eq:St criteria} has \revsix{$\sim0.6\text{--}1$} and we obtained:
\revsec{
\begin{align}
    {\rm St}\lesssim8\sqrt3\,m^2\times[\cos \text{terms}]\sim 10m^2.
\end{align}
}When \revsix{$m\gtrsim0.1\,(R_{\rm atm,fit}=0.22\,R_{\rm Hill})$}, the product of the cosine terms in the right-hand side of \Equref{eq:St criteria} has \revsix{$\sim0.3\text{--}0.6$} and we obtained:
\revsec{
\begin{align}
    {\rm St}\lesssim\frac{8\sqrt3}{3}\times0.4^3\times[\cos \text{terms}]\sim0.1.
\end{align}
}In summary, the outflow could affect the dust motion and suppress the dust accretion onto the planet when ${\rm St}\lesssim\min(10m^2,\,0.1)$. \revsec{This is consistent with the post-processing \revthi{simulation} results where the obtained hydrodynamical simulation data to calculate the dust trajectories \citep{Kuwahara:2020a,Kuwahara:2020b,okamura2021growth}.}


\subsection{Dust velocity influenced by the planet-induced gas flow}\label{sec:Dust velocity influenced by the planet-induced gas flow}
We expect that \revsec{the analytic formulae for the distribution of the $x$-component of the perturbed flow velocity in the radial (\revthi{$v_{x,{\rm out}}(x)$;} \Equref{eq:vx_g_Gaussian}) and vertical (\revthi{$v_{x,{\rm out}}(z)$;} \Equref{eq:vx_gas_z_fit}) directions to the disk can be used for modeling of the dust velocity influenced by the planet-induced gas flow.} \rev{Performing hydrodynamical simulations is a direct way to investigate \revthi{the influence of gas flow on dust motion}, but computationally expensive. Our analytic formulae would allow us to perform an extensive parameter study of the hydrodynamic effect of the outflow on the dust motion.} 
The above attempts for modeling dust velocity will be included in a future study.

\section{Conclusions} \label{sec:Conclusions}
\rev{We introduced the fitting formulae that describe the morphology of gas dynamics around an embedded planet, by performing 3D, nonisothermal hydrodynamical simulations. We then derived an analytic formula for the planet-induced midplane outflow using Bernoulli's theorem.} Our main results are as follows.
\begin{enumerate}
    \item A bound atmosphere forms within the gravitational sphere of the planet, whose size scales with the Bondi or Hill radii of the planet. \revsix{We derived the dimensionless fitting formula for the atmospheric radius, $R_{\rm atm,fit}=\min(0.84\,R_{\rm Bondi}-0.056\,\mathcal{M}_{\rm hw},0.36\,R_{\rm Hill}-0.22\,\mathcal{M}_{\rm hw})$}. Gas from the disk chiefly flows in and out of the twice the atmospheric radius.
    \item The outflow of the gas occurs at the midplane. \revsec{We found that the morphology of the outflow (the shape of the streamline) can be outlined by a simple tangential curve.} 
    \item Based on the obtained quantitative descriptions of the outflow \rev{morphology, we derived} an analytic formula for the outflow speed which is in agreement with the results of hydrodynamical simulations. The outflow speed in the radial direction to the disk has a peak speed of $\sim0.3\text{--}0.4\,c_{\rm s}$ at $m\sim0.3$, corresponding to a super-Earth mass planet at $1$ au for a typical steady accretion disk model. The global pressure force of the disk gas enhances (reduces) the outflow toward the outside (inside) of the planetary orbit, \rev{causing the asymmetry of the outflow speed inside and outside the planetary orbit.}
\end{enumerate}
The outflow of the gas could suppress pebble accretion when ${\rm St}\lesssim\min(10m^2,0.1)$. Our analytic formulae \revthi{can} be used for \rev{\revthi{modeling} the hydrodynamic effects of the outflow on the dust accretion onto the planet and the radial drift of dust,} which could be essential for planet formation via pebble accretion and \rev{the dust ring and gap formation in a disk}.

\begin{acknowledgements}
\revsix{We would like to thank the reviewer, Dr. Ondrej Chrenko for thoughtful comments that have substantially improved the quality of this manuscript.} We thank Athena++ developers. Numerical computations were in part carried out on Cray XC50 at the Center for Computational Astrophysics at the National Astronomical Observatory of Japan. This work was supported by JSPS KAKENHI Grant number 20KK0080, 22H01290, 22H05150, 21H04514, and 21K13976.
\end{acknowledgements}



\newpage
\appendix

\def\thesection{A}
\setcounter{equation}{0}
\def\theequation{A.\arabic{equation}}
\setcounter{figure}{0}
\def\thefigure{A.\arabic{figure}}

\section{Aspect ratio of the steady accretion disk}\label{sec:Aspect ratio of the steady accretion disk}
\revsix{When we convert the dimensionless quantities into dimensional ones, we considered the typical steady accretion disk model with a constant turbulence strength \citep{Shakura:1973}, $\alpha_{\rm acc}$, including viscous heating due to the accretion of the gas and irradiation heating from the central star \citep[e.g.,][]{Ida:2016}.
The aspect ratio of the disk is given by
\begin{align}
    h\equiv\frac{H}{r}=\max(h_{\rm g,vis},\,h_{\rm g,irr}),
\end{align}
where $h_{\rm g,vis}$ and $h_{\rm g,irr}$ are aspect ratios given by \citep{Ida:2016}:
\begin{align}
    h_{\rm g,vis}\simeq&0.027\,\Biggl(\frac{M_{\ast}}{1\,M_{\odot}}\Biggr)^{-7/20}\Biggl(\frac{\alpha_{\rm acc}}{10^{-3}}\Biggr)^{-1/10}\Biggl(\frac{\dot{M}_{\ast}}{10^{-8}\,M_{\odot}/\text{yr}}\Biggr)^{1/5}\Biggl(\frac{a}{1\,\text{au}}\Biggr)^{1/20},\\
    h_{\rm g,irr}\simeq&0.024\,\Biggl(\frac{L_{\ast}}{1\,L_{\odot}}\Biggr)^{1/7}\Biggl(\frac{M_{\ast}}{1\,M_{\odot}}\Biggr)^{-4/7}\Biggl(\frac{a}{1\,\text{au}}\Biggr)^{2/7}.
\end{align} 
In this study, we assume a solar-mass host star, $M_{\ast}=1\,M_{\odot}$, the solar luminosity, $L_{\ast}=1\,L_{\odot}$,  the typical accretion rate of classical T Tauri stars, $\dot{M}_{\ast}=10^{-8}\,M_{\odot}/\text{yr}$, and $\alpha_{\rm acc}=10^{-3}$.
}


\def\thesection{B}
\setcounter{equation}{0}
\def\theequation{B.\arabic{equation}}
\setcounter{figure}{0}
\def\thefigure{B.\arabic{figure}}

\section{Dependence on inner boundary}\label{sec:Dependence on inner boundary}

\begin{figure}[htbp]
    \centering
    \includegraphics[width=\linewidth]{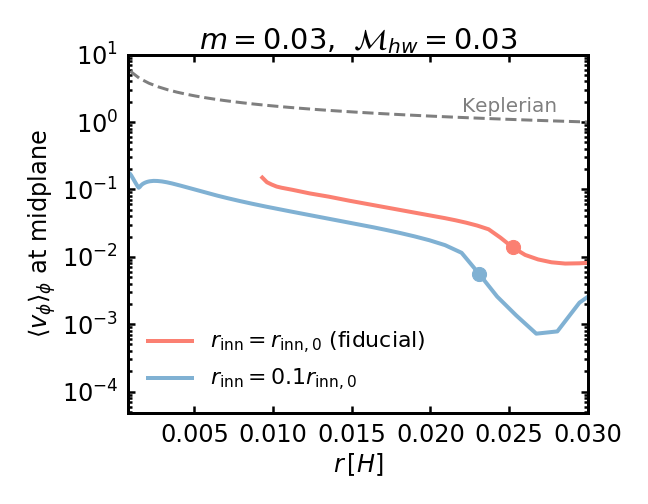}
    \caption{\revsec{Azimuthal gas velocity at the midplane averaged in the azimuthal direction in the spherical polar coordinates centered at the planet. The results were obtained from \texttt{m003-hw003} and \texttt{m003-hw003-01rinn}. Different colors corresponds to different sizes of the inner boundary. The filled circle denotes the atmospheric radius. The dashed line shows the Keplerian speed.}}
    \label{fig:averaged_vphi_rinn}
\end{figure}

\begin{figure}[htbp]
    \centering
    \includegraphics[width=\linewidth]{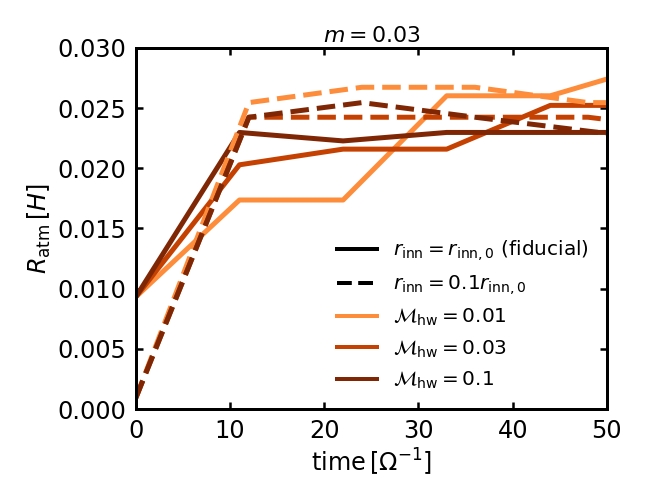}
    \caption{\revsec{Time evolution of the atmospheric radius for different sizes of the inner boundary, $r_{\rm inn}=r_{\rm inn,0}$ (the solid lines) and $r_{\rm inn}=0.1r_{\rm inn,0}$ (the dashed lines). Different colors correspond to different Mach numbers. We set $m=0.03$ and $r_{\rm sm}=0.1R_{\rm grav}$.}}
    \label{fig:Ratm_time_evol_rinn}
\end{figure}

\revsec{In this section, we investigate the influence of the size of the inner boundary (corresponding to the planet location in this study; \Equref{eq:inner boundary}), $r_{\rm inn}$, on the atmospheric radius. We confirmed that the size of the inner boundary does not affect our results. Previous studies reported that the azimuthal gas velocity in the spherical polar coordinates centered at the planet depends on $r_{\rm inn}$ \citep{Ormel:2015b,Bethune:2019,takaoka2023spin}. We also found that the azimuthal gas velocity varies with $r_{\rm inn}$ (\Figref{fig:averaged_vphi_rinn}). Nevertheless, the atmospheric radius did not change significantly. When we changed $r_{\rm inn}$ by an order of magnitude, the relative error between the obtained atmospheric radii with $r_{\rm inn}=r_{\rm inn,0}$ and $0.1r_{\rm inn,0}$ was small: $\lesssim10\%$ (\Figref{fig:Ratm_time_evol_rinn}). This allows the orbital radius to be considered a free parameter regardless of a fixed $r_{\rm inn}$.}


\def\thesection{C}
\setcounter{equation}{0}
\def\theequation{C.\arabic{equation}}
\setcounter{figure}{0}
\def\thefigure{C.\arabic{figure}}

\section{Dependence on smoothing length}\label{sec:Dependence on smoothing length}

\begin{figure}[htbp]
    \centering
    \includegraphics[width=\linewidth]{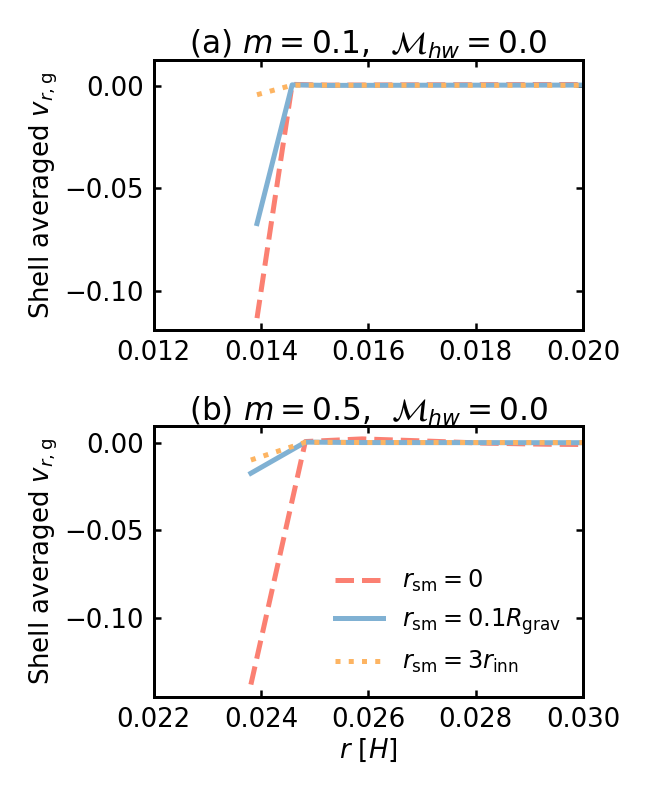}
    \caption{Shell averaged radial velocity of the gas in the spherical polar coordinates centered at the planet for different smoothing lengths, $r_{\rm sm}=0$ (the red dashed line), $r_{\rm sm}=0.1\,R_{\rm grav}$ (the blue solid line), and $r_{\rm sm}=3\,r_{\rm inn}$ (the yellow dotted line). We assumed $m=0.1$ in \textit{panel a} and $m=0.5$ in \textit{panel b}. We set $\mathcal{M}_{\rm hw}=0$.}
    \label{fig:shell_average_vr_2panel}
\end{figure}

\begin{figure}[htbp]
    \centering
    \includegraphics[width=\linewidth]{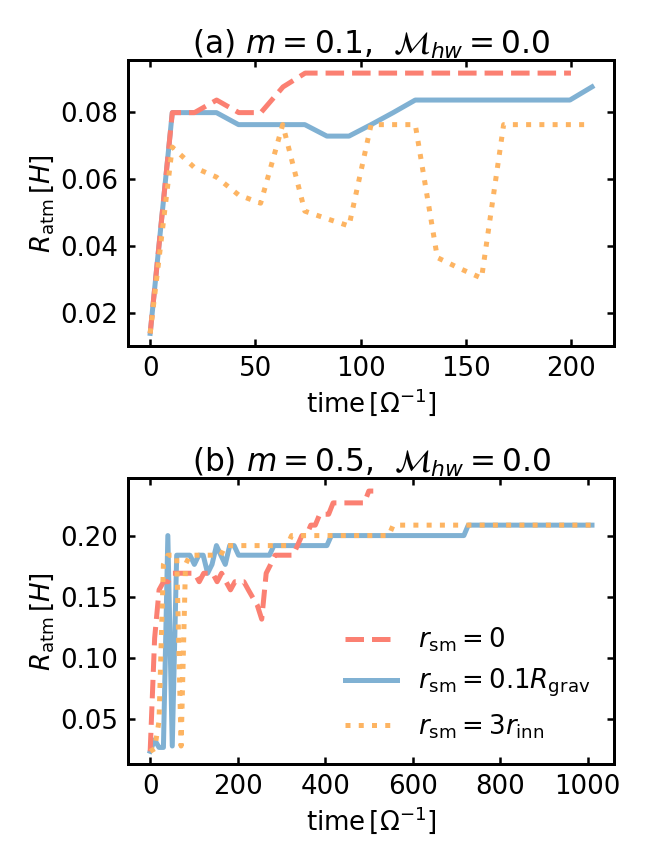}
    \caption{Time evolution of the atmospheric radius for different smoothing lengths, $r_{\rm sm}=0$ (the red dashed line), $r_{\rm sm}=0.1\,R_{\rm grav}$ (the blue solid line), and $r_{\rm sm}=3\,r_{\rm inn}$ (the yellow dotted line). We assumed $m=0.1$ and $\mathcal{M}_{\rm hw}=0$ in \textit{panel a} and $m=0.5$ and $\mathcal{M}_{\rm hw}=0$ in \textit{panel b}.}
    \label{fig:Ratm_time_evol_rsm_2panel}
\end{figure}

\begin{figure}[htbp]
    \centering
    \includegraphics[width=\linewidth]{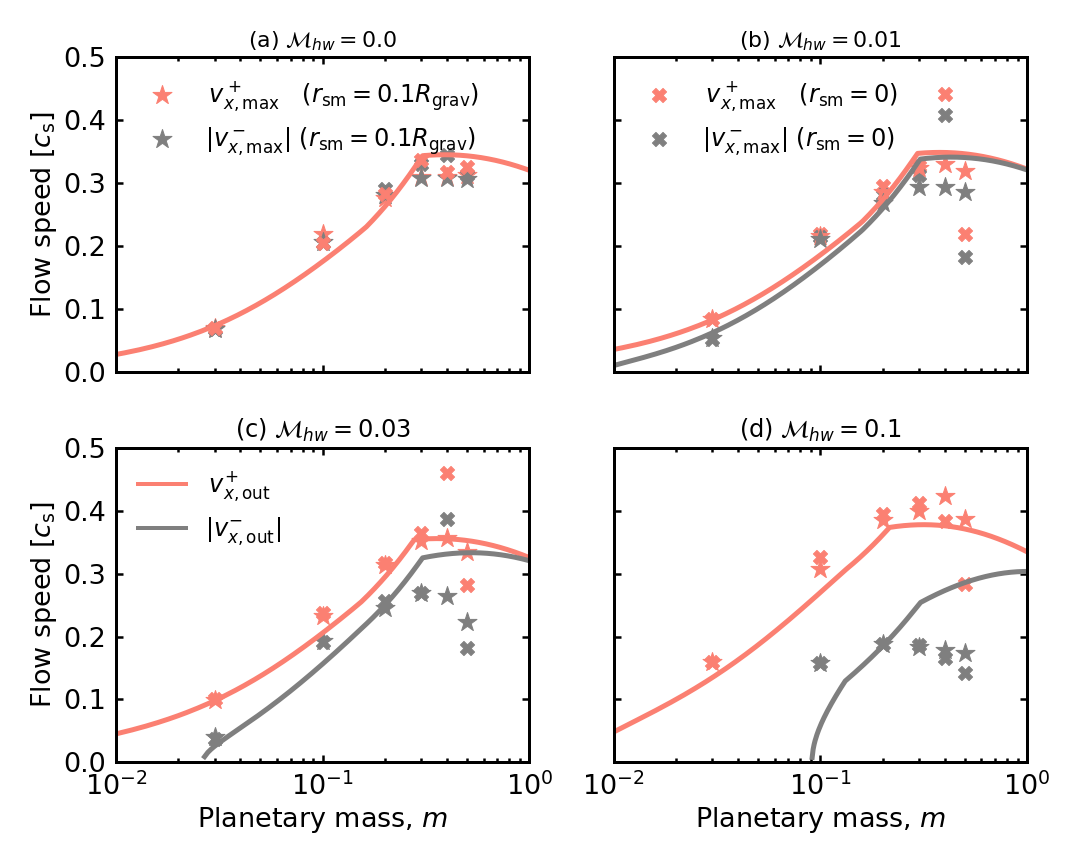}
    \caption{Same as \Figref{fig:vx_with_analytic}, but this figure compares the results obtianed from hydrodynamical simulations with $r_{\rm sm}=0$ (the cross symbols) and $r_{\rm sm}=0.1\,R_{\rm grav}$ (the star symbols).}
    \label{fig:vx_with_analytic_rsm}
\end{figure}

In this section, we investigate the influence of the smoothing length on gas dynamics within the gravitational sphere of the planet, the atmospheric radius, and the outflow speed. 

It is preferable to run the simulations with $r_{\rm sm}=0$ to resolve the local gas flow around the planet. However, through the simulations with $r_{\rm sm}=0$, we found unsteady and unphysical flow pattern within the atmosphere. The simulations were interrupted when we performed long-term simulations, $t\gtrsim500\,\Omega^{-1}$, in particular for higher-mass planets, $m\gtrsim0.3$. To investigate this problem, we plot the shell averaged radial velocity in \Figref{fig:shell_average_vr_2panel}. This figure show that the small but nonzero inward radial gas flow occurs at the region close to the inner boundary, meaning that the hydrostatic equilibrium is not established in this region. The density of the gas increases with time due to the continuous accumulation of the gas in the inner region of the \revsev{envelope}, leading to the numerical errors.

There is a clear correlation between the shell averaged radial gas flow \revthi{near the inner boundary} and the smoothing length. \revsec{In our parameter sets, we always have the following relation: $0.1\,R_{\rm grav}<3\,r_{\rm inn}$.} The larger the smoothing length, the smaller the radial gas velocity at the region close to the inner boundary. This can be seen in previous study and would disappear in a higher-resolution simulation \citep[e.g., Fig. A.1 of][]{Ormel:2015b}.

Figure \ref{fig:Ratm_time_evol_rsm_2panel} shows the time evolution of the atmospheric radius for different smoothing lengths, $r_{\rm sm}=0,\,0.1\,R_{\rm grav},$ and $3\,r_{\rm inn}$. In \Figref{fig:Ratm_time_evol_rsm_2panel}a, where we set $m=0.1$, we found the fluctuation of the atmospheric radius when we adopt a relatively large smoothing length, $r_{\rm sm}=3\,r_{\rm inn}$, corresponding to $\sim0.4\,R_{\rm Bondi}$. The fluctuation does not occur when $r_{\rm sm}=0.1\,R_{\rm grav}=0.1\,R_{\rm Bondi}$. For a higher-mass planet, $m=0.5$, only when we adopt the nonzero smoothing length, the results seem to reach the steady state (\Figref{fig:Ratm_time_evol_rsm_2panel}b).

The choice of the smoothing length does not affect significantly the outflow speed when $m\lesssim0.3$. Figure \ref{fig:vx_with_analytic_rsm} compares the outflow speed in the $x$-direction obtained from hydrodynamical simulations with different smoothing lengths, $r_{\rm sm}=0$ and $0.1\,R_{\rm grav}$.  When $m\lesssim0.3$, the obtained outflow speeds are overlapped. For higher-mass planets, $m\gtrsim0.3$, the outflow speed obtained from hydrodynamical simulations with $r_{\rm sm}=0$ does not match with those obtained with $r_{\rm sm}=0.1\,R_{\rm grav}$ due to the unphysical flow pattern within the atmosphere.

Based on Figs. \ref{fig:shell_average_vr_2panel}--\ref{fig:vx_with_analytic_rsm}, we adopt $r_{\rm sm}=0.1\,R_{\rm grav}$ in our fiducial runs. It should be noted that even when we adopt the nonzero smoothing length, the inward radial gas flow does not completely disappear, indicating that the unsteady and unphysical flow pattern would eventually appear when $t>10^3\,\Omega^{-1}$.

\def\thesection{D}
\setcounter{equation}{0}
\def\theequation{D.\arabic{equation}}
\setcounter{figure}{0}
\def\thefigure{D.\arabic{figure}}

\section{Interpolation of the gas velocity}\label{sec:Interpolation of the gas velocity}
\revsix{We interpolated the gas velocity to plot the streamlines \revsec{in} the local Cartesian coordinates \revsec{(e.g., the thin gray lines in \Figref{fig:vx_m_dependence})} by using the \texttt{matplotlib.pyplot.streamplot} \revsev{library}. Our hydrodynamical simulations were performed in the spherical polar coordinates, $(r,\,\theta,\,\phi)$, centered on the planet.} 

\revsev{Hereafter we only consider the midplane. We set the local Cartesian coordinates, $(x,y)$, where the following relations hold: $r=\sqrt{x^2+y^2}$ and $\phi=\tan^{-1}(y/x)$. We divided the local Cartesian coordinates into $N$ grids in the $x$- and $y$-directions with equal spacing within the view range. We set $N=50$. We confirmed that the result does not change significantly when we doubled the value of $N$. Then, we \reveig{calculated the gas velocity at each Cartesian grid point with the bilinear interpolation method. The interpolated gas velocity at an arbitrary point, $(x,y)$, is given by:
\begin{align}
    v=(1-s)(1-t)v_{i,j}+s(1-t)v_{i+1,j}+stv_{i+1,j+1}+(1-s)tv_{i,j+1},\label{eq:interpolation}
\end{align}
where $v_{i,j,k}$ is the gas velocity at the center of the grid of hydrodynamical simulations}, the subscripts denote the grid number, and $s$ and $t$ are given by 
\begin{align}
    &s=\frac{r-r_i}{r_{i+1}-r_i},\label{eq:s}\\
    &t=\frac{\phi-\phi_j}{\phi_{j+1}-\phi_j}\label{eq:t}.
\end{align}
In Eqs. (\ref{eq:s}) and (\ref{eq:t}), $i$ and $j$ were chosen to satisfy the following relations: $r_i\leq r<r_{i+1}$  and $\phi_j\leq\phi<\phi_{j+1}$.}

\revsev{The number of streamlines depends on the value of the ‘density’ parameter in the \texttt{matplotlib.pyplot.streamplot} library. The default value is '\texttt{density=1}'. The number of streamlines increases as the value of the ‘density’ parameter increases. To show the detailed structure of the gas flow, we set ‘\texttt{density=2}’ in Figs. \ref{fig:vx_m_dependence}, \ref{fig:vx_Mhw_dependence}, and \ref{fig:vx_beta_dependence_m02}.}

\def\thesection{E}
\setcounter{equation}{0}
\def\theequation{E.\arabic{equation}}
\setcounter{figure}{0}
\def\thefigure{E.\arabic{figure}}

\section{Widest horseshoe streamline and the shear streamline passing closest to the planet}\label{sec:Widest horseshoe streamline and the shear streamline passing closest to the planet}
\reveig{We calculated the widest horseshoe streamline and the streamline of the background shearing flow passing closest to the planet with the bisection method and plotted them in Figs. \ref{fig:vx_m_dependence}, \ref{fig:vx_Mhw_dependence}, \ref{fig:horizontal_m01_m05}, \ref{fig:vx_max_point}, and \ref{fig:vx_beta_dependence_m02} \reveig{without using the \texttt{matplotlib.pyplot.streamplot} library}. The following sections describe the dedicated calculation method for these characteristic streamlines and how to determine the boundary between the widest horseshoe streamline and the shear streamline passing closest to the planet.}

\subsection{\reveig{Dedicated calculation method for the characteristic streamlines}}\label{sec:Streamline calculation}
\reveig{This section describes the dedicated calculation method for the two characteristic streamlines: the widest horseshoe streamline and the streamline of the background shearing flow passing closest to the planet.} \revsev{\reveig{A fifth-order Runge-Kutta-Fehlberg method was used to calculate the streamlines.} We set the starting point of the streamline, $(x_0,y_0)$, at the edge of the calculation domain of hydrodynamical simulation. The $x$-coordinate of the starting point, $x_0$, is a parameter. The $y$-coordinate of the starting point is given by:}
\begin{empheq}
    [left={y_0=\empheqlbrace}]{alignat=2}
    &\sqrt{r_{\rm out}^2-x_0^2}\quad(x_0\geq x_{\rm cor,g}),\\
    &-\sqrt{r_{\rm out}^2-x_0^2}\quad(x_0< x_{\rm cor,g}).
\end{empheq}
\reveig{The position of the gas parcel at the $i$-th time step, $(x_i,y_i)$, was calculated by the fifth-order Runge-Kutta-Fehlberg scheme, where the gas velocity at an arbitrary position was calculated by \Equref{eq:interpolation}. Following \cite{Kuwahara:2020a}, the maximum relative error tolerance was set to $10^{-8}$, where the authors adopted the same scheme described in this section to compute the trajectories of dust particles influenced by the planet-induced gas flow.}

\revsev{\reveig{We terminated the above-mentioned streamline calculation when one of the following conditions was met: (1) the gas parcel reached the edge of the computational domain of hydrodynamical simulations,} $\sqrt{x_i^2+y_i^2}\geq r_{\rm out}$ $(i=1,2,\dots)$, or (2) the gas parcel reached a stagnation point. We judged that a parcel of the gas reached the stagnation point when 
\begin{align}
    \frac{|v^{i+1}-v^i|}{\delta t}\leq\epsilon.\label{eq:stagnation criterion}
\end{align}
\reveig{In \Equref{eq:stagnation criterion}, $v^i=\sqrt{v_{x,i}^2+v_{y,i}^2}$ is the gas speed at $(x_i, y_i)$ where the $x$- and $y$-components of the gas velocity, $v_{x,i}$ and $v_{y,i}$ are calculated by \Equref{eq:interpolation} and we set $\epsilon=10^{-10}$.}}

\subsection{Determination of the boundary between the widest horseshoe streamline and the shear streamline passing closest to the planet}\label{sec:Determination of the boundary between the widest horseshoe streamline and the shear streamline passing closest to the planet}
\revsev{Based on the \reveig{dedicated calculation method for the characteristic streamlines} described in Appendix \ref{sec:Streamline calculation}, we determined the boundary between the widest horseshoe streamline and the shear streamline passing closest to the planet \reveig{with the bisection method}. We calculated the streamlines with the initial spatial intervals of $\delta x_0=10^{-2}\,H$ in the $x$-direction. \reveig{ We denote $x_0^j$ and $x_{\rm e}^j$ as the $x $-coordinates of the starting and end points of the $j$-th streamline, respectively.} We judged that the obtained streamline was classified as a horseshoe streamline when $x_{\rm e}^j< x_{\rm cor,g}$ and $x_0^j>x_{\rm cor,g}$, or $x_{\rm e}^j> x_{\rm cor,g}$ and $x_0^j<x_{\rm cor,g}$, where $x_{\rm e}^j$ was the $x$ -coordinate of the $j$-th streamline at the edge of the computational domain of hydrodynamical simulation, otherwise classified as a shear streamline.}

\reveig{When the $j$-th and $(j+1)$-th streamlines were determined to be the shear and horseshoe streamlines, respectively, we set $x_0^j\equiv a_0$, $x_0^{j+1}\equiv b_0$, and $c_0\equiv (a_0+b_0)/2$. We calculated a streamline whose $x$-coordinate of the starting point was $c_0$, and then set $a_1\equiv c_0$, $b_1\equiv b_0$ ($b_1\equiv c_0$, $a_1\equiv a_0$), and $c_1\equiv (a_1+b_1)/2$ when the streamline was determined to be a shear (horseshoe) streamline. We repeated the above steps until the following condition was satisfied: $|a_k-b_k|<0.01\,\delta x_0=10^{-4}\,H$ ($k=0,1,\dots$). }


\def\thesection{F}
\setcounter{equation}{0}
\def\theequation{F.\arabic{equation}}
\setcounter{figure}{0}
\def\thefigure{F.\arabic{figure}}

\section{Numerical convergence}\label{sec:Numerical convergence}
\begin{figure}[htbp]
    \centering
    \includegraphics[width=\linewidth]{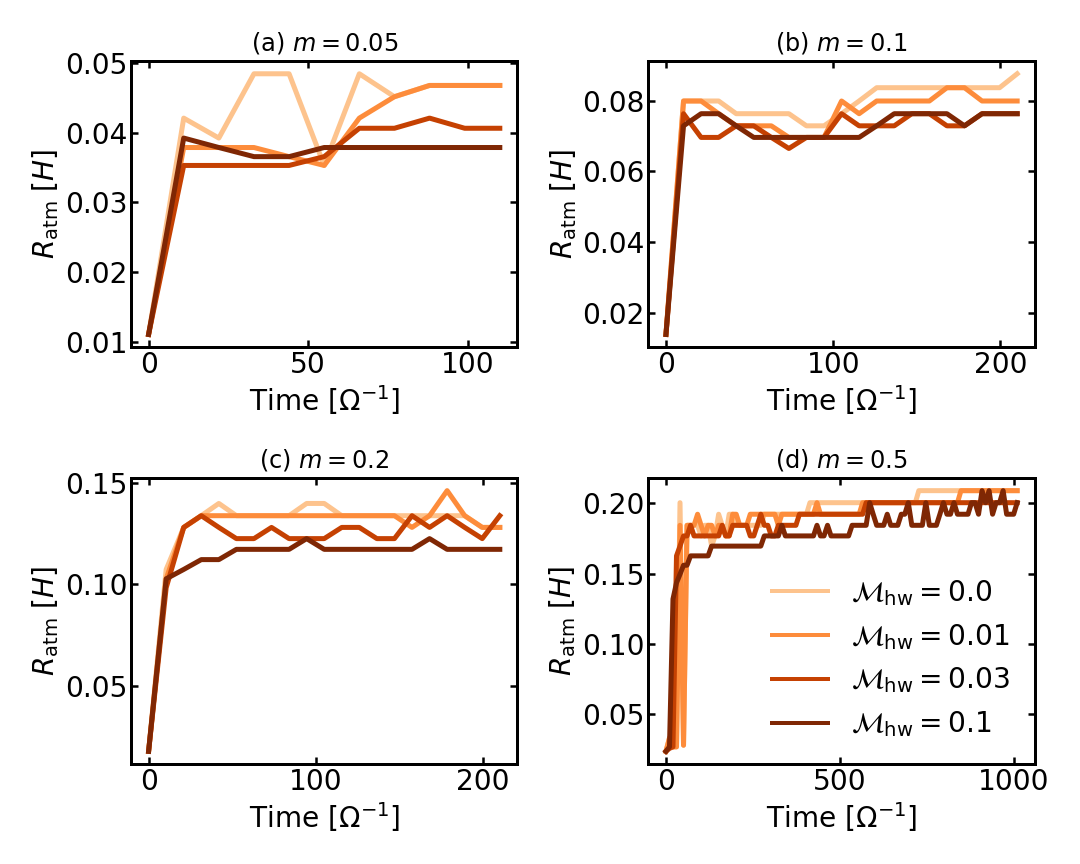}
    \caption{Time evolution of the atmospheric radius for different planetary masses, $m=0.05$ (\textit{panel a}), $m=0.1$ (\textit{panel b}), $m=0.2$ (\textit{panel c}), and $m=0.5$ (\textit{panel d}). Different colors correspond to different Mach numbers. We set $r_{\rm sm}=0.1\,R_{\rm grav}$ in this figure.}
    \label{fig:Ratm_time_evol_4panel}
\end{figure}

To obtain the atmospheric radius, we used the final state of the hydro-simulations data (Sect. \ref{sec:Atmospheric radius}). Figure \ref{fig:Ratm_time_evol_4panel} shows the time evolution of the atmospheric radius for different planetary masses, $m=0.05,\,0.1,\,0.2,$ and $0.5$, and Mach numbers, $\mathcal{M}_{\rm hw}=0\text{--}0.1$. We set $r_{\rm sm}=0.1\,R_{\rm grav}$. The atmospheric radius increases in the early stage of the time evolution and reaches a steady state after $t\gtrsim30\text{--}100\,\Omega^{-1}$.

\def\thesection{G}
\setcounter{equation}{0}
\def\theequation{G.\arabic{equation}}
\setcounter{figure}{0}
\def\thefigure{G.\arabic{figure}}

\section{Numerically-calculated widths of the horseshoe and outflow region\revsev{s}}\label{sec:Numerically calculated widths of the horseshoe and outflow region}

\begin{figure}[htbp]
    \centering
    \includegraphics[width=\linewidth]{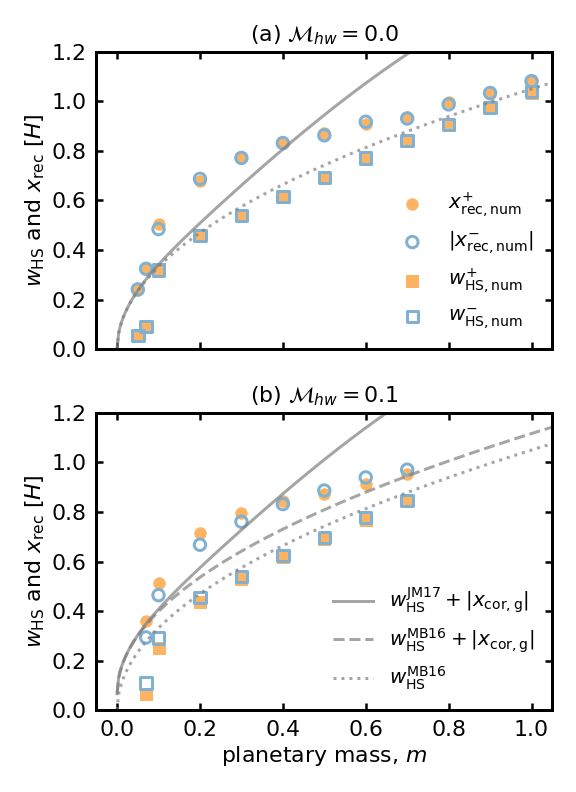}
    \caption{\revsev{Width of the horseshoe flow and the $x$-coordinate of the critical recycling streamline at the edge of the computational domain of hydrodynamical simulations. The gray solid, dotted, and dashed lines are given by the analytic formulae for the width of the horseshoe region (Eqs. (\ref{eq:w_HS MB16}) and (\ref{eq:w_HS})). The numerically-calculated width of the outflow region ($x_{\rm rec,num}$, the circle symbols) is in better agreement with the analytic estimate based on \Equref{eq:w_HS} ($w_{\rm HS}^{\rm JM17}$, the solid lines) than \Equref{eq:w_HS MB16} (the dashed line). The numerically-calculated width of the horseshoe region ($w_{\rm HS,num}$, the squares symbols) is in good agreement with \Equref{eq:w_HS MB16} ($w_{\rm HS}^{\rm MB16}$, the dotted lines).}}
    \label{fig:w_HS_x_rec_2panel}
\end{figure}

\revsev{Based on the streamline calculation method described in Appendix \ref{sec:Widest horseshoe streamline and the shear streamline passing closest to the planet}, we calculated numerically the $x$-coordinate of the critical recycling streamline at the midplane, $x_{\rm rec,num}^\pm$, and the half-width of the horseshoe region, $w_{\rm HS,num}^\pm$. We defined $x_{\rm rec,num}^\pm$ as the $x$-coordinate of the shear streamline passing closest to the planet at the edge of the computational domain of our hydrodynamical simulations. We defined the numerically-calculated half-width of the horseshoe flow as:
\begin{align}
    w_{\rm HS,num}^+=x_{\rm OL,HS}^{\rm num}+|x_{\rm cor,g}|,\\
    w_{\rm HS,num}^-=|x_{\rm IT,HS}^{\rm num}|-|x_{\rm cor,g}|,
\end{align}
where $x_{\rm OL,HS}^{\rm num}$ and $x_{\rm IT,HS}^{\rm num}$ are the $x$-coordinates of the widest outer-leading and inner-trailing horseshoe flows at \revsev{the edges of the computational domain of our hydrodynamical simulations} (the starting points of the streamlines in the first and third quadrants in the $x$-$y$ plane).
}

\revsev{Figure \ref{fig:w_HS_x_rec_2panel} compares the numerically-calculated quantities, $x_{\rm rec,num}^\pm$ and $w_{\rm HS,num}^\pm$, with the analytic formulae.
We found that $w_{\rm HS}^{\rm JM17}$ (\Equref{eq:w_HS}) is a better indicator than $w_{\rm HS}^{\rm MB16}$ (\Equref{eq:w_HS MB16}) in describing the width of the outflow region when $m\lesssim 0.6$, though a mismatch between the numerical results and the analytic estimate is prominent when $m\sim0.2$ (the circle symbols and the solid lines in \Figref{fig:w_HS_x_rec_2panel}).}

\revsev{Since our analytic model for the gas flow is not valid in the higher-mass regime ($m\gtrsim m_{\rm iso}\simeq0.6$; see Sect. \ref{sec:Morphology of the critical recycling streamline at the midplane}), here we restrict our attention to the limited range of the planetary mass, $m<0.6$. We speculate that the significant deviation between $x_{\rm rec,num}^\pm$ and $w_{\rm HS}^{\rm JM17}$ around $m\sim0.2$ is related to the dependence of $v_{x,{\rm out}}$ on the planetary mass. As shown \Figref{fig:vx_with_analytic}, $v_{x,{\rm out}}$ has a peak when $m\sim0.3$, which means that a parcel of the gas moves significantly in the $x$-direction and $x_{\rm rec,num}^{\pm}$ increases. \reveig{We consider that the deviation between $x_{\rm rec,num}^\pm$ and the analytic estimate around $m\sim0.2$ is likely caused by the existence of the recycling flow revealed by our} local 3D hydrodynamical simulations which have a higher-resolution within the gravitational sphere of the planet, $R_{\rm grav}$.}

\revsev{\reveig{The influence of the recycling flow may not be included} in both Eqs. (\ref{eq:w_HS MB16}) and (\ref{eq:w_HS}). Equation (\ref{eq:w_HS}) was based on \Equref{eq:w_HS MB16}, and \Equref{eq:w_HS MB16} was derived by a combination of linear theory and the results of 3D global hydrodynamical simulations with the moderate resolution at $\lesssim R_{\rm grav}$ \citep{Masset:2016}. Nevertheless, we keep ourselves conservative for the estimation of $x_{\rm rec}$ and assume that \Equref{eq:w_HS} is valid in deriving the width of the outflow region.}

\revsev{While the width of the outflow region is better described by $w_{\rm HS}^{\rm JM17}$ (\Equref{eq:w_HS}), we found that the half-width of the horseshoe flow agrees with $w_{\rm HS}^{\rm MB16}$ (\Equref{eq:w_HS MB16}) regardless of the assumed Mach number of the headwind, except when $m<0.1$ (the square symbols and the dotted lines in \Figref{fig:w_HS_x_rec_2panel}). When $m<0.1$, the horseshoe region might not be well established in our hydrodynamical simulations.}

\revsev{Although a discrepancy between $w_{\rm HS,num}^\pm$ and  $w_{\rm HS}^{\rm JM17}$ (\Equref{eq:w_HS}) does not affect an analytic derivation of the outflow speed, we discuss the reason of this discrepancy below. The discrepancy between $w_{\rm HS,num}^\pm$ and  $w_{\rm HS}^{\rm JM17}$ (\Equref{eq:w_HS}) could be caused by the limitation of our hydrodynamical simulations. Our local simulation cannot handle the gas-gap opening in the high-mass regime, $m\gtrsim m_{\rm iso}$. Gas-gap formation reduces the Lindblad torque, leading to an extension of the horseshoe region into the Lindblad resonance region, $\sim H$ \citep{paardekooper2009width}. Then the dependence of the horseshoe width on the planetary mass deviates from $w_{\rm HS}^{\rm MB16}\propto\sqrt{m}$.  Moreover, our local hydrodynamical simulations were performed on the spherical polar coordinates with a logarithmic grid for the radial coordinate, which has a lower resolution at the region far from the planet. We fixed the gas velocity at the outer edge of the computational domain of hydrodynamical simulations. These simulation settings could affect the width of the horseshoe flow.}

\end{document}